\documentclass[11pt, a4paper]{article}
\pdfoutput=1

\usepackage{amsmath,color,multicol}
\usepackage{amsfonts}
\usepackage{amssymb}
\usepackage{graphicx}
\usepackage{mathrsfs}
\usepackage{epsfig}
\usepackage{latexsym}
\usepackage{graphicx}
\usepackage{xcolor}
\usepackage{cite}
\usepackage{slashed, cancel}
\usepackage{hyperref}
\usepackage{datetime}
\usepackage{cancel}
\usepackage{multirow}

\def\circa#1{\,\raise.3ex\hbox{$#1$\kern-.75em\lower1ex\hbox{$\sim$}}\,}

\newcommand{\beq}{\begin{equation}}
\newcommand{\eeq}{\end{equation}}
\newcommand{\beqn}{\begin{eqnarray}}
\newcommand{\eeqn}{\end{eqnarray}}
\def\app#1#2{%
  \mathrel{%
    \setbox0=\hbox{$#1\sim$}%
    \setbox2=\hbox{%
      \rlap{\hbox{$#1\propto$}}%
      \lower1.1\ht0\box0%
    }%
    \raise0.25\ht2\box2%
  }%
}

\numberwithin{equation}{section}

\font\tenrsfs=rsfs10 at 12pt

\font\sevenrsfs=rsfs7
\font\fiversfs=rsfs5
\newfam\rsfsfam
\textfont\rsfsfam=\tenrsfs
\scriptfont\rsfsfam=\sevenrsfs
\scriptscriptfont\rsfsfam=\fiversfs
\def\mathscr#1{{\fam\rsfsfam\relax#1}}

\hypersetup{colorlinks, citecolor=bluscuro, linkcolor=black, urlcolor=bluscuro}
\definecolor{rossos}{rgb}{0.8,0.2,0.3}
\definecolor{bluscuro}{rgb}{0.15, 0.2, .85}
\definecolor{bluchiaro}{cmyk}{1,.3,0.,0.1}

\makeatother   % Cancel the effect of \makeatletter

 \def\be   {\begin{equation}}   \def\ee   {\end{equation}}
 \def\ba   {\begin{array}}      \def\ea   {\end{array}}
 \def\bea  {\begin{eqnarray}}   \def\eea  {\end{eqnarray}}
 \def\bean {\begin{eqnarray*}}  \def\eean {\end{eqnarray*}}

\begin{document}

\vspace{0.0cm}

\begin{center}

{\LARGE \textbf {
EFT triangles in the same-sign $WW$ scattering process at the HL-LHC and HE-LHC
}}
\\ [1cm]
%\\ [1.5cm]
{\large {Geetanjali Chaudhary,$^a$
Jan Kalinowski,$^{b} $
Manjit Kaur,$^a$
Pawe{\l} Koz{\'o}w,$^{bc}$
Kaur Sandeep,$^a$
Micha{\l} Szleper$^d$
and
S{\l}awomir Tkaczyk$^e$
}}
\\[1cm]
\end{center}

\hspace{1cm}
\begin{minipage}{14cm}
\textit{
\noindent 
$^a$ Panjab University, Chandigarh-160014, India \\[0.1cm]
$^b$ Faculty of Physics, University of Warsaw, ul.~Pasteura 5, 02-093 Warsaw, Poland\\[0.1cm]
$^c$ CAFPE and Departamento de F\'isica Te\'orica y del Cosmos, Universidad  \\ 
\phantom{m} de Granada, Campus de Fuentenueva, 18071 Granada, Spain\\[0.1cm]
$^d$  National Center for Nuclear Research, High Energy Physics Department,\\
\phantom{m} ul.~Pasteura 7, 02-093 Warsaw, Poland \\[0.1cm]
$^e$  Fermi National Accelerator Laboratory, Batavia, IL 60510, USA \\[0.1cm]
}
\end{minipage}

%\vspace{-0.3cm}
\vspace{0.7cm}
\begin{center}
\textbf{Abstract}
\begin{quote}
We investigate the Beyond Standard Model discovery potential in the framework of 
the Effective Field Theory (EFT) for the same-sign $WW$ scattering process in 
purely leptonic $W$ decay modes at the High-Luminosity and High-Energy phases
of the Large Hadron Collider (LHC).  
%{\MScut The commonly applied procedure of studying the $WWWW$ quartic coupling by testing dimension-8 operators of the EFT one at a time is used in this work. }
The goal of this paper is to examine the applicability of the EFT approach, with one dimension-8 operator varied at a time, to describe a hypothetical new physics signal in the $WWWW$ quartic coupling.
In the considered process there is no experimental handle on the $WW$ invariant mass, and it has previously been shown that the discovery potential at 14 TeV is rather slim.
In this paper we report the results calculated for a 27 TeV machine and compare them 
with the discovery potential obtained at 14 TeV.
We find that while the respective discovery regions shift to lower values
of the Wilson coefficients, the overall discovery potential of this
procedure does not get significantly larger with a higher beam energy.
\end{quote}
%\today
\end{center}
\vspace{10mm}

\def\thefootnote{\arabic{footnote}}
\setcounter{footnote}{0}

\section{Introduction}

The Large Hadron Collider (LHC) has completed data taking for Run II.  
While a lot of collected data still awaits to be analyzed, 
no physics Beyond the Standard Model (BSM) has been announced until now. 
 The lack of direct indications for the presence of new physics (NP) makes indirect searches more interesting. 
The High Luminosity LHC (HL-LHC) upgrade will eventually collect an integrated luminosity of 3 ab$^{-1}$ 
of data in $pp$ collisions 
at a center-of-mass (c.o.m.) energy of 14 TeV, which should maximize the LHC potential to uncover new phenomena.
It may however well be that the NP degrees of freedom are at  higher masses  making it difficult  at the LHC  to identify experimentally  
 new particles, or  new paradigms.
These considerations have been driving, in the last few years, intense activity worldwide to assess the future of 
collider experiments beyond the HL-LHC. Several proposals and studies have been performed. 
The prospects of pushing the LHC program  further with the  LHC tunnel and the whole CERN infrastructure, 
together with future magnet technology, is an exciting possibility that could push the energy up into an 
unexplored region with the 27 TeV High Energy LHC (HE-LHC), that could collect an integrated luminosity of 15 ab$^{-1}$. 

Precision measurements provide an important tool to search for heavy BSM dynamics, associated 
with mass scales beyond the LHC direct energy reach, exploiting the fact that such dynamics can still 
have an impact on processes at smaller energy, via virtual effects. In this context the well-established 
framework of effective field theories (EFTs) allows to systematically parameterize BSM effects and elucidate how 
they modify SM processes.
The BSM contributions are effectively parametrized in terms of higher
dimension operators ${\cal O}^{(n)}_i$, with some effective couplings  $C^{(n)}_i$ suppressed by appropriate 
powers of an unknown energy
scale  $\Lambda$ at which new physics sets in,

\begin{equation}
{\cal L}={\cal L}_{SM} +\Sigma_i\frac{C_i^{(6)}}{\Lambda^2}{\cal O}_i^{(6)} +\Sigma_i\frac{C_i^{(8)}}{\Lambda^4}{\cal O}_i^{(8)} +...,
\label{lagrangian}
\end{equation}

\noindent
where  the superscript $n=6,8$ indicates the dimensionality of the corresponding operator. 
 Following the usual notation, we introduce a
set of Wilson coefficients $f_i^{(n)}$, defined as

\begin{equation}
f_i^{(6)}=\frac{C_i^{(6)}}{\Lambda^2} ,  ~~~~f_i^{(8)}=\frac{C_i^{(8)}}{\Lambda^4},....,
\label{efy}
\end{equation}
which are free parameters since neither $C_i^{(n)}$ nor $\Lambda$ of  the full theory are known. 
Eq.~(\ref{lagrangian}) represents in principle an infinite and
model-independent 
%{\MScut expansion}
parametrization of possible BSM effects, valid by construction up to the cutoff 
value $\Lambda$ in the energy scale of the studied process.
%{\MScut For practical reasons, however, data analysis
%is destined to be restricted to }
The EFT approach is based on the 
%justified 
assumption that a limited number of chosen operators contribute
%{\MScut ;
%it is therefore implicitly assumed that other operators
%will not play a role for}
to the given process in the studied energy range.
In particular, vector boson scattering (VBS) processes are widely recognized as the best laboratory to
study 
%{\MScut dimension-8}
the operators which modify only the $VVVV$ quartic couplings. In the SM EFT framework these operators start at dimension-8.
%{\MScut In addition to skipping the dimension-6 ($n=6$) operators,} 
The usual ATLAS and CMS procedure in
VBS data analyses to date involves testing one such dimension-8 operator at a time.
%{\MScut Such a procedure breaks model independence of the EFT approach
%and it is interesting to investigate its physics usefulness.}
The goal of this paper is to investigate the sensitivity of this approach by identifying for each operator the corresponding reach in the $(f_i,\Lambda)$ plane at the HE-LHC.\\

In a recent paper \cite{wwpaper}  the  physics potential of the single dimension-8 operator
EFT approach has been tested  
on  a hypothetical new physics signal observed in the same-sign
$WW$ scattering process at the HL-LHC.  
The analysis was focused on  the ``gold-plated" purely leptonic $W$ decay modes:  

\begin{equation}
pp \rightarrow  2 jets+l^+\nu +l'^+\bar\nu^\prime
\label{process}
\end{equation}

\noindent
where $l$ and $l'$ stand for any combination of electrons and muons.  
In this process the $M_{WW}$ invariant mass cannot be reconstructed experimentally on the event-by-event basis, leading to a restricted space in the $(f_i, \Lambda)$ plane (the ``EFT triangle") for which the single-operator EFT description of the data is viable.
 In this note we extend such investigations to the HE-LHC energy and expected luminosity domain.

The paper is organized as follows. In Sec.~\ref{triangle} we remind the concept of  ``EFT triangles".
In Sec.~\ref{analysis}. we detail our procedure of event simulation and subsequent treatment of generated events. In Sec.~\ref{conclusion} we summarize our findings and conclude.
Three appendices contain supplementary material. In the first appendix the definitions of dimension-8 operators are recalled. In Appendix B we argue that qualitative features of the full $WW$ scattering process, including off-shell effects, can be inferred from considering the on-shell $WW$ scattering amplitudes and  we discuss  helicity amplitudes after adding the higher dimension operators. In the third  
%Two Appendices contain supplementary material. In the first appendix we argue that qualitative features of the full $WW$ scattering process, including off-shell effects, can be inferred from considering the on-shell $WW$ scattering amplitudes and  we discuss  helicity amplitudes after adding the higher dimension operators. In the second 
appendix we address the question of  what values of BSM couplings can be drawn from the discovery regions.

\section{EFT triangles}
\label{triangle}
Since the truncation of the expansion in Eq.~(\ref{lagrangian}) introduces model dependence, 
in Ref.~\cite{wwpaper} the concept of ``EFT models" has been introduced where they are defined by the choice of operators ${\cal O}_i^{(n)}$ and the  values of Wilson coefficients $f_i^{(n)}$.  
The EFT description is valid up to a cutoff energy $\Lambda$ at which new states are expected to appear; the cutoff value is unknown a priori.  However, in the presence of higher dimension operators the scattering amplitudes grow with energy and eventually break the perturbative unitarity limit $M^U$.  The condition $\Lambda < M^U(f_i)$ defines the upper bound on the range of possible values of $\Lambda$ as a function of $f_i$.
 
A specific feature
of the  process in Eq.(\ref{process})  is that the scale $M\equiv M_{WW}$, i.e., the  invariant mass of the scattered $WW$ bosons, is not experimentally
accessible, making it impossible to properly apply the
cutoff $\Lambda$ on the data. Any BSM signal, $S$, is defined as the deviation from the SM prediction in the differential  distributions $d\sigma/dx_i$
of some observable $x_i$.

\begin{equation}
\frac{dS}{dx_i}=\left(\frac{d\sigma}{dx_i}\right)^{BSM}-\left(\frac{d\sigma}{dx_i}\right)^{SM},
\label{N}
\end{equation}

Any collected data sample will in general
be a sum of the contributions from $M < \Lambda$ and $M > \Lambda$ (unless $\Lambda$
happens to be out of kinematic reach).
To cope with events with $M > \Lambda$, different solutions have been advocated in the literature, e.g.:
from discarding these events at the level of simulation, to invoking unitarization
procedures (usually assuming $\Lambda = M^U$)~\cite{Alboteanu:2008my},~\cite{Kilian:2014zja}, to ignoring the cutoff altogether.
Any of the above prescriptions is related to additional arbitrariness of choices and
therefore affects the physics interpretation of the results.
Genuine data interpretation in the EFT language requires its successful description
without any additional assumptions as to the nature of BSM physics at the scale
above $\Lambda$.
This is only possible if
the bulk of the total observed BSM signal 
originates indeed from the EFT-controlled range.

The EFT-controlled signal reads:
\begin{equation}
\left(\frac{d\sigma}{dx_i}\right)^{EFT}=\int^\Lambda_{2M_W}\left(\frac{d^2\sigma}{dx_idM}\right)^{EFT} dM 
+\int_\Lambda^{M_{max}}\left(\frac{d^2\sigma}{dx_idM}\right)^{SM} dM, 
\label{dsigma}
\end{equation}

Here $M_{max}$ is the kinematic limit of the $WW$ invariant mass. 
Eq.~(\ref{dsigma}) defines signal coming uniquely from the ``EFT model" in its range
of validity and assumes only the SM contribution in the region $M>\Lambda$. 
%}

The additional contribution from the region above $\Lambda$ may enhance the signal, but it may also 
preclude proper description of the data within the EFT.
The total BSM signal can be estimated without detailed knowledge of the 
UV completion from the expected asymptotic behavior for $M \to \infty$, i.e.,
by assuming that all the helicity amplitudes above $\Lambda$
remain constant at their respective values they reach at $\Lambda$ (hence superscript $A=const$):
\begin{equation}
\left(\frac{d\sigma}{dx_i}\right)^{BSM}=\int^\Lambda_{2M_W}\left(\frac{d^2\sigma}{dx_idM}\right)^{EFT} dM 
+\int_\Lambda^{M_{max}}\left(\frac{d^2\sigma}{dx_idM}\right)^{A=const} dM, 
\label{unitarized}
\end{equation}
%}
%

For every value of $\Lambda$, BSM observability imposes some minimum value of $f_i$
for which the total BSM signal defined by Eq.(\ref{unitarized}) has enough statistical
significance, 5$\sigma$ in our example.  Successful description in the EFT framework 
imposes some maximum value of $f_i$ such that signal
estimates computed from Eqs.(\ref{dsigma}) and (\ref{unitarized}) remain 
statistically consistent, e.g., within 2$\sigma$.  
The ``EFT triangle" is the region in the $(f_i, \Lambda)$ plane for which
a statistically significant BSM signal can be successfully described with
a chosen higher dimension operator ${\cal O}_i$.  It is bounded from three sides:
\begin{itemize}
\item from above by the unitarity limit $M^U(f_i)$, 
\item from the left by the signal significance of 5$\sigma$, computed according to Eq.~(\ref{unitarized}), 
\item and from the right by the consistency within 2$\sigma$ with Eq.~(\ref{dsigma}).
\end{itemize}

In the HL-LHC case, for all the individual dimension-8 operators that 
affect the $WWWW$ quartic coupling  
such triangles were found to be rather narrow or even entirely empty (for
${\cal O}_{S1}$) \cite{wwpaper}.
In this paper we
extend this analysis to the HE-LHC case, in an attempt to verify if
an increased beam energy and integrated luminosity will translate into larger EFT triangles.

Throughout this work we follow the MadGraph convention
for the definition of dimension-8 operators (implemented therein via public
UFO files~\cite{Eboli:2006wa}), in which the field strength tensors $W_{\mu\nu}\equiv\frac{i}{2}g\tau^i(\partial_\mu W^i_\nu - \partial_\nu W^i_\mu + 
g\epsilon_{ijk}W^j_\mu W^k_{\nu})$ are replaced
with $\hat{W}_{\mu\nu} \equiv \frac{1}{ig}W_{\mu\nu}$. Such conversion factors are equivalent to absorbing the electroweak coupling constants $g$, 
each one explicitly factored out for each occurrence of the field stress tensor, in the effective couplings $C_i$.
For the reader's convenience,  Appendix \ref{app_op8}  lists dimension-8 operators; more details can be found in  Ref.~\cite{Degrande:2013rea}.

\section{Analysis}
\label{analysis}

In this section we present a generator-level study aimed at finding the EFT triangles
for the individual $n=8$ operators at the HE-LHC.
Event samples of the process
$pp \to jj\mu^+\mu^+\nu\nu$ at 27 TeV were generated 
for each $n=8$
operator ${\cal O}_i$ that modifies the $WWWW$ quartic coupling, 
$i = S0, S1$ (so called scalar operators), $T0, T1, T2$ (transverse), and $M0, M1, M6, M7$ (mixed\footnote{$M6$ is redundant: $\mathcal{O}_{M6}=\frac{1}{2}\mathcal{O}_{M0}$; we omit this operator in further analysis.} ones). Generation has been done at LO using 
MadGraph5\_aMC@NLO v5.2.6.2 generator \cite{MadGraph},
with the appropriate UFO files
containing additional vertices involving the desired $n=8$ operators.
A scan of $f_i$ values for each operator was made using the MadGraph 
{\tt reweight} command, including $f_i=0$ to represent the SM case.
The Pythia package v6.4.1.9 \cite{Pythia6} was used for hadronization 
as well as initial and final state
radiation processes. Unitarity limits were determined using the VBFNLO \cite{VBFNLO} calculator v1.4.0,
after applying appropriate Wilson coefficients conversion factors. Cross sections at the
output of MadGraph were multiplied by a factor 4 to
account for all the lepton (electron and/or muon) combinations in the 
final state (in this work only positively charged leptons are
taken into account, although the same analysis can be done for the negative
charges).
Only signal samples were generated and the SM case was treated as irreducible
background in the study of possible BSM effects.  
%{\MScut Reducible background for this 
%process is known to be strongly
%detector dependent and was not considered in this analysis.}

\begin{figure}[h] 
  \begin{tabular}{cc}
      \includegraphics[width=0.5\linewidth]{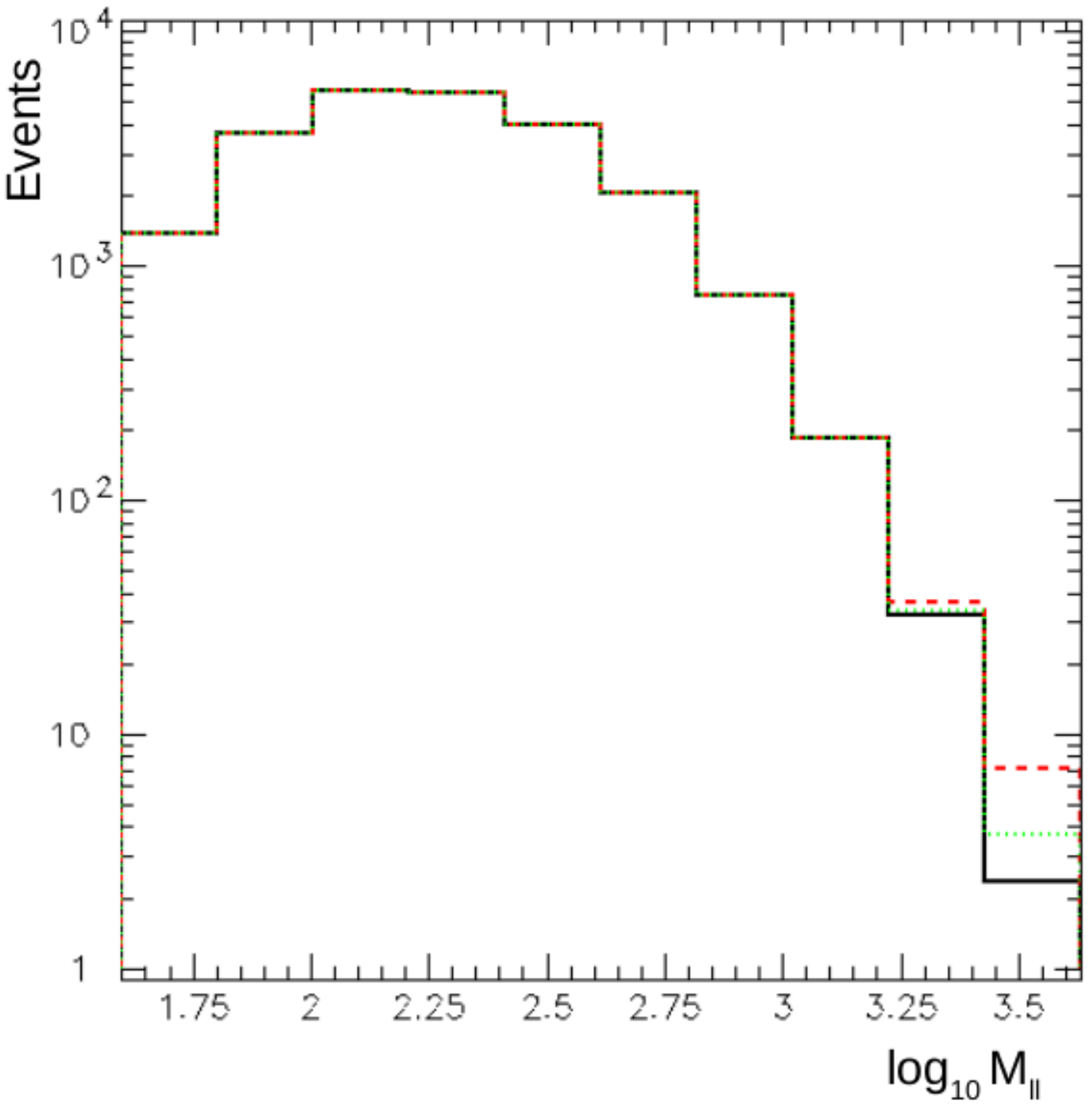}  & \includegraphics[width=0.5\linewidth]{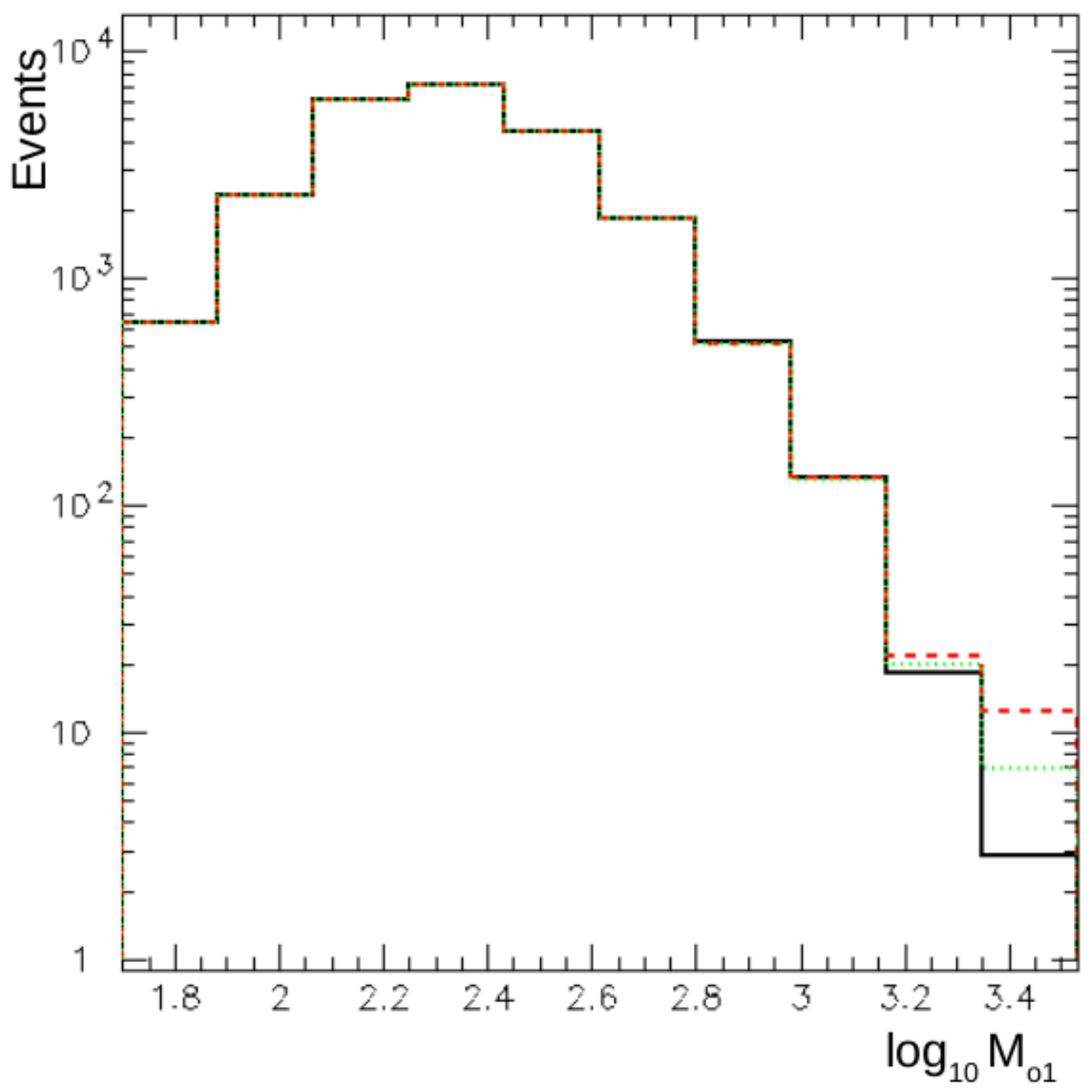} \\
\includegraphics[width=0.5\linewidth]{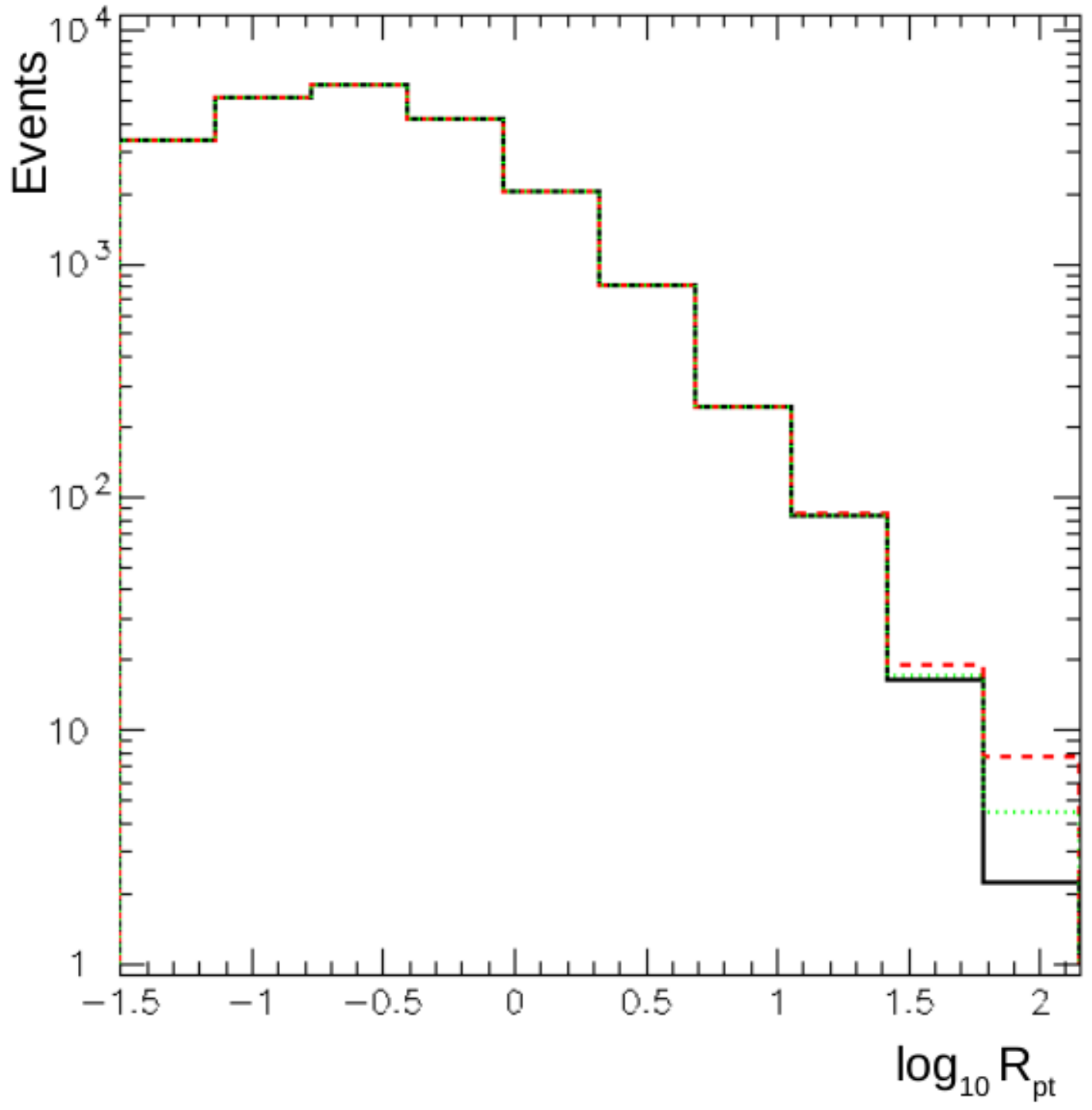} & \includegraphics[width=0.5\linewidth]{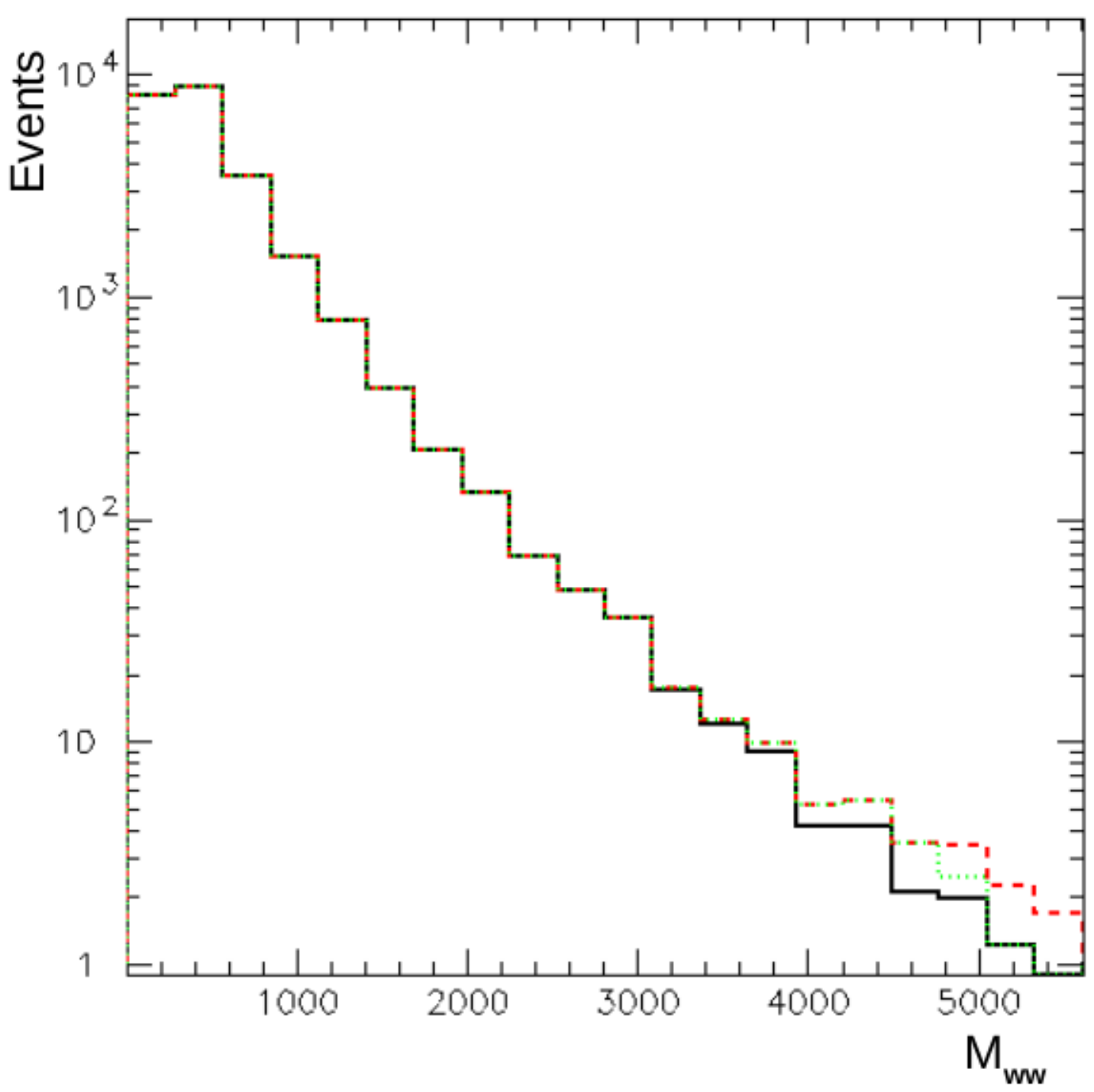} 
  \end{tabular}
\caption{
Typical examples of kinematic distributions used for the assessment of BSM signal significances. Shown are the distributions of $M_{ll}$, $M_{o1}$ and $R_{pT}$ (in log scale): in the Standard Model
(black), with $f_{M1}=0.2$ TeV$^{-4}$ and the high-$M_{WW}$ tail treatment according to Eq.~\eqref{unitarized} (red), and with $f_{M1}=0.2$ TeV$^{-4}$ and the high-$M_{WW}$ tail treatment according to Eq.~\eqref{dsigma} (green). In addition the lower-right plot shows  the distribution in the invariant mass of the WW system, $M_{WW}$ with $f_{M1}=0.2$ TeV$^{-4}$. In all the plots the scale $\Lambda$ was chosen as: $\Lambda = M^U = 4.9$ TeV. Assumed is $\sqrt{s} = 27 $ TeV and an integrated luminosity of 3 ab$^{-1}$.
}
\label{fig:ObsVarDistributions}
\end{figure}

The applied analysis chain to the events generated for HE-LHC is a carbon copy of the one described in detail in
Ref.~\cite{wwpaper}, here we only briefly outline the main points,
with special emphasis on the differences.  
%{\MScut All event selection criteria are kept the same}
 Standard VBS selection criteria were applied, namely we
require at least two reconstructed jets and exactly two leptons (muons or electrons)
satisfying the following conditions: $M_{jj} >$ 500 GeV, 
$\Delta\eta_{jj} >$ 2.5, $p_T^{~j} >$ 30 GeV,
$|\eta_j| <$ 5, $p_T^{~l} >$25 GeV and $|\eta_l| <$ 2.5.
%{\MScut We disregard the fact that an increased beam energy may lead to a
%re-optimization of the detector geometries and of selection criteria in order to accommodate
%for the slightly different event topologies.}
Although we have not optimized our selection criteria for 27 TeV, one can safely suppose they will not be much different. A detailed optimization will be done with full knowledge of the future detector geometry.
Like before, the total BSM signal is estimated according to Eq.(\ref{unitarized}) by suppressing the 
high-mass tail 
above the calculated value of $\Lambda$. This is achieved  by applying an additional weight
of the form $(\Lambda/M_{WW})^4$ to each generated event in this region.  
The EFT-controlled signal is calculated according to Eq.(\ref{dsigma})  by replacing the 
generated high-mass tail with the one
expected in the SM (known as ``clipping  method").
Signal significances are computed as the square root of a $\chi^2$
resulting from a bin-by-bin comparison of the event yields, with statistical
errors such as expected from the data at 3 ab$^{-1}$, in the
distribution of the most sensitive kinematic variables.  Compared to
the 14 TeV analysis, the binning of histograms was changed so
that in the highest bin the SM prediction normalized to 3/ab
is still between 2-3 events. In Fig.~\ref{fig:ObsVarDistributions} as an 
example shown are distributions of four chosen variables:  
\begin{itemize}
\item invariant mass of two leptons $M_{ll}$, 
\item ratio of transverse momenta of leptons 
and jets $R_{p_T} \equiv p_T^{~l1}p_T^{~l2}/(p_T^{~j1}p_T^{~j2})$, 
\item $M_{o1} \equiv \sqrt{(|\vec{p}_T^{~l1}|+|\vec{p}_T^{~l2}|
+|\vec{p}_T^{~miss}|)^2 - (\vec{p}_T^{~l1}+\vec{p}_T^{~l2}+\vec{p}_T^{~miss})^2}$ 
\item and the (true) invariant mass $M_{WW}$
\end{itemize}
in the SM and in the case $f_{M1} = 0.2$ TeV$^{-4}$ and $\Lambda =$ 4.9 TeV.
As it was for 14 TeV,
we found  
$R_{p_T} $ to be the most sensitive kinematic variable for
${\cal O}_{S0}$ and ${\cal O}_{S1}$, and 
$M_{o1} $
for the remaining operators.

As the unitarity limit we take always the lower of the two values
between on-shell $W^+W^+$ and on-shell $W^+W^-$ scattering,
calculated from T-matrix diagonalization in the helicity space. Indeed,
both processes probe the same quartic coupling and are governed by the same Wilson coefficients, as further explained in Appendix \ref{app:unitarity}. 

\begin{figure}[h] 
  \begin{tabular}{cc}
      \includegraphics[width=0.5\linewidth]{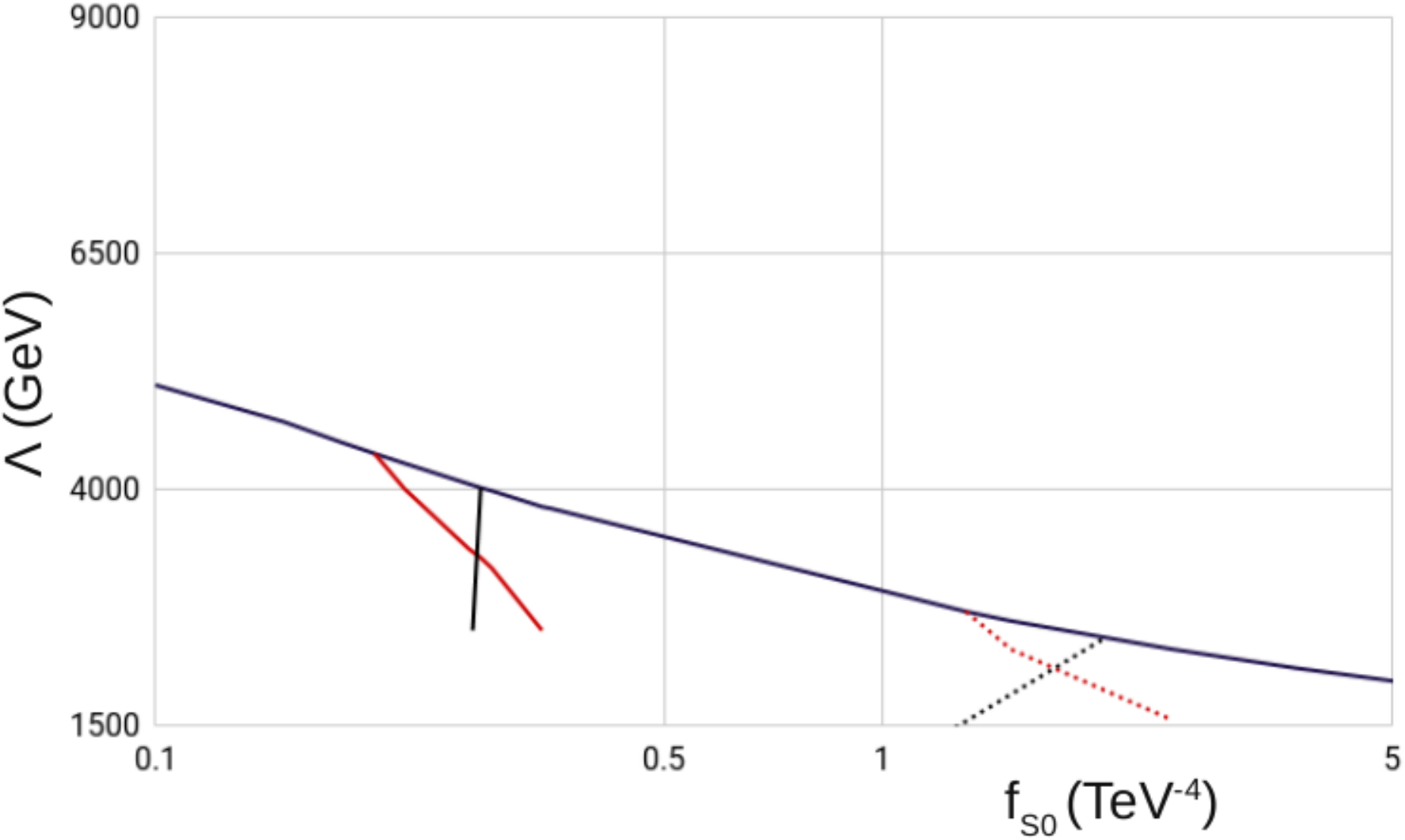} 
& \includegraphics[width=0.5\linewidth]{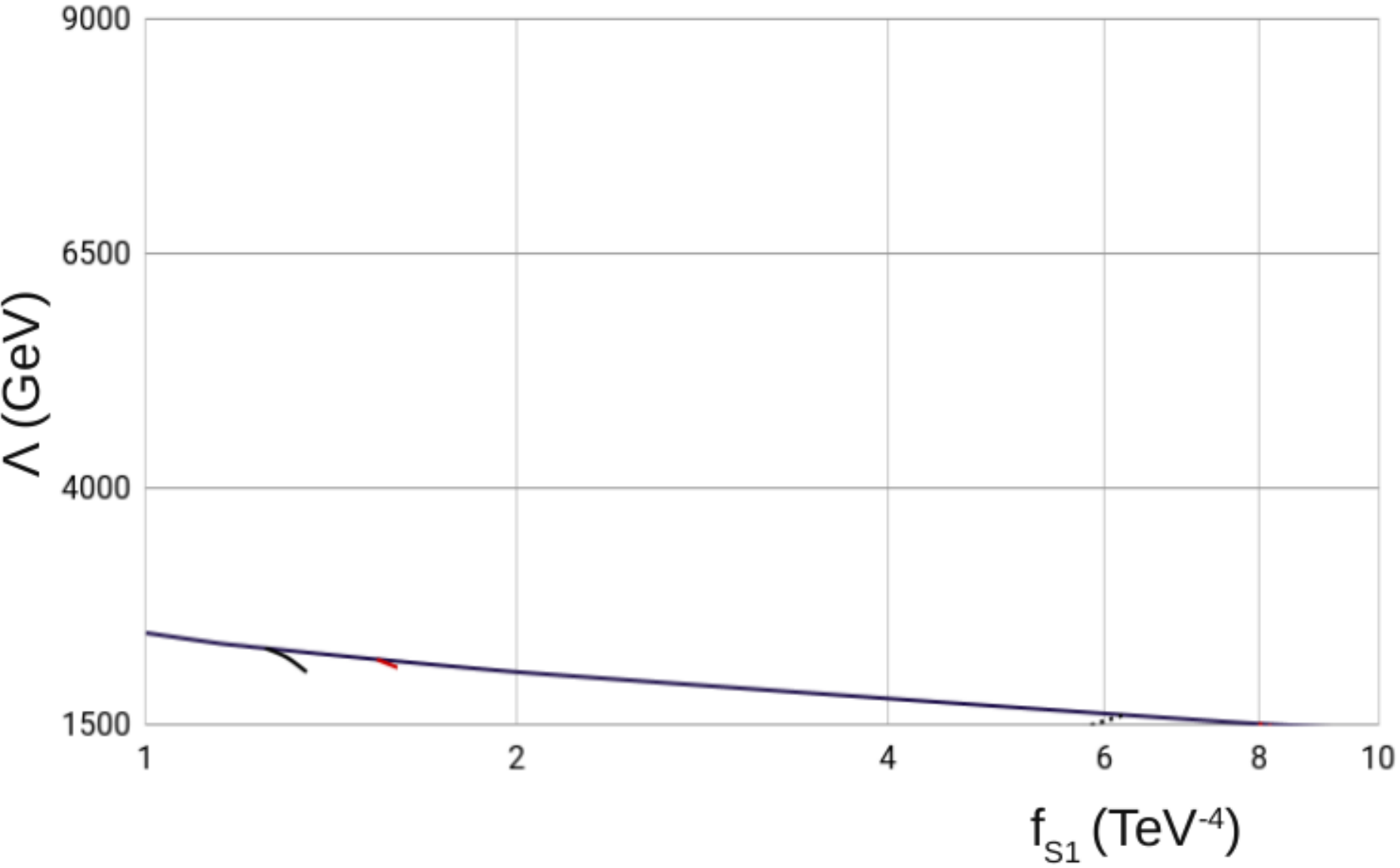} \\
\includegraphics[width=0.5\linewidth]{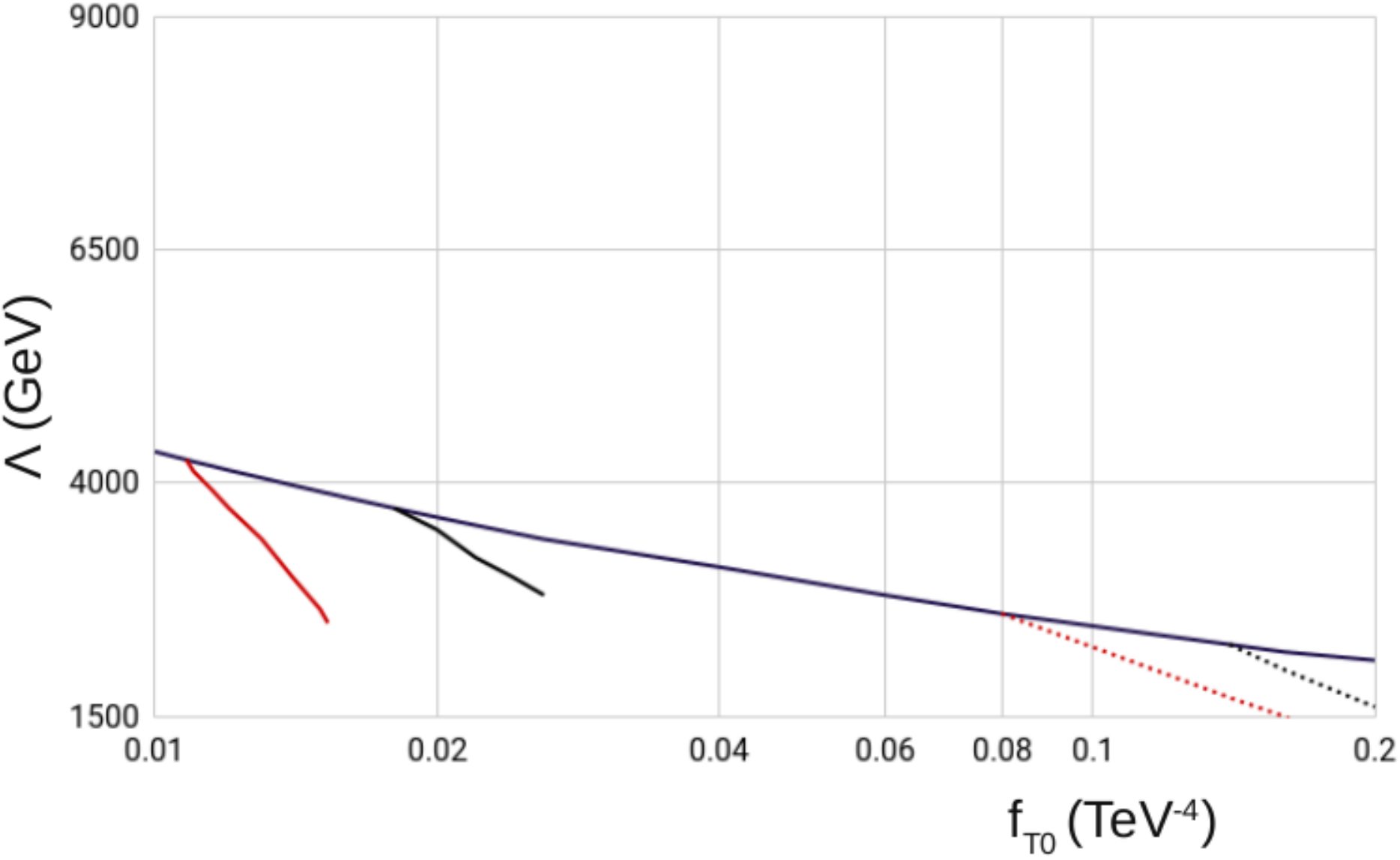} 
& \includegraphics[width=0.5\linewidth]{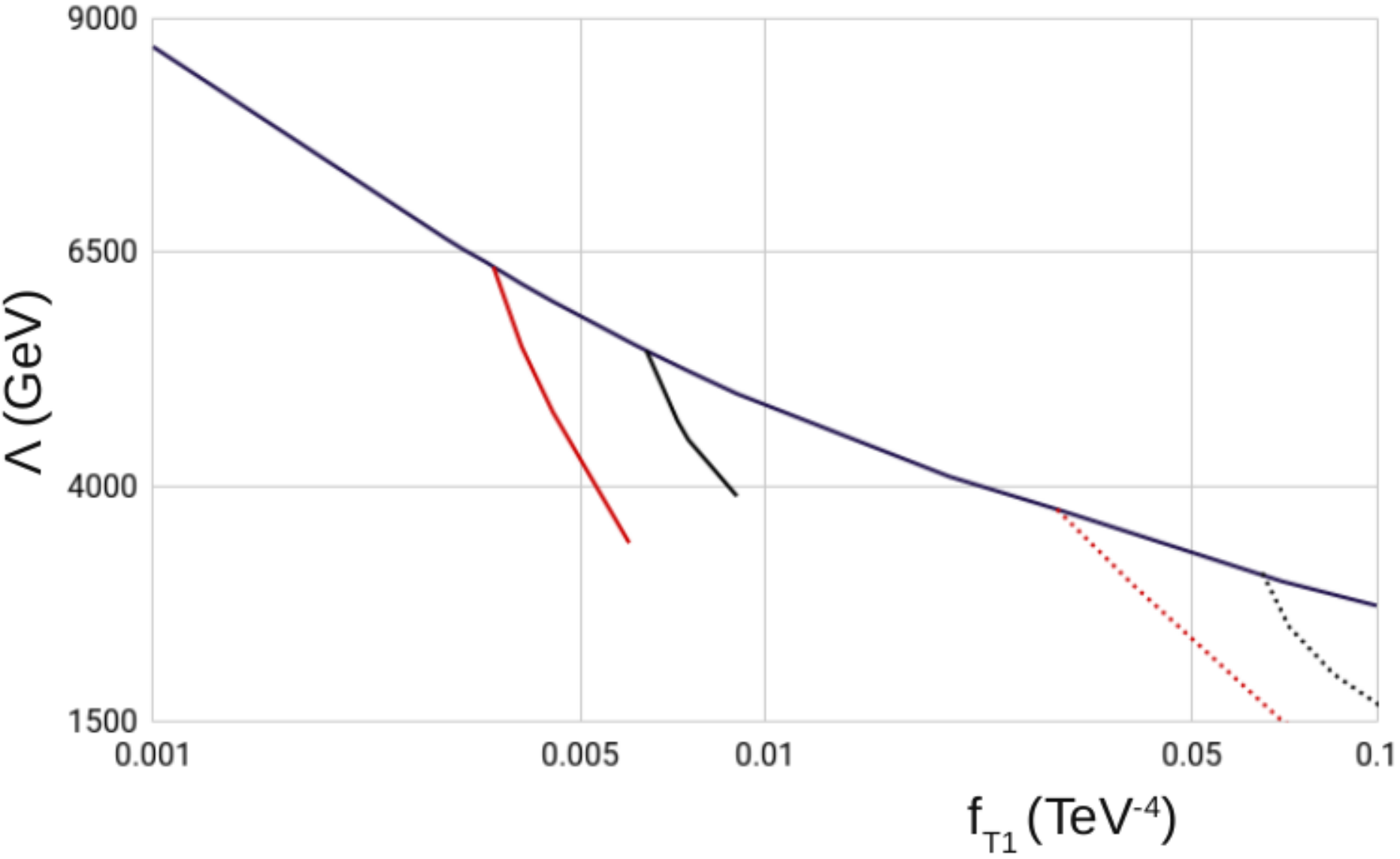} \\
\includegraphics[width=0.5\linewidth]{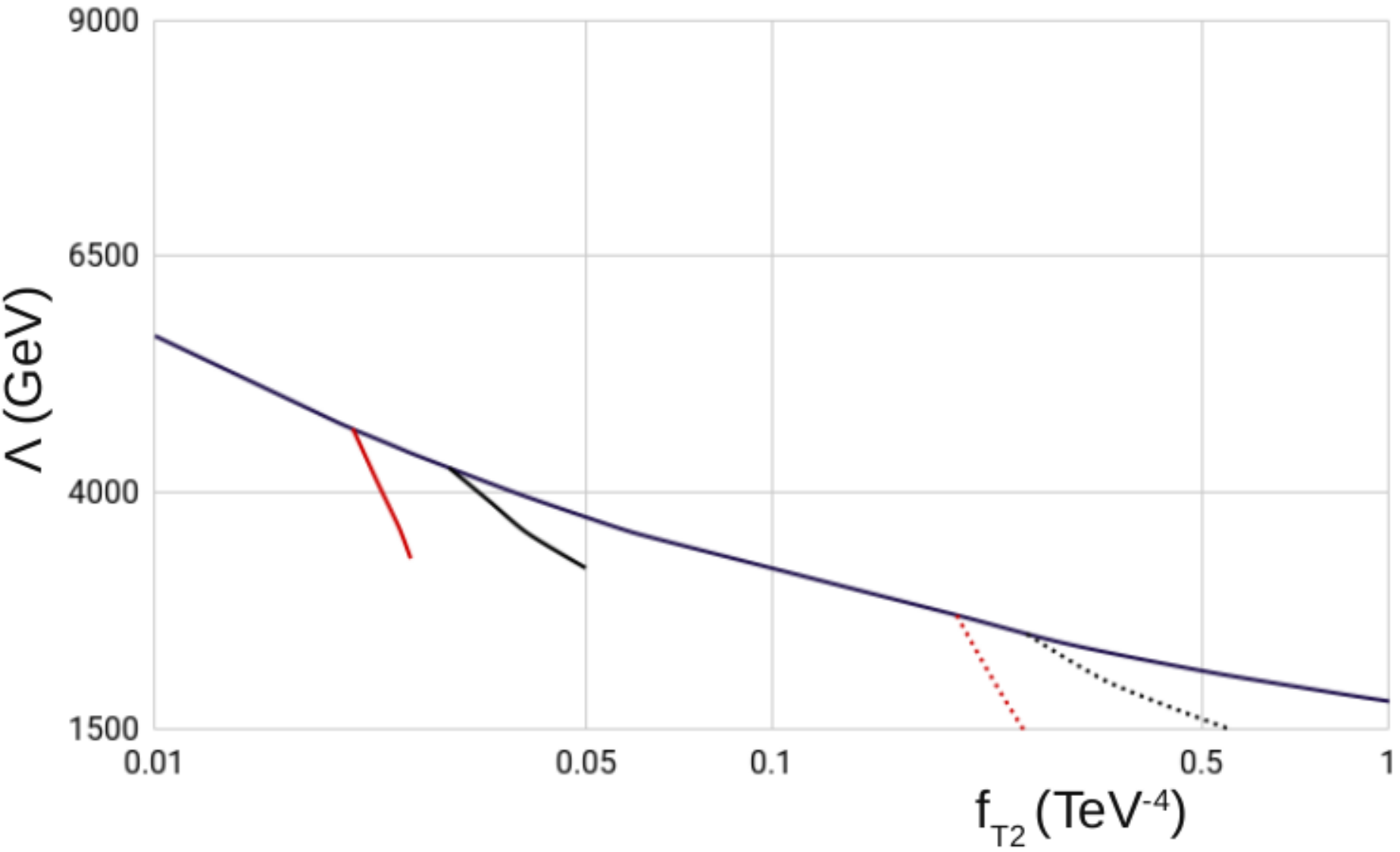} 
& \includegraphics[width=0.5\linewidth]{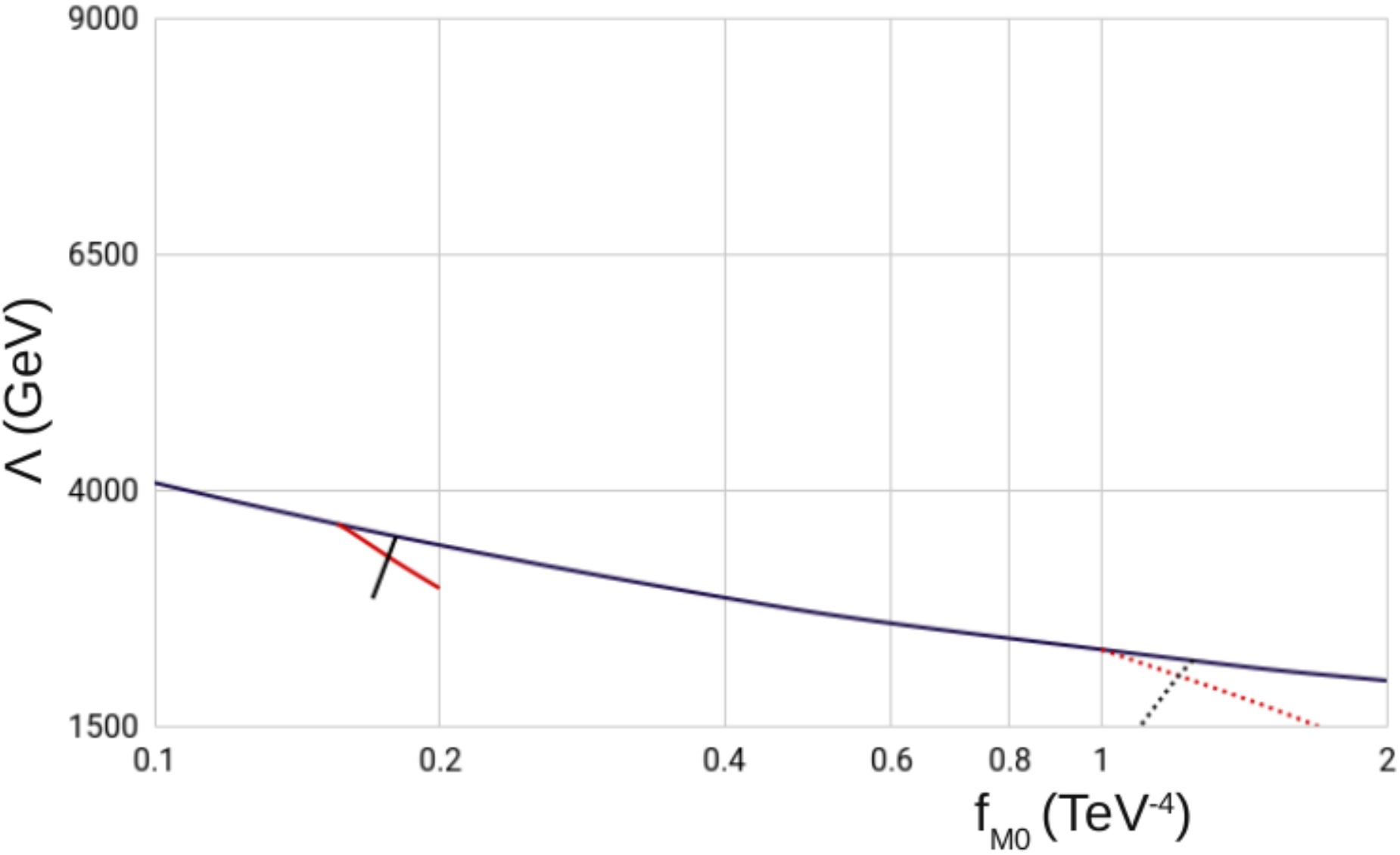} \\
\includegraphics[width=0.5\linewidth]{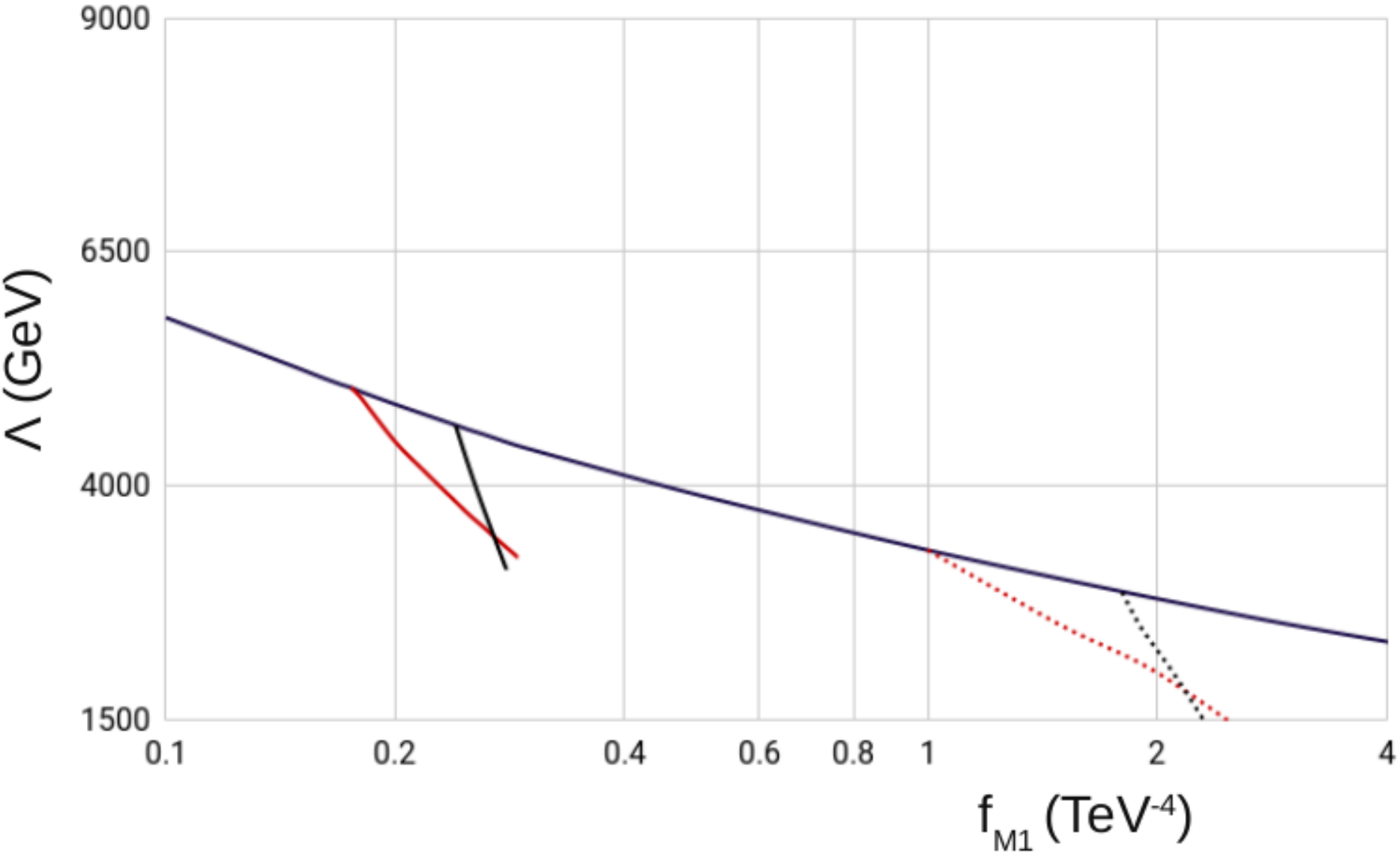} 
& \includegraphics[width=0.5\linewidth]{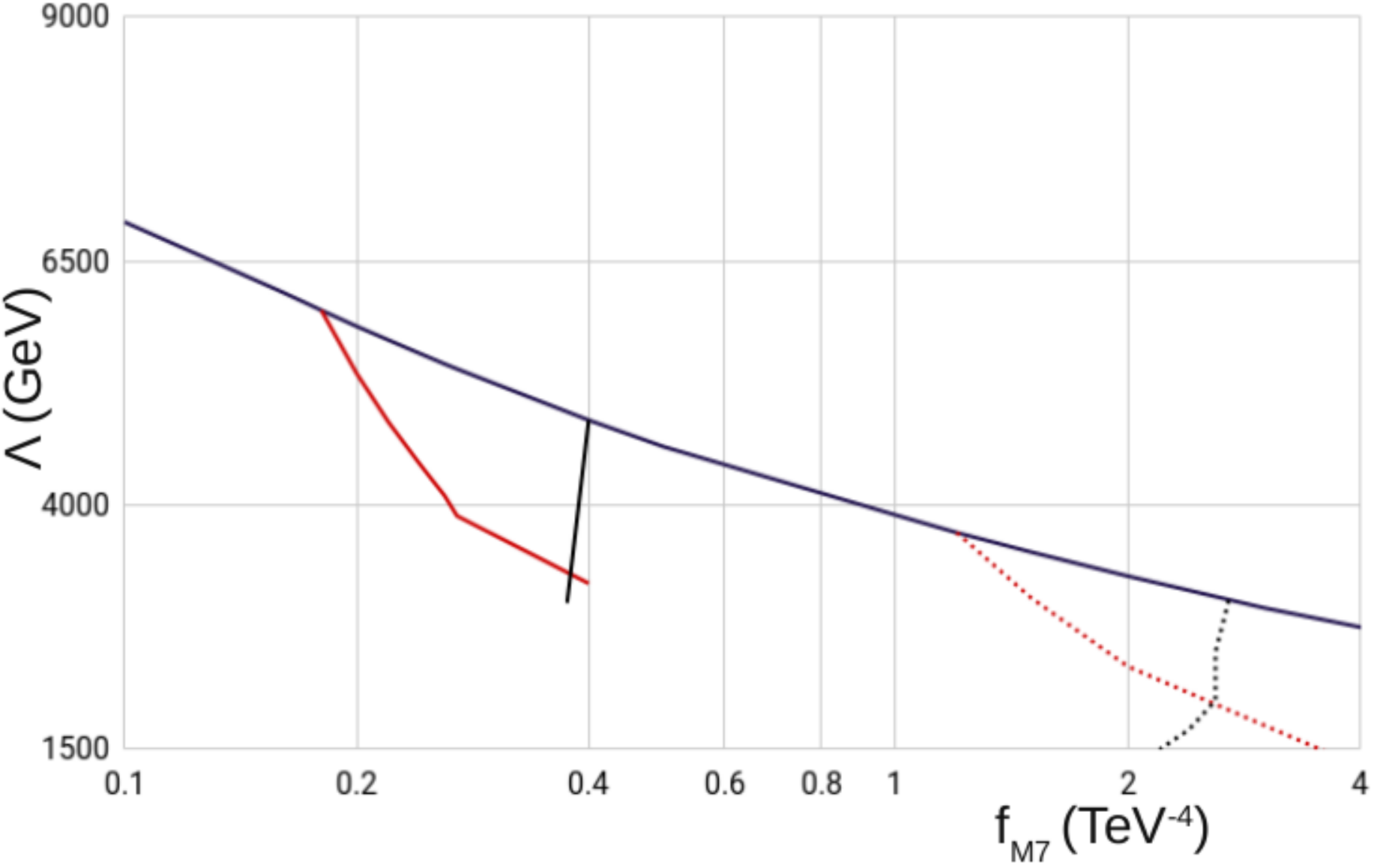} 
  \end{tabular}
\caption{Regions in the $\Lambda$ vs $f$ (positive $f$ values) space for dimension-8 operators in which a $5\sigma$
BSM signal can be observed and the EFT is applicable. The unitarity limit is shown in blue; the lower limits for a $5\sigma$ signal significance from Eq.~\eqref{unitarized} (red) and the upper
limit on $2\sigma$ EFT consistency (black). The solid (dotted) lines correspond to $\sqrt{s} = 27\ (14)$ TeV. 
Assumed is the integrated luminosity of 3 ab$^{-1}$.}
\label{fig:compPos}
\end{figure}%tutaj
Fig.~\ref{fig:compPos} shows the results for the individual operators $S0,\, S1,\, T0,\, T1,\, T2,\, M0,\, M1$
and $M7$, in comparison with results at 14 TeV (for positive $f$ values).
Not unexpectedly, all the triangles are shifted to lower $f$ values compared
to 14 TeV, the shift being as large as almost an order of magnitude.  However,
the total area of the triangles does not get significantly larger as we increase
the energy.  This is because the EFT consistency criterion pushes the effective
upper limits on $f$ in a similar manner as does the BSM observability criterion
for the lower limits.  Overall, the shapes and sizes of all the EFT triangles
are remarkably similar for 27 TeV as for 14 TeV, only their respective positions
differ.  

\begin{figure}[h] 
  \begin{tabular}{cc}
\includegraphics[width=0.5\linewidth]{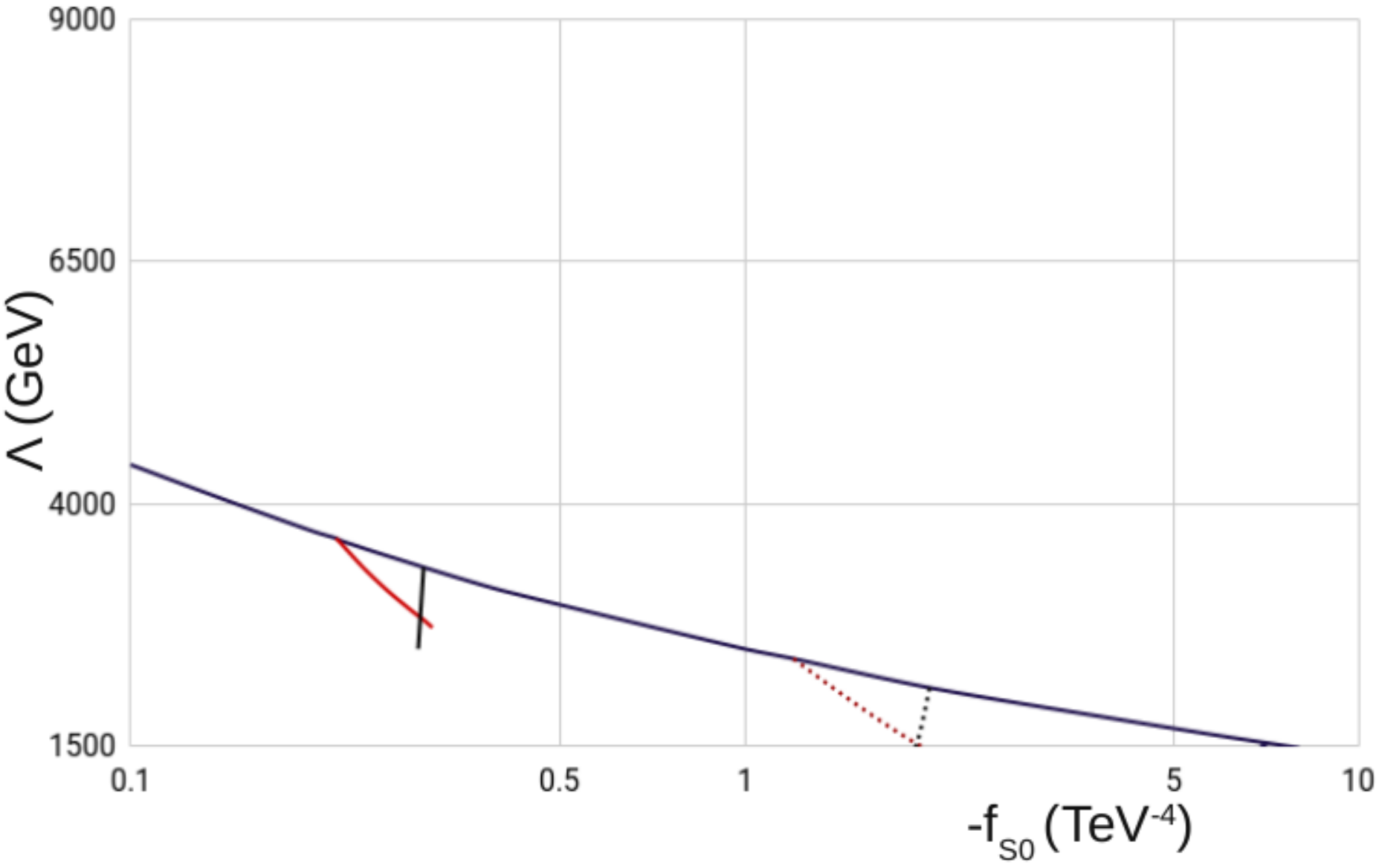} 
& \includegraphics[width=0.5\linewidth]{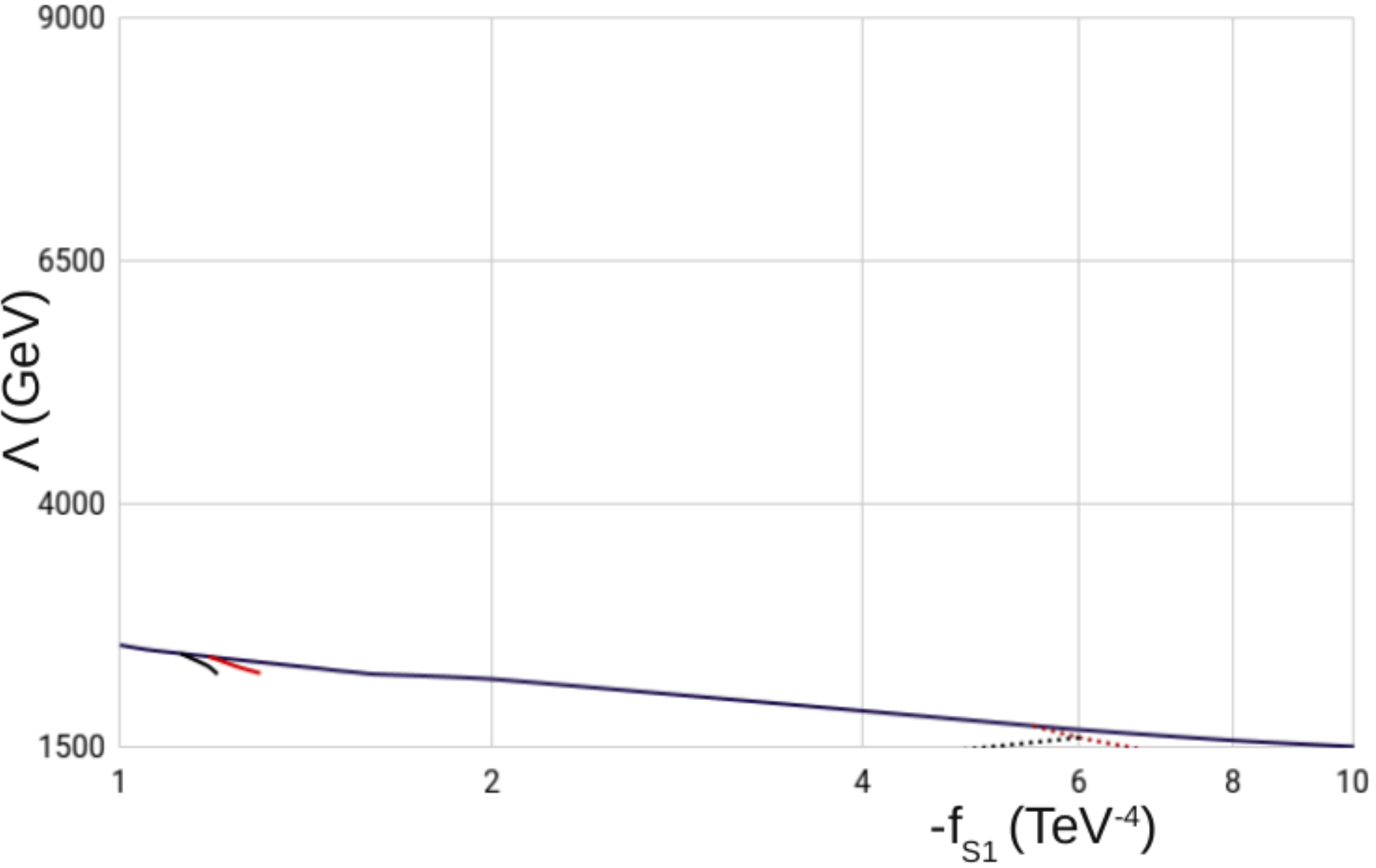} \\     
\includegraphics[width=0.5\linewidth]{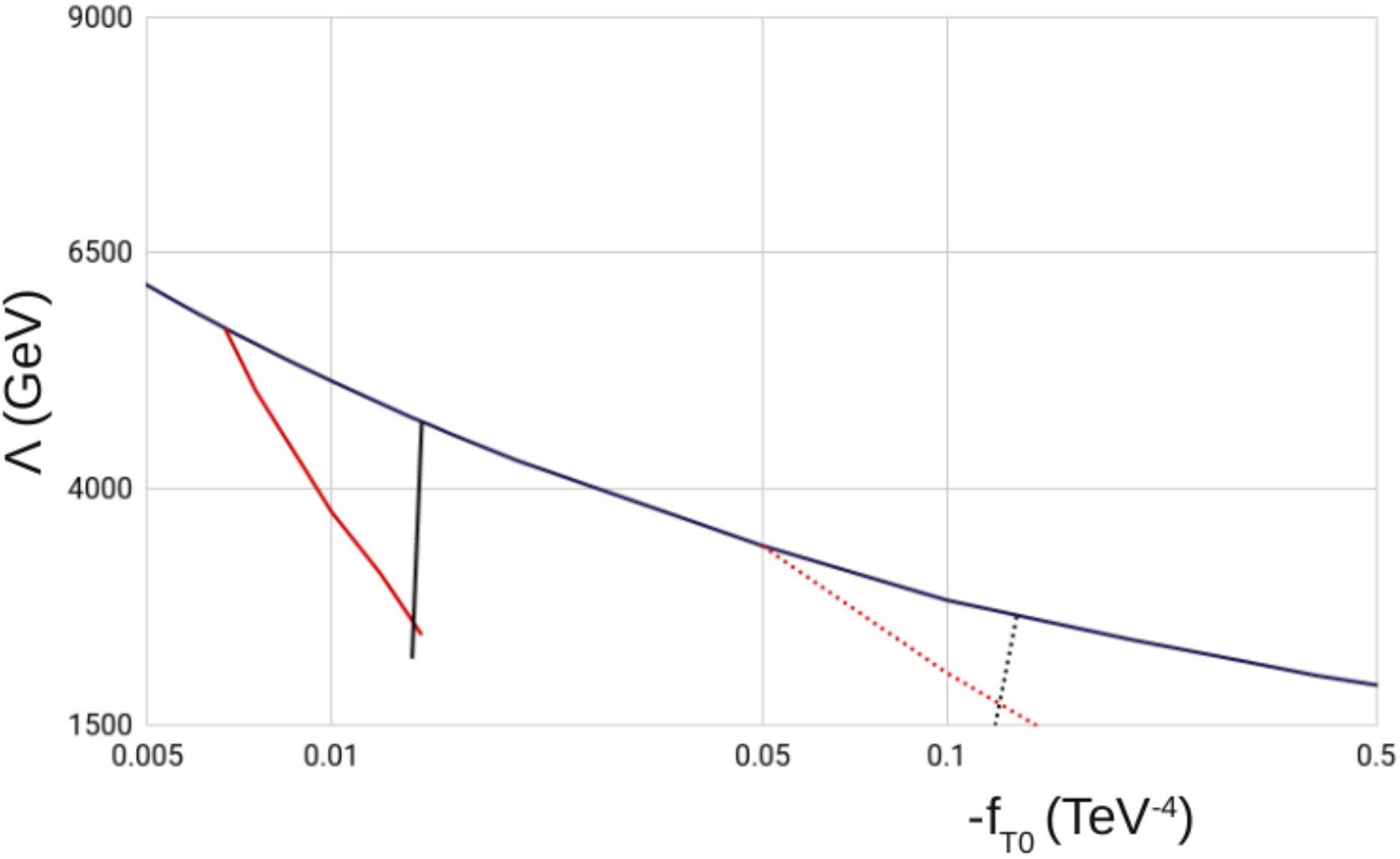} 
& \includegraphics[width=0.5\linewidth]{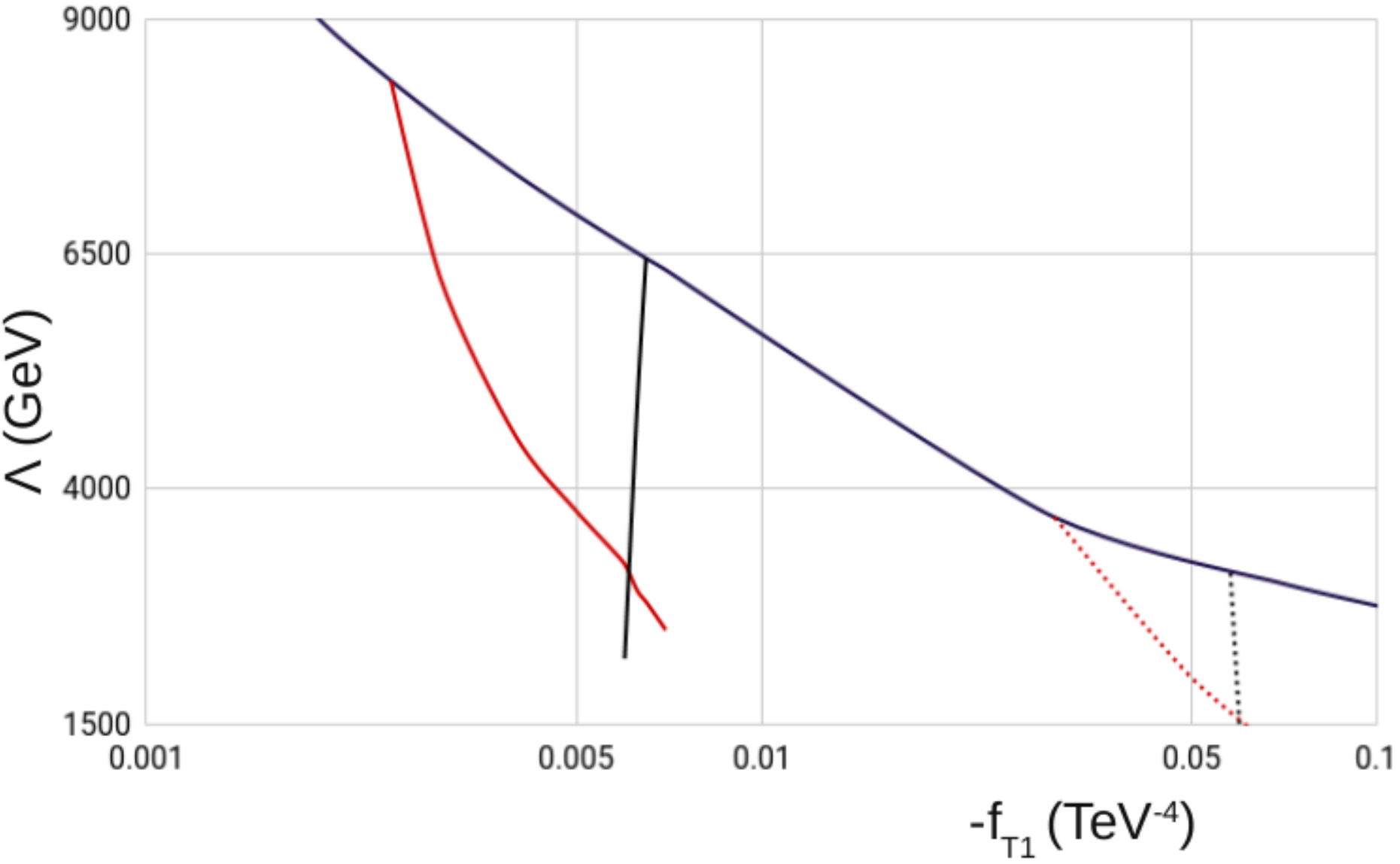} \\
\includegraphics[width=0.5\linewidth]{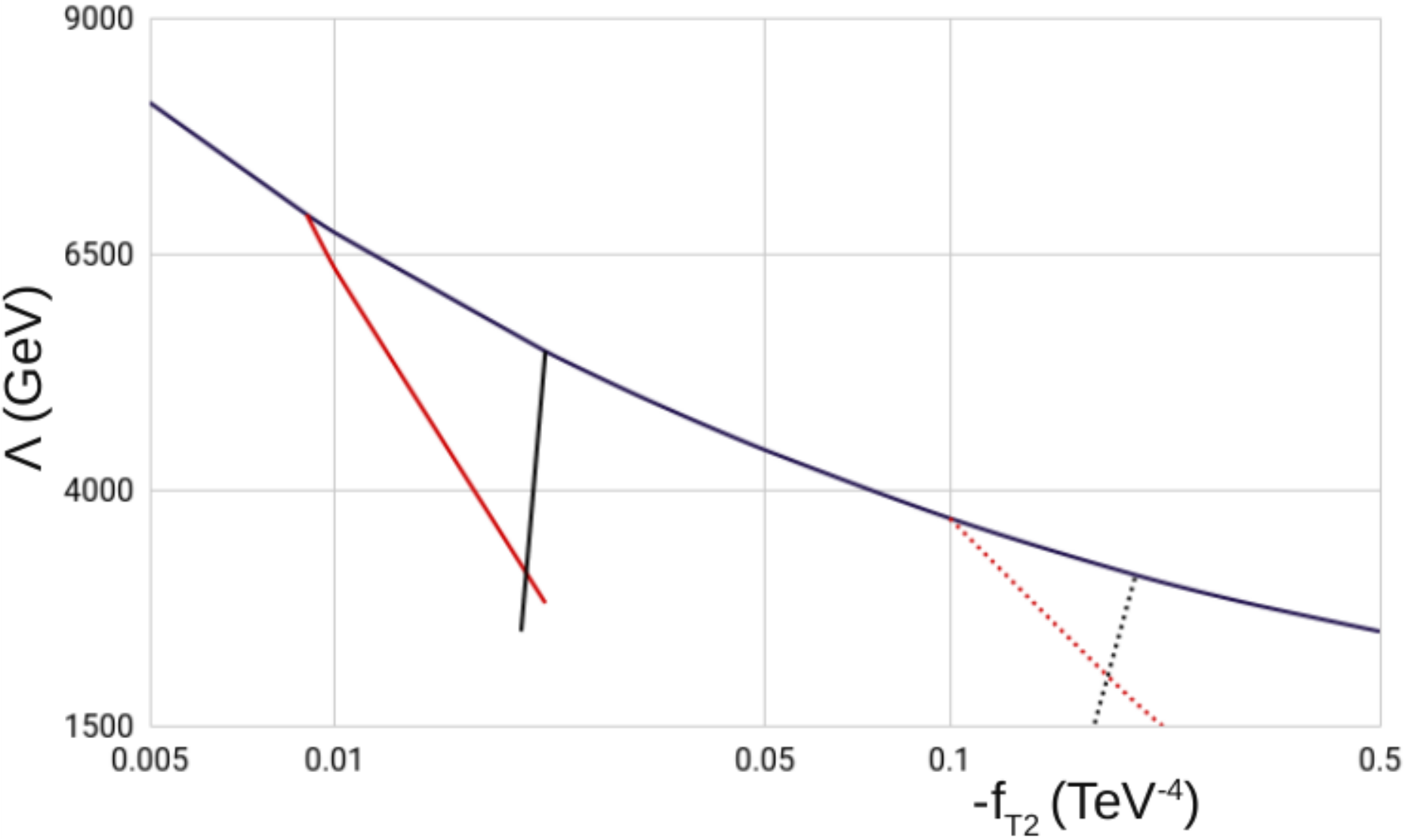} 
& \includegraphics[width=0.5\linewidth]{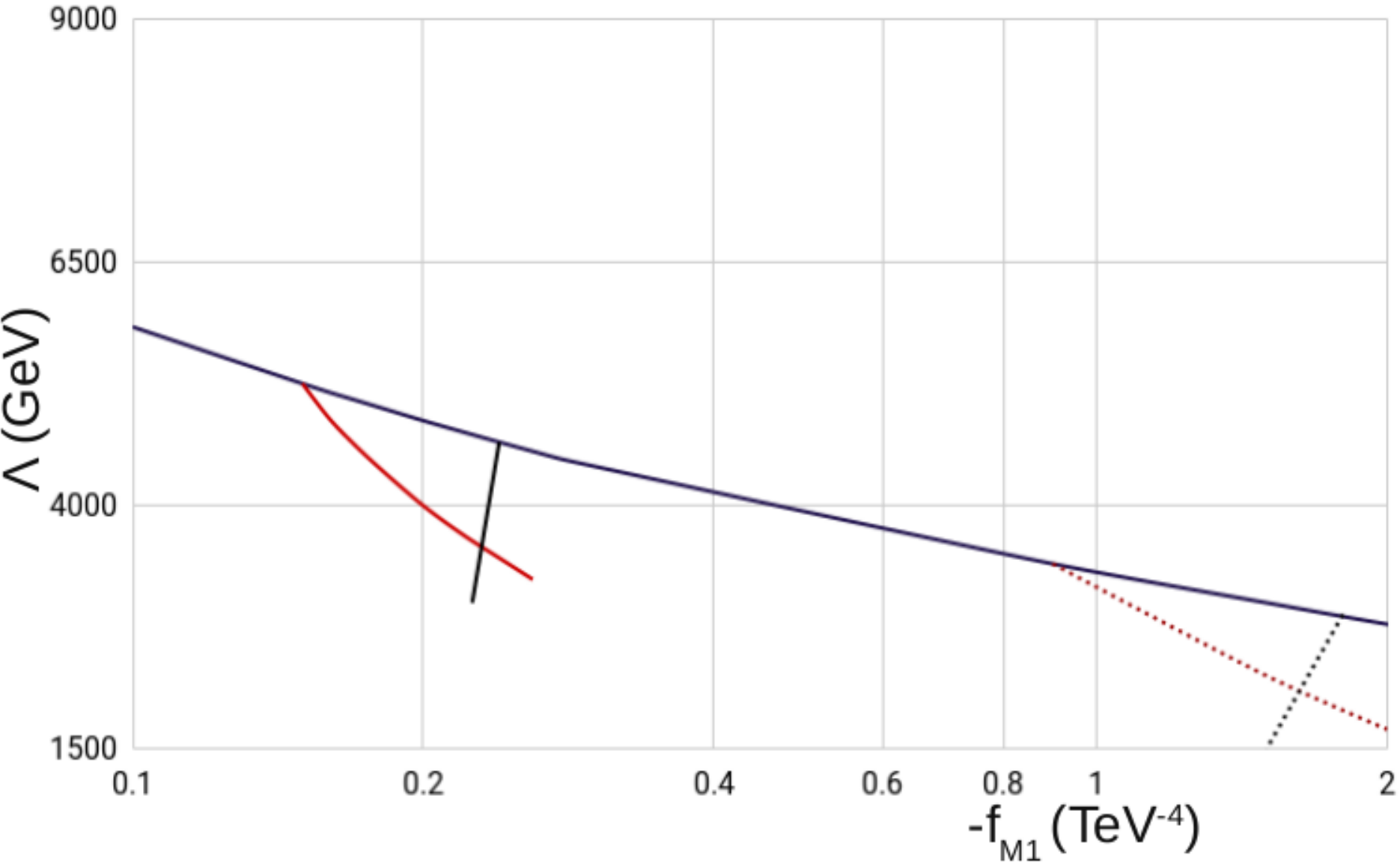} \\
\includegraphics[width=0.5\linewidth]{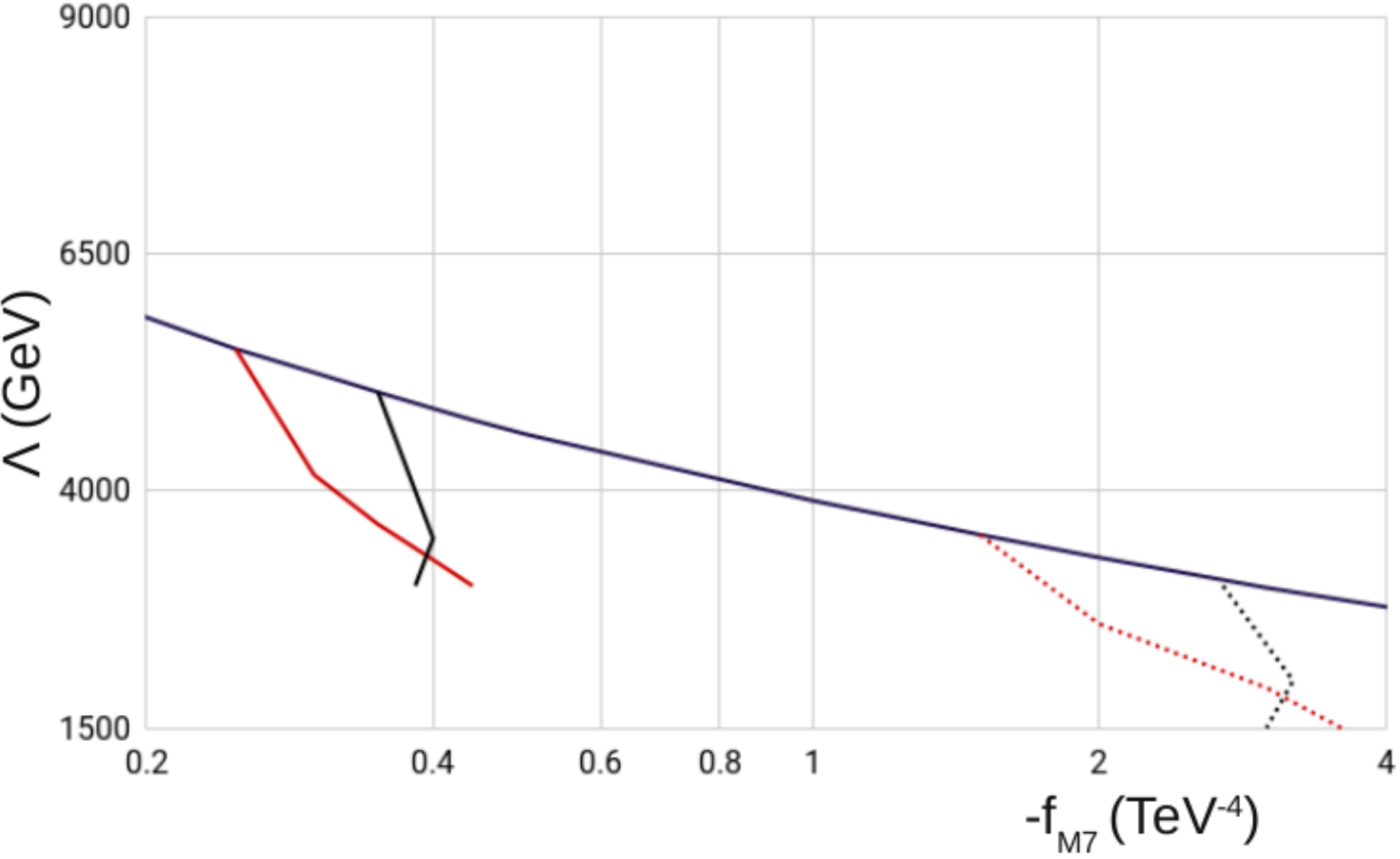} 
  \end{tabular}
\caption{Regions in the $\Lambda$ vs $f$ (negative $f$ values) space for dimension-8 operators in which a $5\sigma$
BSM signal can be observed and the EFT is applicable. The unitarity limit is shown in blue; the lower limits for a $5\sigma$ signal significance from Eq.~\eqref{unitarized} (red) and the upper
limit on $2\sigma$ EFT consistency (black). The solid (dotted) lines correspond to $\sqrt{s} = 27\ (14)$ TeV. 
Assumed is the integrated luminosity of 3 ab$^{-1}$.}
\label{fig:compNeg}
\end{figure}

Fig.~\ref{fig:compNeg} shows the respective results for negative $f$ values
of $S0,\, S1,\, T0,\, T1,\, T2,\, M1$ and $M7$.  
Here exactly the same observations can be made again.
The negative $f$ values of $M0$  look virtually identical to their
positive counterparts, since for these operators the SM-BSM interference term in the
total amplitude calculation is practically negligible (see the Appendix 
for details), and so we do not show them here.
There is no triangle at all for $S1$, for which the overall lower limit for
BSM observability is about 1.2~TeV$^{-4}$ and the upper limit for EFT 
consistency is 1.4 ~TeV$^{-4}$.
Here as well we observe a similar behavior as for 14 TeV.

Both analyses, at 14  and  27 TeV, were done at the generator level. Reducible backgrounds were not
simulated, since they are known to be strongly detector dependent
(for the different compositions of reducible background as measured at 13 TeV by CMS and ATLAS, see Figs. 2-a and 2-b in Ref.~\cite{Sirunyan:2017ret} and Fig. 3 in Ref.~\cite{Aaboud:2019nmv}, respectively, where it has been found that 'non-prompt' leptons (from hadron decay and fakes) is the largest reducible background and WZ is small in CMS, while in ATLAS WZ is the largest background and 'non-prompt' is small). Such simulations have to be carried by each experiment  for its specific detector performances. 
For a realistic estimate of the sensitivity
limits to new physics effects the reader is referred to  literature,
e.g., Ref.~\cite{Snowmass2013}.
%{\MScut Such as for 14 TeV,}
Full detector simulation 
%{\MScut with reducible backgrounds included
%can only make the picture worse.}
will move the EFT triangles to larger Wilson coefficients, but it
cannot make them larger, hence all our conclusions still hold.

\begin{figure}[h] 
\begin{center}
\includegraphics[width=0.7\linewidth]{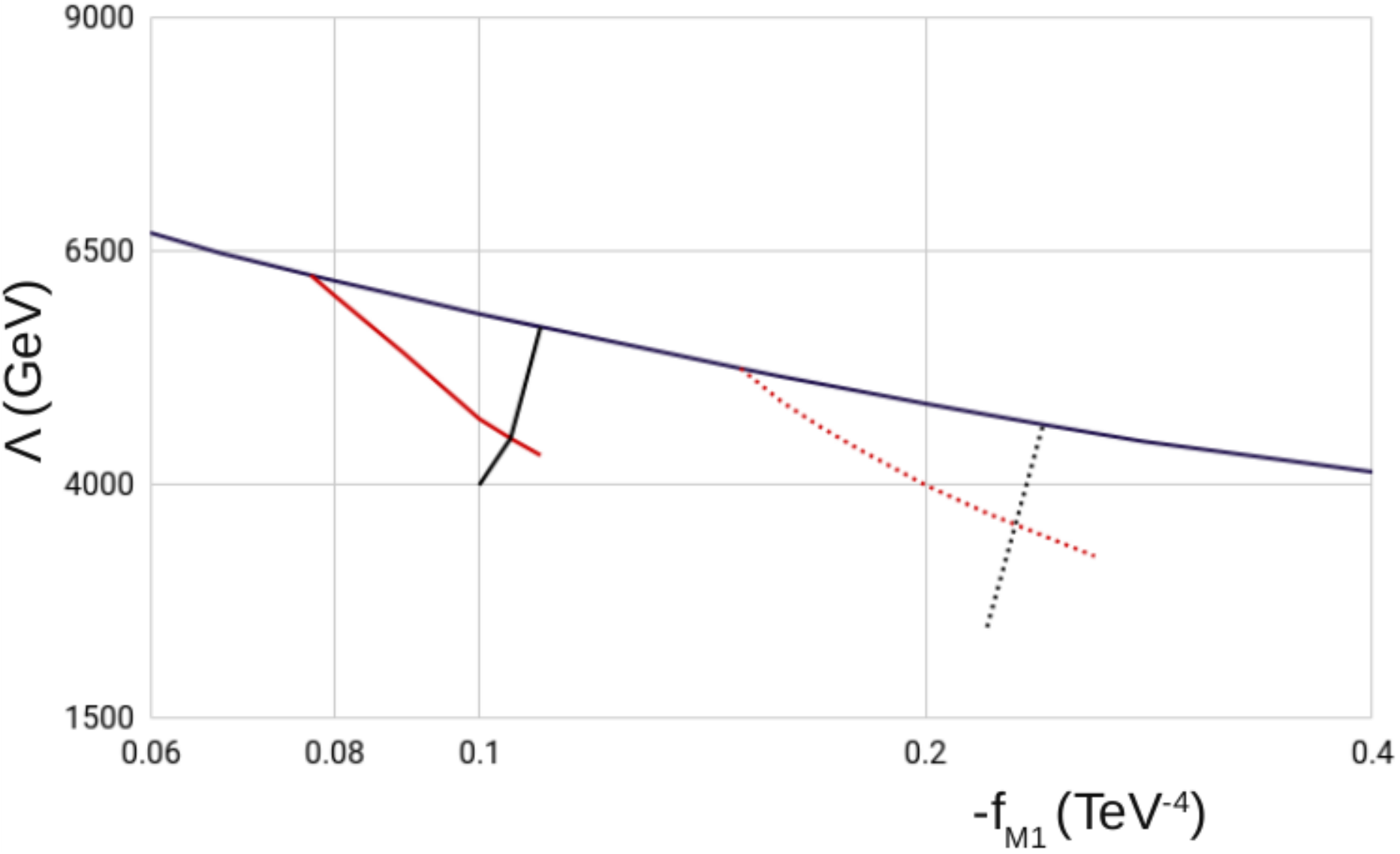} 
\end{center}
\caption{Regions in the $\Lambda$ vs $f$ (negative $f$ values) space for $M1$ operator in which a $5\sigma$
BSM signal can be observed and the EFT is applicable. The unitarity limit is shown in blue; the lower limits for a $5\sigma$ signal significance from Eq.~\eqref{unitarized} (red) and the upper
limit on $2\sigma$ EFT consistency (black). The solid (dotted) lines correspond to 15 ab$^{-1}$ (3 ab$^{-1}$ ). 
Assumed is $\sqrt{s} = 27$ TeV.}
\label{fig:compLumis}
\end{figure}

For the sake of a convenient comparison between the respective results
at two different $pp$ beam energies, in the bulk of this study
we have always assumed the same integrated
luminosity of 3 ab$^{-1}$ for both cases.  This number is appropriate for the
HL-LHC stage, but underestimates the expected statistical power of the
HE-LHC.  However, it is trivial to recalculate all the results to 15/ab
in order to get the true expected discovery reach of the HE-LHC, taking 
into account
its actual expected luminosity.  An increase of statistics by a factor 5
will lead to a further shift of all the EFT triangles by a factor close 
to $\sqrt{5}$,
both in the 5$\sigma$ discovery and the 2$\sigma$ consistency curves
(in fact, somewhat less than that because of non-linear dependence of 
the BSM signal on the value of the individual Wilson coefficients).  It 
will not significantly change either the shape nor the size of triangles.  
A comparison of results calculated for the same $pp$ beam energy of
27 TeV and two different integrated luminosities is exemplified for 
the $M1$ operator in Fig.~\ref{fig:compLumis}.

Our simple procedure to suppress the high-mass tails
by applying a { $(\Lambda/M_{WW})^4$ }weight to events generated above the scale
of $\Lambda$ works reasonably well in the vicinity of the unitarity limit.
In this region 
it produces a tail falling approximately like $1/M_{WW}^2$, which is the expected
asymptotic (i.e., for $M_{WW}>>\Lambda$) behavior of the total cross section after regularization.
It nonetheless becomes too strong as we go to $\Lambda << M^U$, where the
total cross section is still dominated by the SM contribution which does
not require any further suppression.
Moreover, for low values of $\Lambda$ the tail itself becomes large, leading
to large uncertainties due to the details of its modeling.
We have discontinued the curves on Figs.~\ref{fig:compPos} and~\ref{fig:compNeg} below the values
at which we find the method lead to the unphysical result of
signal being suppressed below the SM level itself.
For this reason the EFT triangles for $T0, T1$ and $T2$ do not close.
For the remaining cases, however, they
are completely contained in the region where our simple method is 
still viable.

According to Ref.~\cite{Biedermann:2017bss}, full NLO (EW+QCD) corrections to the SM same-sign WW scattering process at 13 TeV lower the total cross section within a fiducial volume defined by standard VBS cuts by 17\%; the relative effect increases in the high invariant mass region. Corresponding corrections at 27 TeV were reported in~\cite{Azzi:2019yne}. They are relatively a few percent larger than in the former case. Therefore, accounting for higher-order corrections is well motivated in further more dedicated studies.

Positivity constraints applied to VBS, derived e.g. in Ref.~\cite{Zhang:2018shp},  require that   certain linear combinations of  Wilson coefficients of dimension-8 operators to be  positive definite.  
Since
we choose one operator at a time, the bounds would determine what signs of the Wilson coefficients 
should be considered for each operator. Therefore accounting for these bounds using our results is straightforward.

\section{Conclusion and Outlook}
\label{conclusion}

Although an increase of the LHC energy vastly improves the sensitivity to
new physics effects in VBS processes, the question of EFT applicability is
a different one and cannot be solved by changing the energy.
The same-sign $WW$ process with its purely leptonic $W$ decays is often
considered ``gold-plated" due to its relatively good signal to background ratio,
but the lack of experimental access to the $WW$ invariant mass poses a severe
problem in describing the data in terms of the EFT.  Despite reasonable
sensitivity to BSM effects, such effects if observed will most likely not be
possible to interpret using the data from this process alone and applying
the usual framework of testing one dimension-8 operator at a time.
This conclusion holds regardless of the actual proton beam energy.

The present results reinforce the former conclusion that future VBS
data analysis, both at the LHC experiments as well as future proton-proton 
colliders, should evolve in the direction of multidimensional fits with
many higher dimension operators varied at a time.  This in turn may require
global simultaneous fits to many processes (including $ WZ,\, ZZ$ and semi-leptonic
$WV$, if not other processes) to help disentangle the correlations between
signals originating from different operators. Independently of the goal and results of this paper, one always has to remember that the approach of varying one operator at-a-time is not the most appropriate in any case -- such a procedure breaks the model independence of the EFT description and therefore EFT requires ultimately a global approach.

Helpful in disentangling the effects of different operators may be also the 
polarizations of the outgoing $W$ bosons, as different operator subsets, S, T and M, affect different polarizations.  There are new theoretical ideas how to project the total VBS cross sections onto individual polarizations without invoking the rather crude $W$ on-shell approximation~\cite{Maina:1710}.  $WW$
polarizations can be extracted from the data by fitting simulated templates
of the corresponding polarized distributions.  Unfortunately,
purely leptonic $WW$ decays do not offer the possibility to reconstruct the $W$ decay angle, which is the only strictly model-independent signature of $W$ polarization.  While many other distributions exhibit qualitative differences between the different polarizations, they are usually also strongly model-dependent. 
Consequently, SM templates cannot be used in the BSM case without the risk of losing
sensitivity to the BSM signal.  If, however, a set of observables is identified for which sufficiently model-independent
templates for $W_LW_L$, $W_TW_T$ and $W_TW_L$ can be constructed, it could
vastly improve the perspectives of future VBS data analysis in the framework
of the EFT.

%%%%%%%%%%%%%%%%%%%%%%%%%
\section*{Acknowledgments}
%%%%%%%%%%%%%%%%%%%%%%%%%
We would like to thank Adam Falkowski and Luca Merlo for valuable comments and discussions. The work of PK is supported by the Spanish MINECO project FPA2016-78220-C3-1- P (Fondos FEDER) and by National Science Centre, Poland, the PRELUDIUM project under contract 2018/29/N/ST2/01153. JK was supported in part by the National Science Centre, Poland, the HARMONIA project under contract UMO-2015/18/M/ST2/00518 (2016-2020).  JK and MS were partly supported by the COST Action CA16108 and are grateful to all the members of the Action for inspiring discussions. ST is supported by Fermi Research Alliance, LLC under Contract No. De-AC02-07CH11359 with the United States Department of Energy.

\appendix

\section{Dimension 8 operators}
\label{app_op8}

The following dimension eight operators  contribute to the $WWWW$ vertex, without affecting tri-linear couplings:
\begin{equation}
\begin{aligned}
  {\cal O}_{S0} &= \left [ \left ( D_\mu \Phi \right)^\dagger
 D_\nu \Phi \right ] \times
\left [ \left ( D^\mu \Phi \right)^\dagger
D^\nu \Phi \right ],
\\
  {\cal O}_{S1} &= \left [ \left ( D_\mu \Phi \right)^\dagger
 D^\mu \Phi  \right ] \times
\left [ \left ( D_\nu \Phi \right)^\dagger
D^\nu \Phi \right ],
\\
 {\cal O}_{M0} &=   \hbox{Tr}\left [ {W}_{\mu\nu} {W}^{\mu\nu} \right ]
\times  \left [ \left ( D_\beta \Phi \right)^\dagger
D^\beta \Phi \right ],
\\
 {\cal O}_{M1} &=   \hbox{Tr}\left [ {W}_{\mu\nu} {W}^{\nu\beta} \right ]
\times  \left [ \left ( D_\beta \Phi \right)^\dagger
D^\mu \Phi \right ],
%\\
  %{\cal O}_{M6} &= \left [ \left ( D_\mu \Phi \right)^\dagger {W}_{\beta\nu}
%{W}^{\beta\nu} D^\mu \Phi  \right ],
\\
  {\cal O}_{M7} &= \left [ \left ( D_\mu \Phi \right)^\dagger {W}_{\beta\nu}
{W}^{\beta\mu} D^\nu \Phi  \right ],
\\
 {\cal O}_{T0} &=   \hbox{Tr}\left [ {W}_{\mu\nu} {W}^{\mu\nu} \right ]
\times   \hbox{Tr}\left [ {W}_{\alpha\beta} {W}^{\alpha\beta} \right ],
\\
 {\cal O}_{T1} &=   \hbox{Tr}\left [ {W}_{\alpha\nu} {W}^{\mu\beta} \right ] 
\times   \hbox{Tr}\left [ {W}_{\mu\beta} {W}^{\alpha\nu} \right ],
\\
 {\cal O}_{T2} &=   \hbox{Tr}\left [ {W}_{\alpha\mu} {W}^{\mu\beta} \right ]
\times   \hbox{Tr}\left [ {W}_{\beta\nu} {W}^{\nu\alpha} \right ] ,
\end{aligned}
\end{equation}
where 
$\Phi$ is  the Higgs doublet field, 
the covariant derivative 
%\begin{equation}
$ D_\mu \equiv \partial_\mu + i \frac{g'}{2} B_\mu  + i g W_\mu^i \frac{\tau^i}{2} $
%\end{equation}
and the field strength tensor  
%\begin{equation}
% \begin{aligned}
$W_{\mu\nu} =  \frac{1}{2} \tau^i (\partial_\mu W^i_\nu - \partial_\nu W^i_\mu
       + g \epsilon_{ijk} W^j_\mu W^k_\nu )$.

\section{$WW$ scattering: off-shell versus on-shell}
\label{sec:app1}
In this Appendix we investigate what can be said  about the VBS subprocess  in the full $pp\rightarrow j j l l' \nu_l \nu_{l'}$ reaction 
from the analysis of the on-shell $WW$ scattering process. We start with the discussion of the $WW$ scattering in full $pp$ process, then identify the helicity amplitudes that dominate the high-energy behavior in the presence of dimension-8 operators and discuss the question of determining the unitarity limits.

\subsection{ $WW$ scattering in the full $pp$ reaction}
 In the physical process $pp\to jjll'\nu\nu_{l'}$ the $W$ bosons  are off-shell. Nevertheless, in this subsection we would like to show that 
 qualitative conclusions on the influence of dimension-8 operators  on the full process  can be drawn from 
 the analysis of on-shell $WW$ scattering. To this end, let us employ the identity~\cite{Maina:1710}:
\begin{equation}
g_{\mu\nu}+\frac{k_\mu k_\nu}{M^2_W} = \sum_{\lambda=1}^{4} \epsilon_\lambda^\mu(k)\left(\epsilon_\lambda^\nu(k)\right)^\ast.
\label{eq:identity}
\end{equation}
to express the numerator of the off-shell vector boson as a sum over polarization vectors $ \epsilon_\lambda^\mu(k)$. In the frame in which the spatial component of  $k_\mu$ is in the $z$ direction, $k_\mu=(E,0,0,k)$, the explicit form of each polarization vector reads:
\begin{equation}
\begin{array}{llll}
\epsilon^\mu_- &=& \frac{1}{\sqrt{2}}(0,+1-i,0) &\qquad \mathrm{(left),}\\
\epsilon^\mu_+ &=& \frac{1}{\sqrt{2}}(0,-1-i,0) &\qquad \mathrm{(right),}\\
\epsilon^\mu_0 &=& (k,0,0,E)/\sqrt{k^2} &\qquad \mathrm{(longitudinal),}\\
\epsilon^\mu_A &=& (E,0,0,k)/\sqrt{\frac{k^2-M^2_W}{k^2M^2_W}}&\qquad (\mathrm{auxiliary}),
\end{array}
\label{eq:pols}
\end{equation}
where $k^2\equiv k_\mu k^\mu$. In the on-shell limit $k^2\rightarrow M^2_W$ the auxiliary polarization vanishes and $\epsilon_0$ approaches the exact on-shell form of longitudinal polarization. 
With the help of eq.~\eqref{eq:identity} one can then rewrite each of the 4 $W$ propagators in each of the diagram that has VBS topology, as
\begin{equation}
\frac{-i\sum_{\lambda=1}^4\epsilon_\lambda^\mu\left(\epsilon_\lambda^\nu\right)^\ast}{k^2 - M_W^2}.
\label{eq:identityy}
\end{equation}
Then the parton-level amplitude $qq\rightarrow qqll'v_lv_l'$ with VBS topology can be decomposed as follows
\begin{eqnarray}
M&\equiv&\frac{\sum_{\lambda_1\lambda_2\lambda_3\lambda_4} M_{\lambda_1}^{q1}M_{\lambda_2}^{q2} M^{WW}_{\lambda_1\lambda_2\lambda_3\lambda_4}M_{\lambda_3}^{l1}M_{\lambda_4}^{l2}}{(k_1^2-M_W^2)(k_2^2-M_W^2)(k_3^2-M_W^2)(k_4^2-M_W^2)}, \qquad \lambda_i\in\{\epsilon_{-},\epsilon_{+},\epsilon_{0},\epsilon_{A}\}.
\label{eq:fullamp}
\end{eqnarray}
The $M_{\lambda_i}^{qi}$ ($M_{\lambda_i}^{li}$)  terms are the trilinear $qqW$ ($llW$) vertices  contracted with  $\epsilon^\ast$ ($\epsilon$) of eq.~\eqref{eq:identityy}, while the  $M^{WW}_{\lambda_1\lambda_2\lambda_3\lambda_4}$ term is  the (off-shell) $WW$ elastic scattering amplitude. 
The sum over $i$ includes necessarily polarization configurations in which  the $W$  polarizations are auxiliary.
Now, the effect of dimension-8 operators grows with the scattering energy  $M_{WW}>>M_W$ and modifies significantly helicity amplitudes so that  deviations from the SM behavior become non-negligible. Since the off-shellness $k^2_i$   are  suppressed dynamically by propagators $1/(k_i^2-M^2_W)$, in this kinematic  limit  the scattered vector bosons must be fast, $|\vec{k_i}|\sim E_i>>M_W$,    Therefore in the high $M_{WW}$ region $\epsilon^\mu_0 \sim \epsilon^\mu_A$ and approach the on-shell form of the longitudinal polarization vector. As a result,  the sum in eq.~(\ref{eq:identityy}) runs effectively over $\epsilon_i=\epsilon_0,\, \epsilon_+,\, \epsilon_-$ and the off-shell helicity amplitude can be approximated by the on-shell one, accounting corrections of order  $\sqrt{(k^2-M^2_W)/(k^2M^2_W)}$ or $1/\sqrt{k^2}$. Therefore in the following subsections we will discuss in detail the high-energy behavior of on-shell $WW$ scattering in the presence of contributions from dimension-8 operators and the unitarity bound.

\subsection{The on-shell $WW$ scattering  and the helicity amplitudes}
Let us consider  the elastic on-shell $W^+W^+\rightarrow W^+W^+$ in the presence of BSM part represented by  a single dimension-8 operator, as in an ''EFT model''.   The scattering amplitude $iM$ can be written as:
\begin{equation}
iM = A_{SM}+A_{BSM},
\label{eq:pk1}
\end{equation}
where $A_{SM}$ denotes the SM part and $A_{BSM}$ represents the BSM part that depends on the Wilson coefficients $f_i$. 

For the on-shell $W$ bosons we choose to work in the helicity basis in which the polarizations are $\epsilon^\mu_i$ with $i=+,\,-,\,0$.. There are in total $3^4=81$ helicity amplitudes $iM(ij\rightarrow kl)$ corresponding to helicity configurations $(ijkl)$ in the $WW\to WW$ scattering process.  The total unpolarized on-shell $WW$ cross section can schematically be written as:
\begin{equation}
\sigma\sim\frac{1}{9}\ \sum_{i,j,k,l}\ \ \left|A_{SM}(ij\rightarrow kl)\right|^2 + (A_{SM}(ij\rightarrow kl)A_{BSM}(ij\rightarrow kl)^{\ast} + h.c.) + \left|A_{BSM}(ij\rightarrow kl)\right|^2
\label{eq:pk3}
\end{equation}
Since  there are orders of magnitude differences concerning contributions of different helicity amplitudes  to the total cross section it is convenient, using  discrete symmetries $\mathcal{P}$ and $\mathcal{T}$ and Bose statistics, to divide 
 81 polarization amplitudes into classes. Amplitudes from the same class yield the same contribution to (polarized) cross sections. 
Hence, in practice one can consider a reduced number of 13 independent polarization classes, taking into account their multiplicities when computing the cross section.  It turns out that 
only a few helicity configurations contribute non-negligibly at high $WW$ scattering energy. We refer to such helicities as saturating helicities. 

\begin{figure}[h]
\begin{center}
\includegraphics[width=0.8\linewidth]{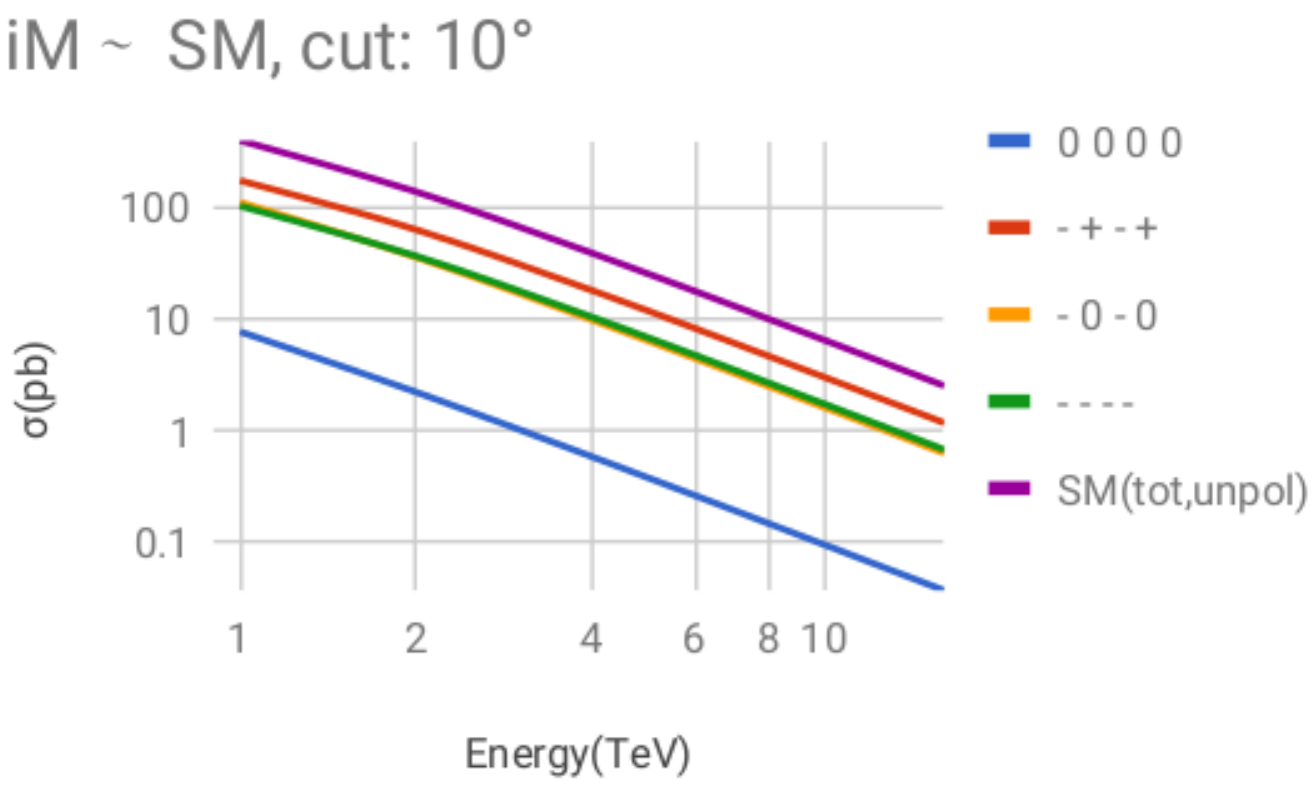} 
\end{center}
\caption{Contributions of various helicitity configurations (multiplicity taken into account) to the total unpolarized cross section as a function of the center-of-mass collision energy ($E_{CM} \equiv \sqrt{s}$, in TeV) in the SM.  The total unpolarized cross section is shown in violet.}
\label{fig:polSM}
\end{figure}

For the case of the SM the contribution from  the saturating helicities  to the total unpolarized cross-section   is shown in Fig.~\ref{fig:polSM}. 
The four saturating helicity configurations  are the only ones whose scattering amplitude is asymptotically constant in energy. The remaining helicity configurations behave asymptotically at most as $1/s$, hence their contribution is strongly suppressed at large $s=M^2_{WW}$. All on-shell $WW$ cross sections are computed with a $10^{\circ}$ cut in the forward and backward scattering regions (which is explicitly written in the plots).

\begin{figure}[h] 
  \begin{tabular}{cc}
      \includegraphics[width=0.5\linewidth]{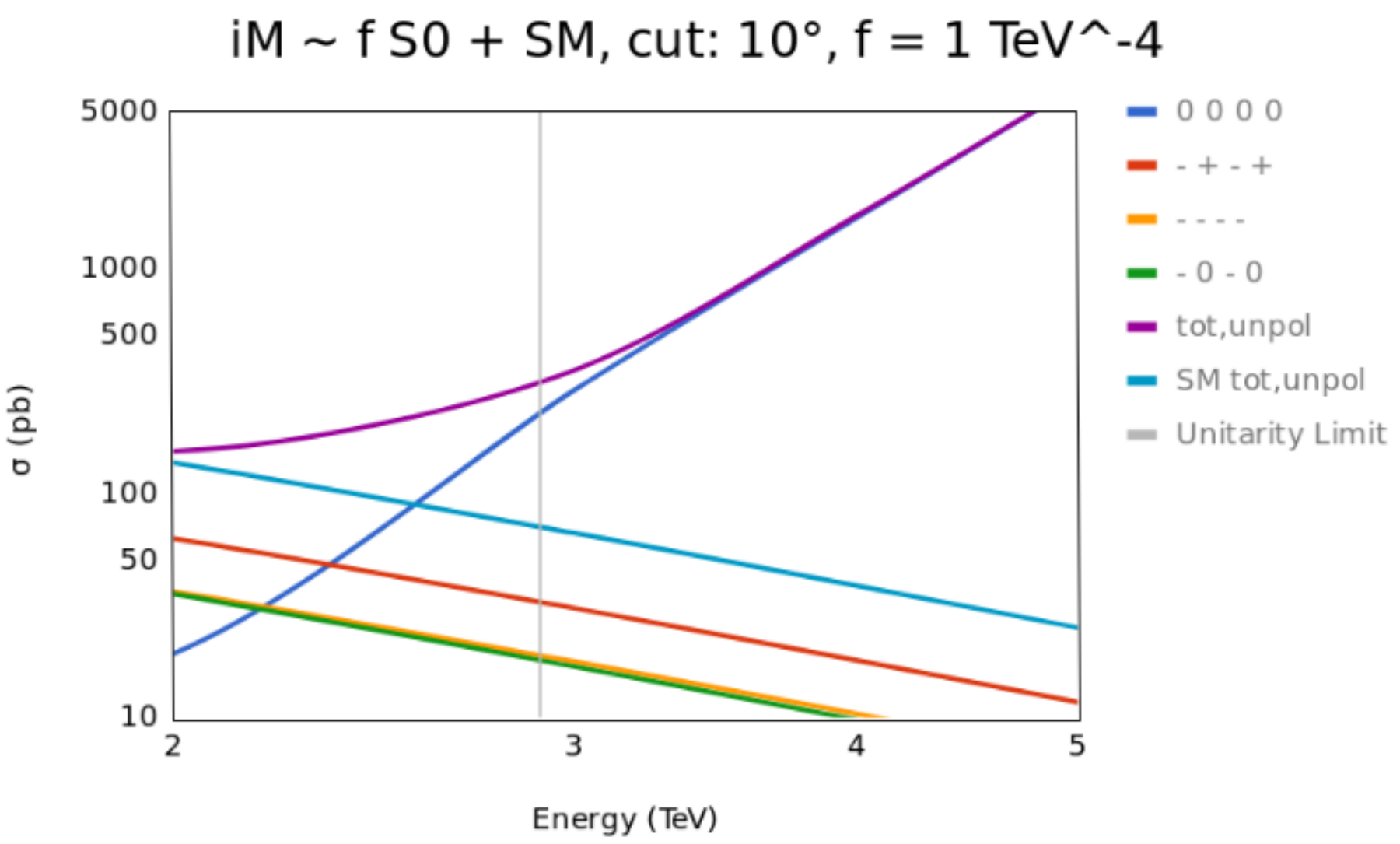} 
& \includegraphics[width=0.5\linewidth]{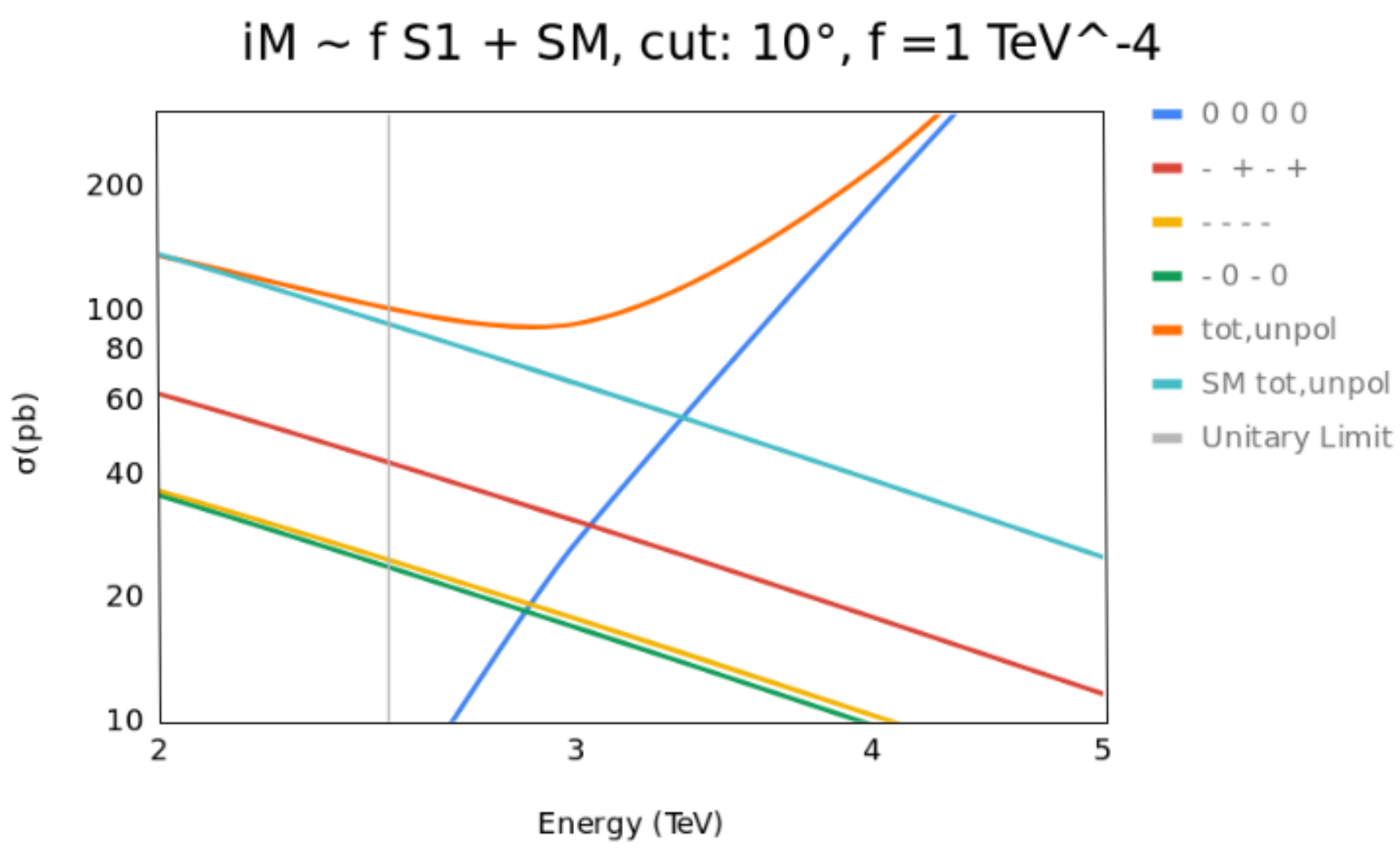} \\
\includegraphics[width=0.5\linewidth]{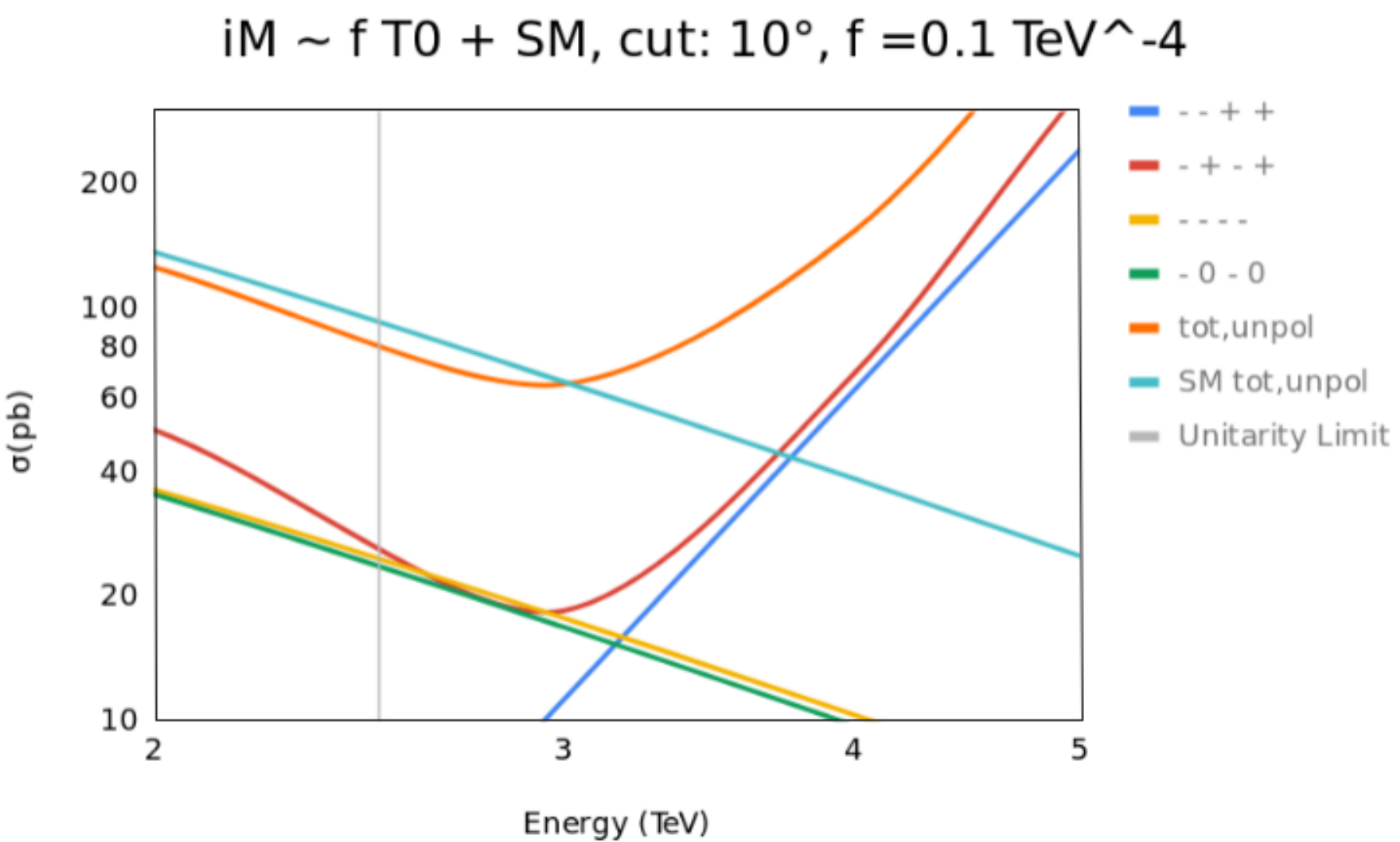} 
& \includegraphics[width=0.5\linewidth]{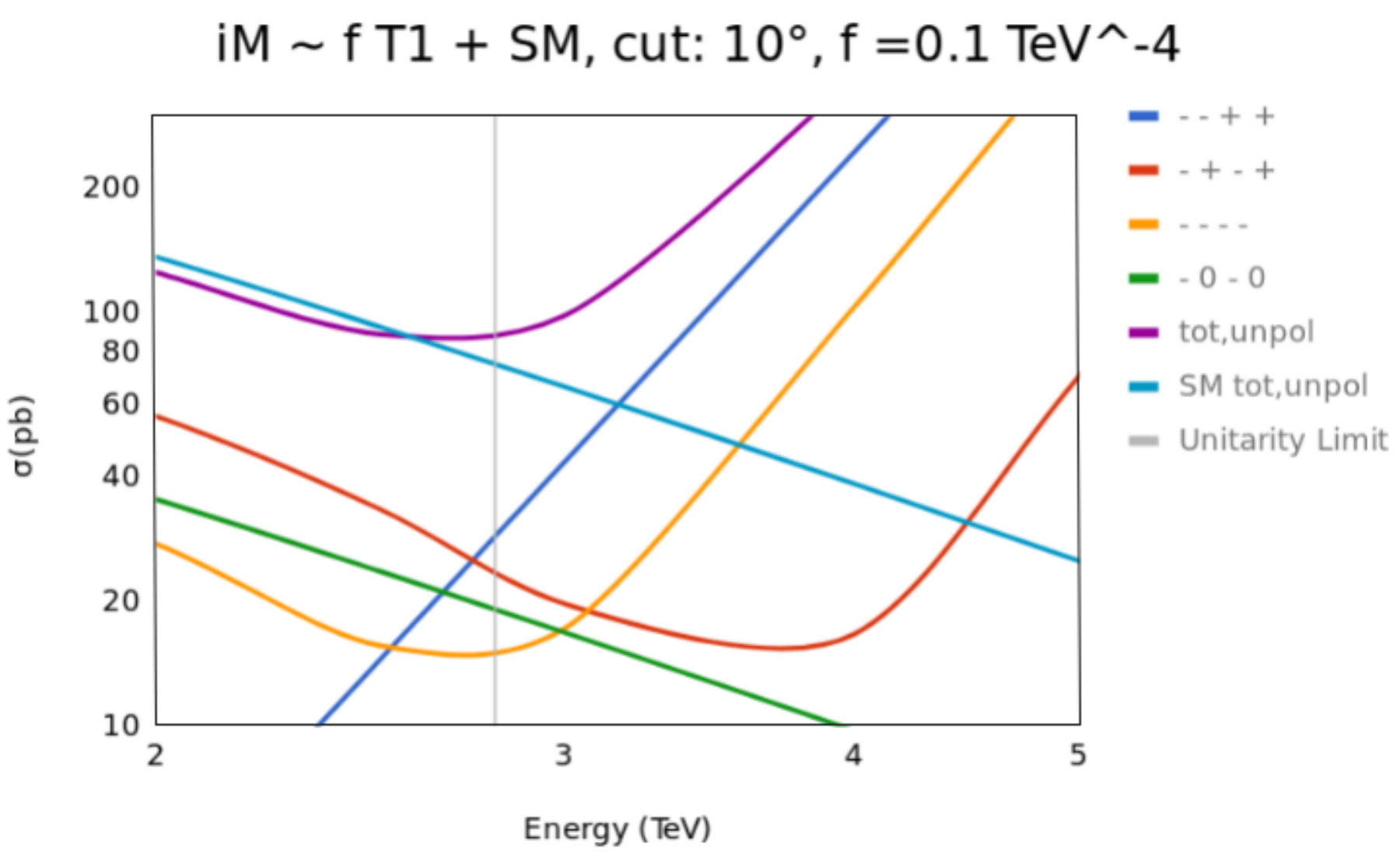} \\
\includegraphics[width=0.5\linewidth]{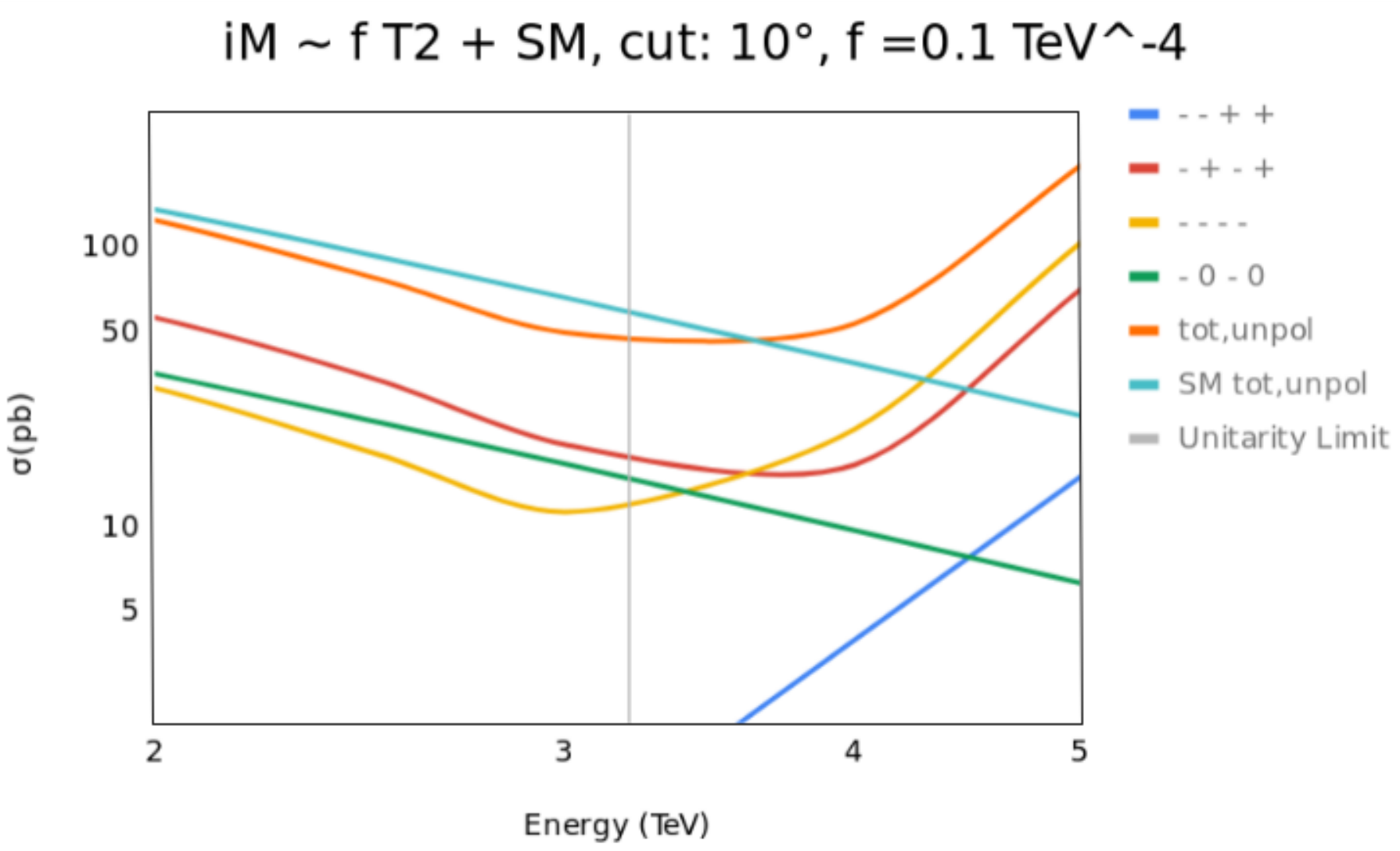} 
& \includegraphics[width=0.5\linewidth]{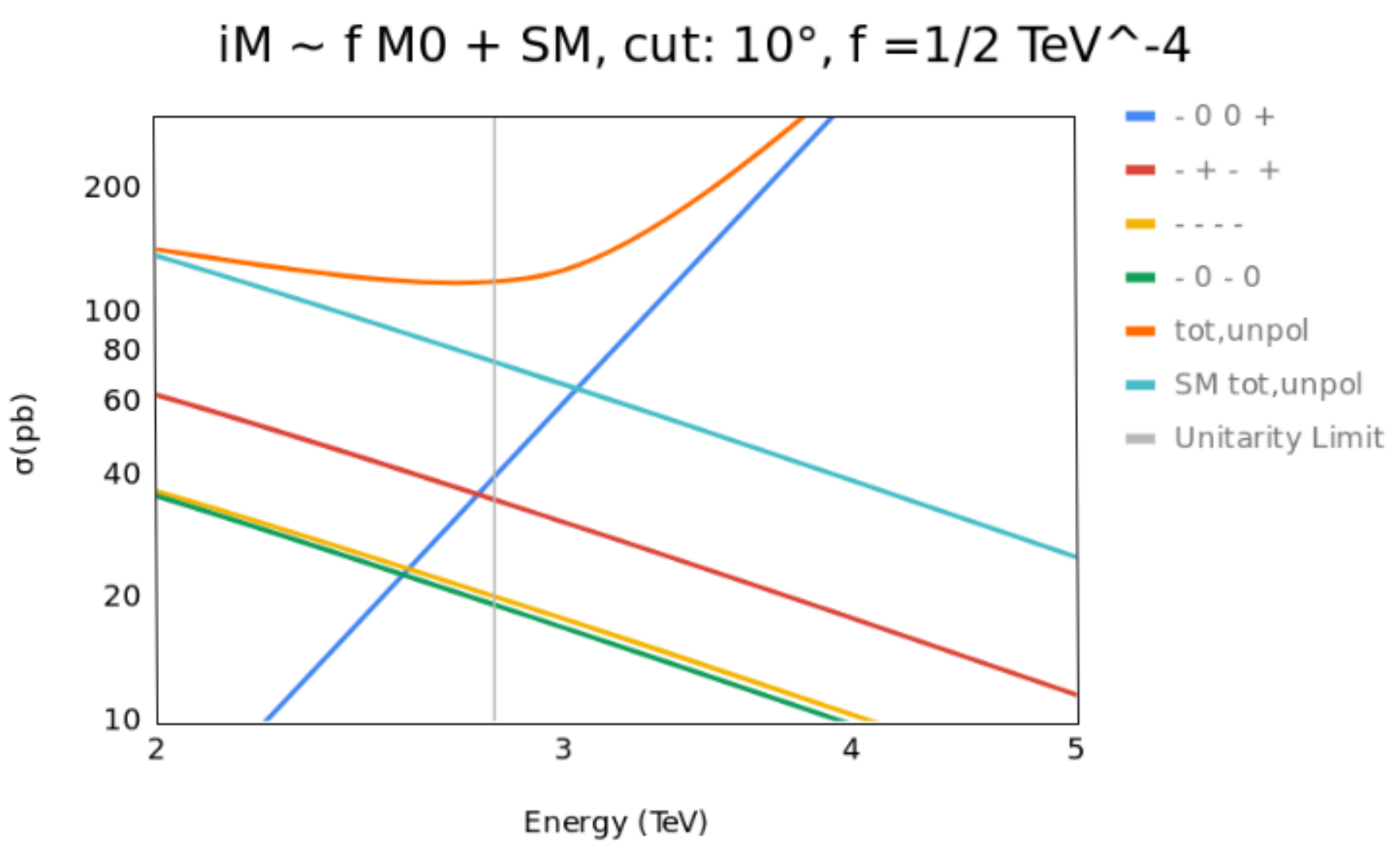} \\
\includegraphics[width=0.5\linewidth]{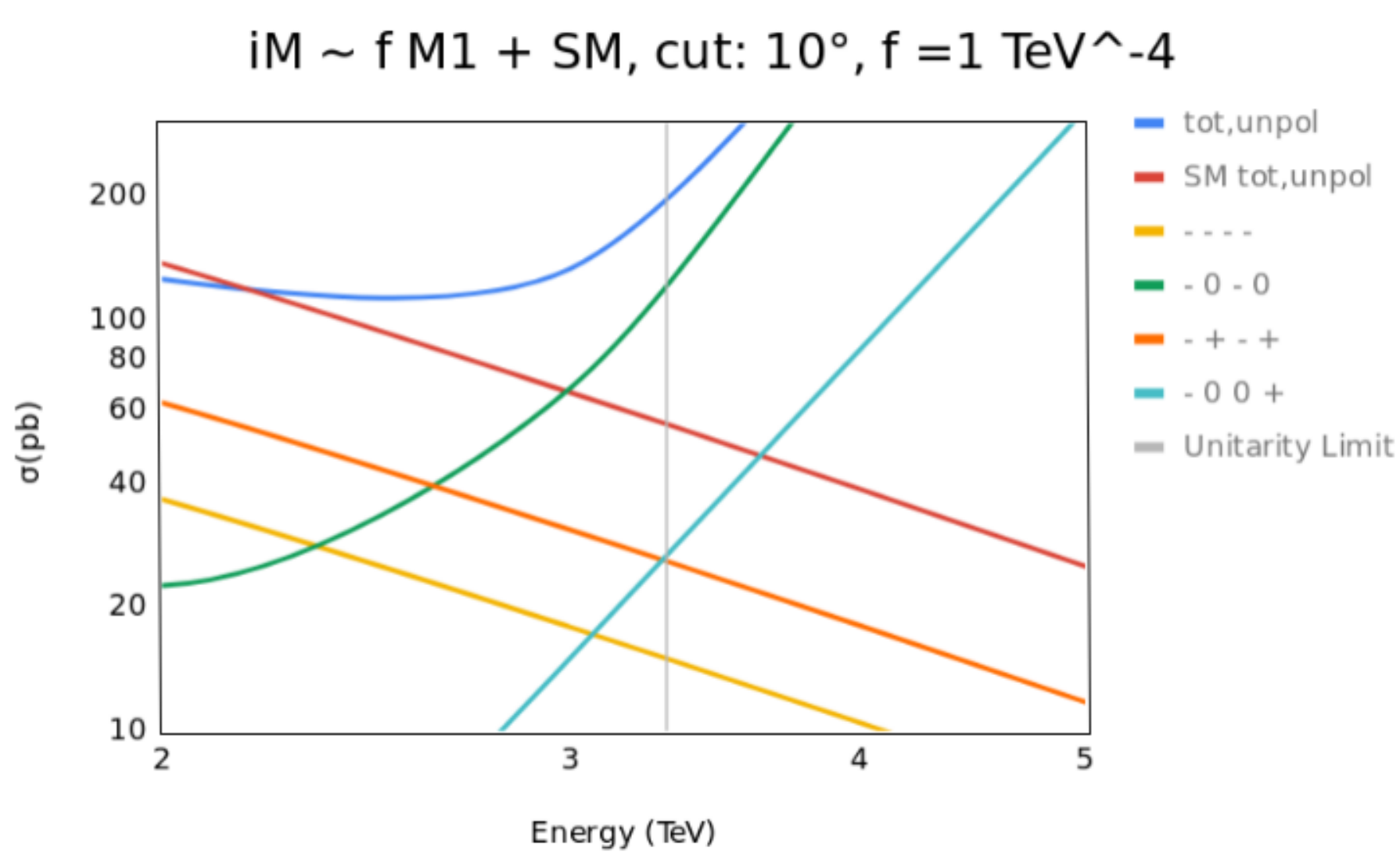} 
& \includegraphics[width=0.5\linewidth]{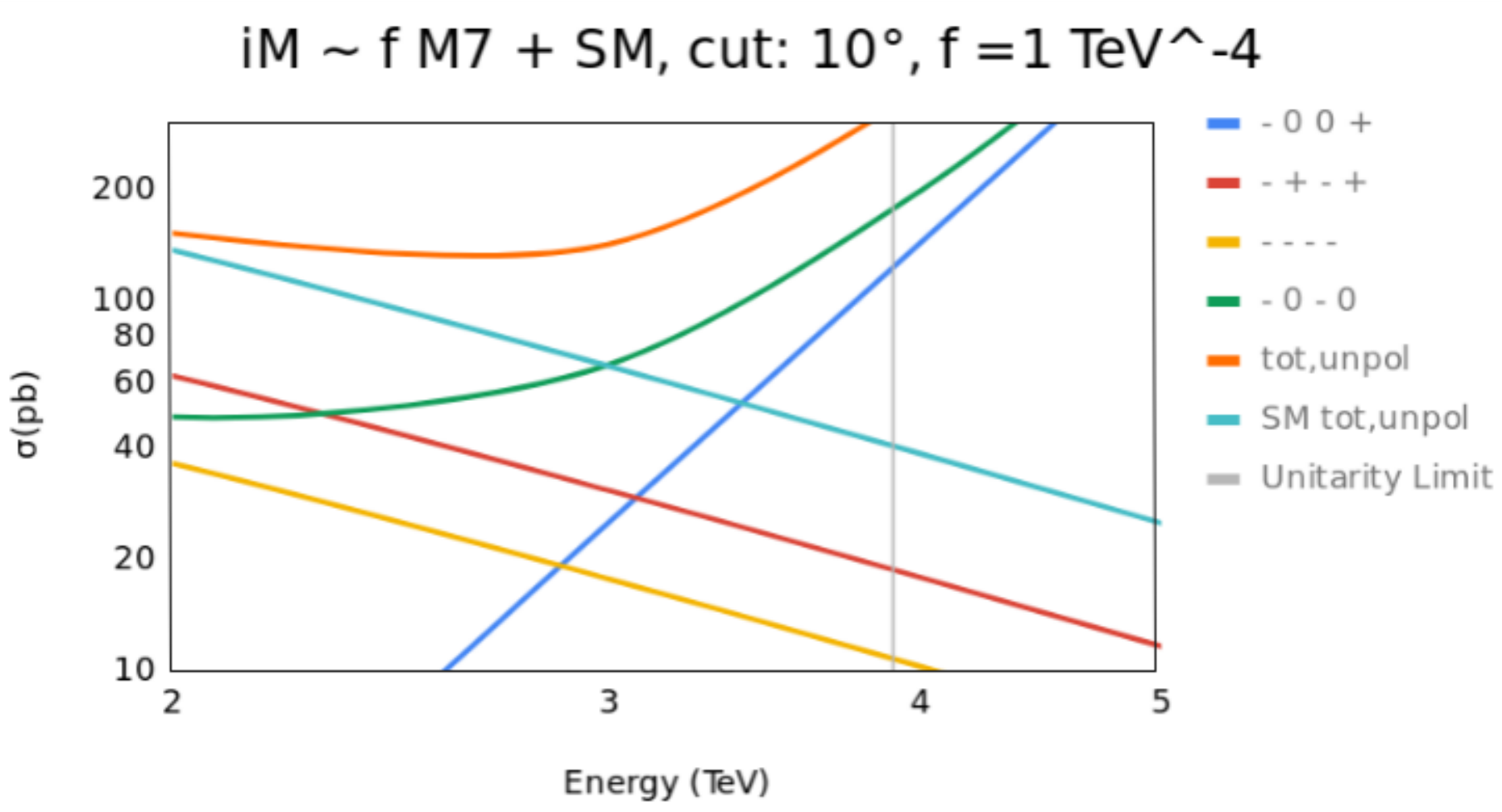} 
  \end{tabular}
\caption{Contributions of the polarized cross sections (multiplicity taken into account) as functions of the center-of-mass collision energy
($E_{CM} \equiv \sqrt{s}$, in TeV) for chosen values of $f_i>0$. The remaining (not shown) polarized cross sections are negligibly small. In each plot shown is in addition the total cross section of a EFT ''model'' and the total cross section in the SM.}
\label{fig:polsPos}
\end{figure}

\begin{figure}[h] 
  \begin{tabular}{cc}
      \includegraphics[width=0.5\linewidth]{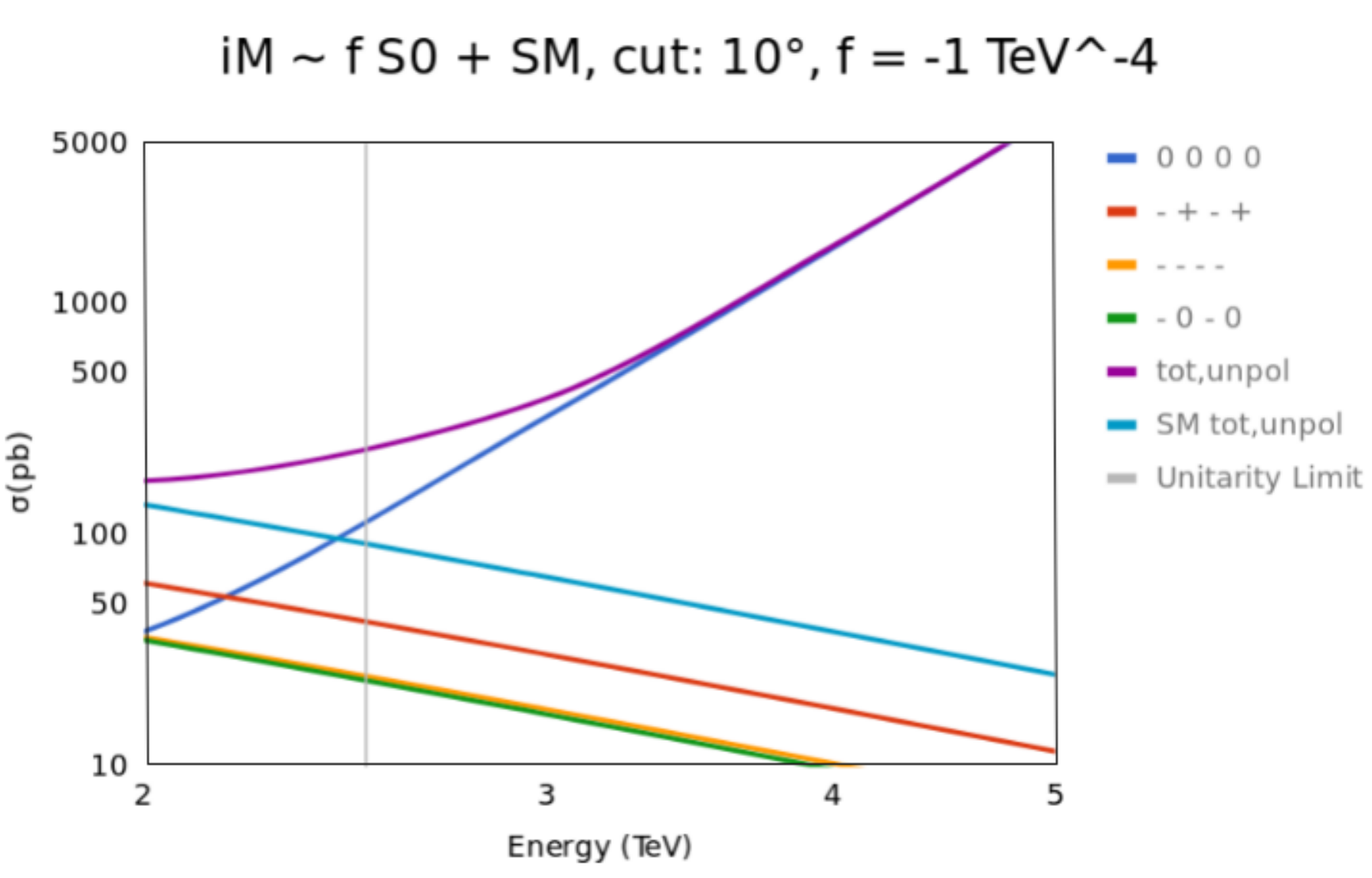} 
& \includegraphics[width=0.5\linewidth]{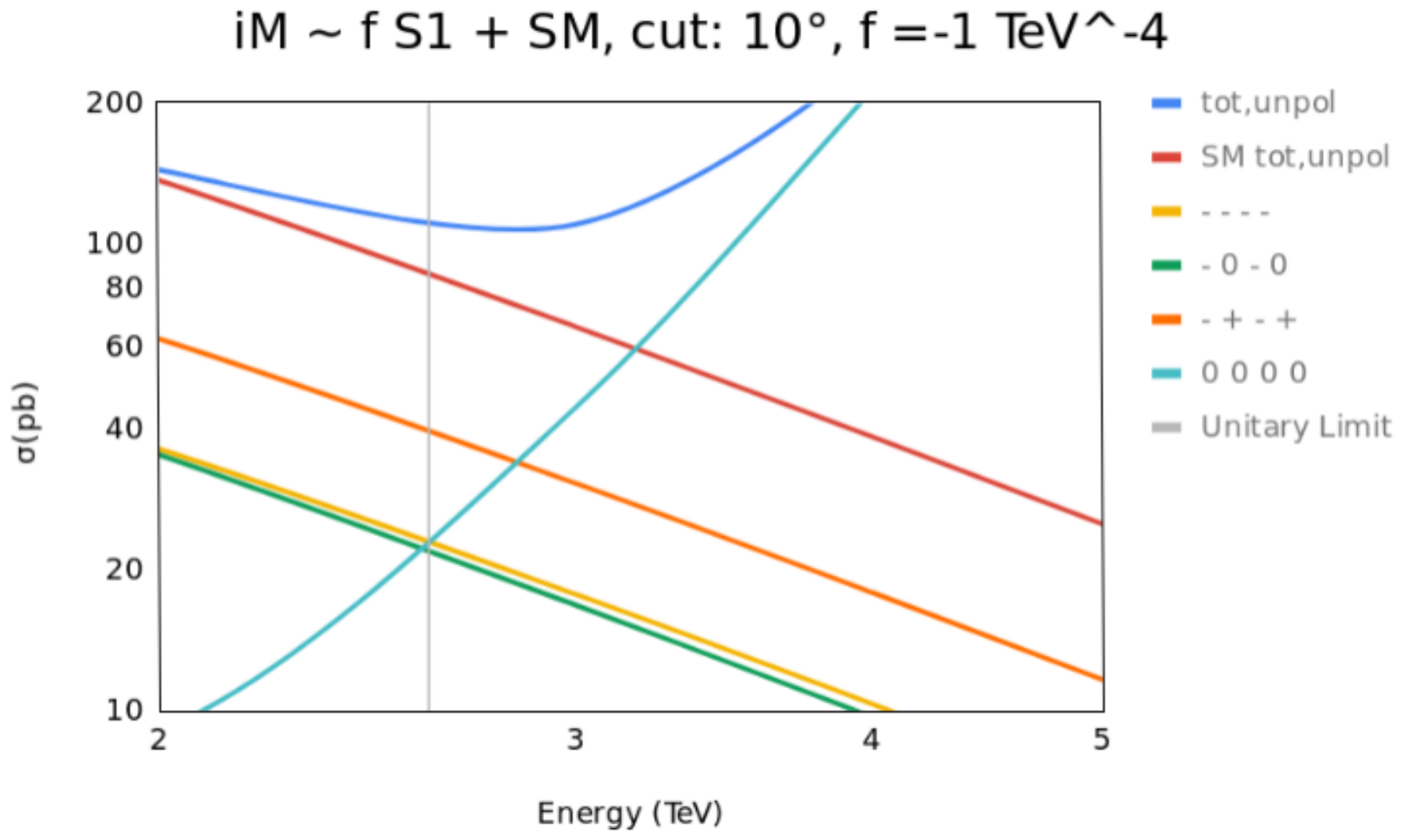} \\
\includegraphics[width=0.5\linewidth]{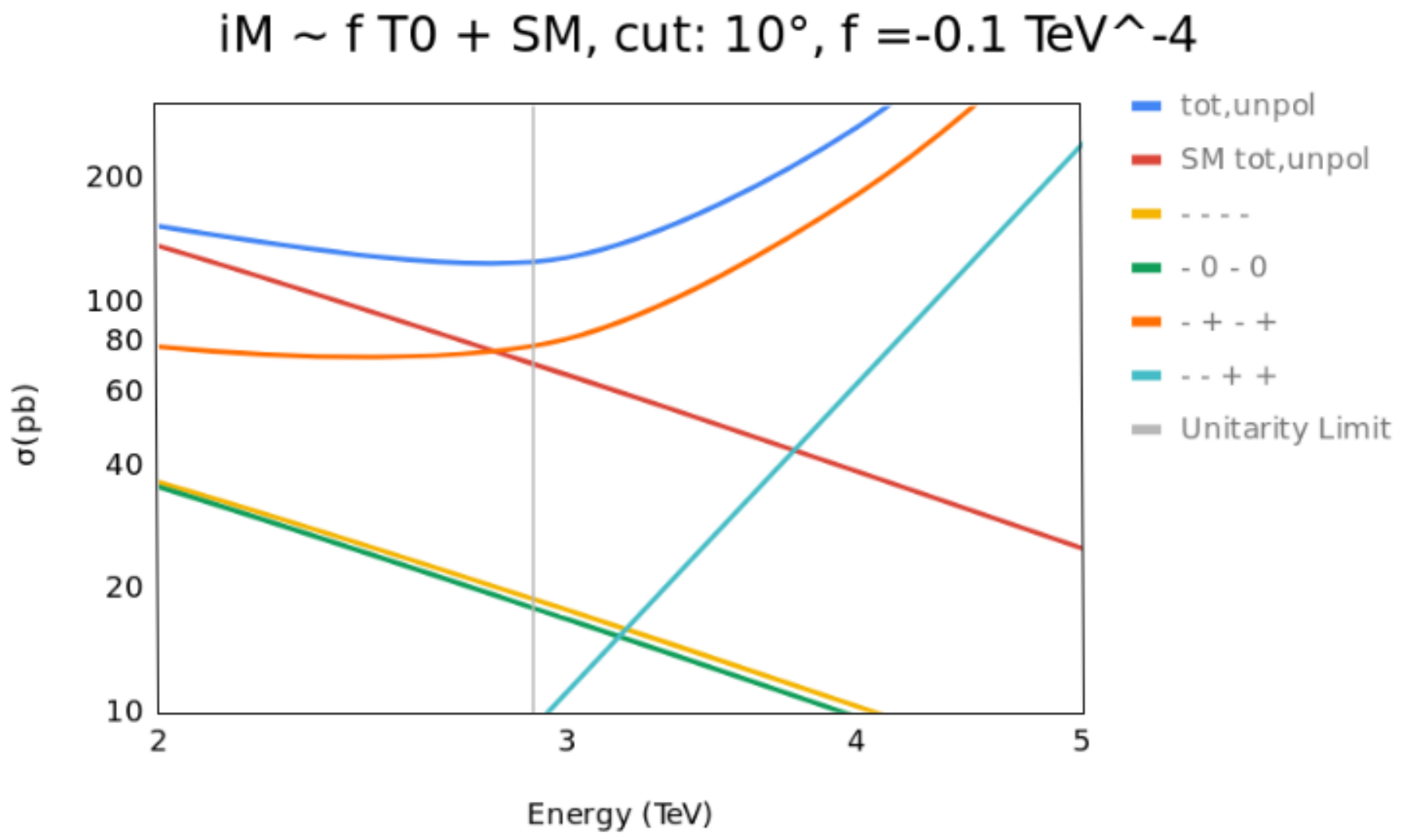} 
& \includegraphics[width=0.5\linewidth]{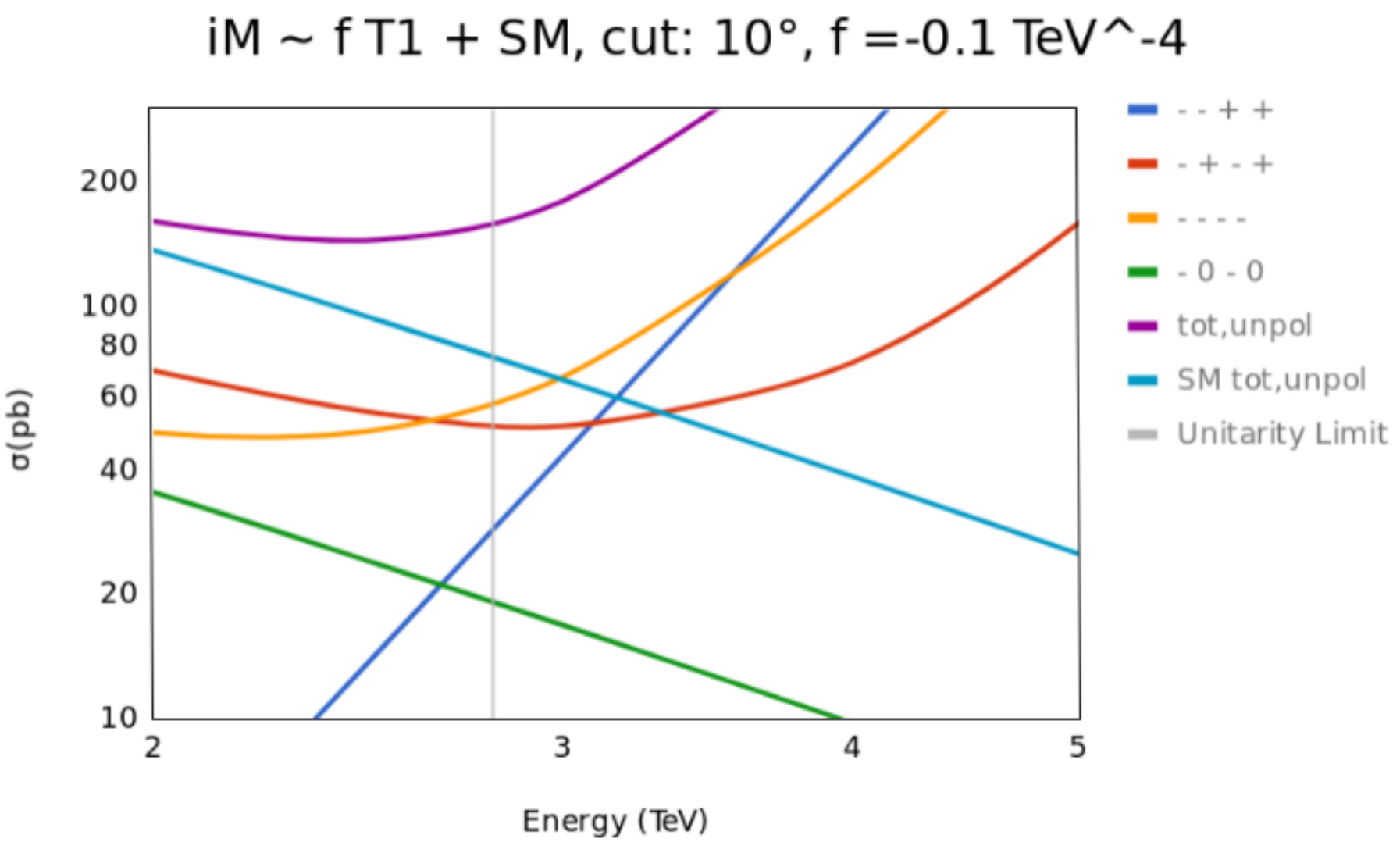} \\
\includegraphics[width=0.5\linewidth]{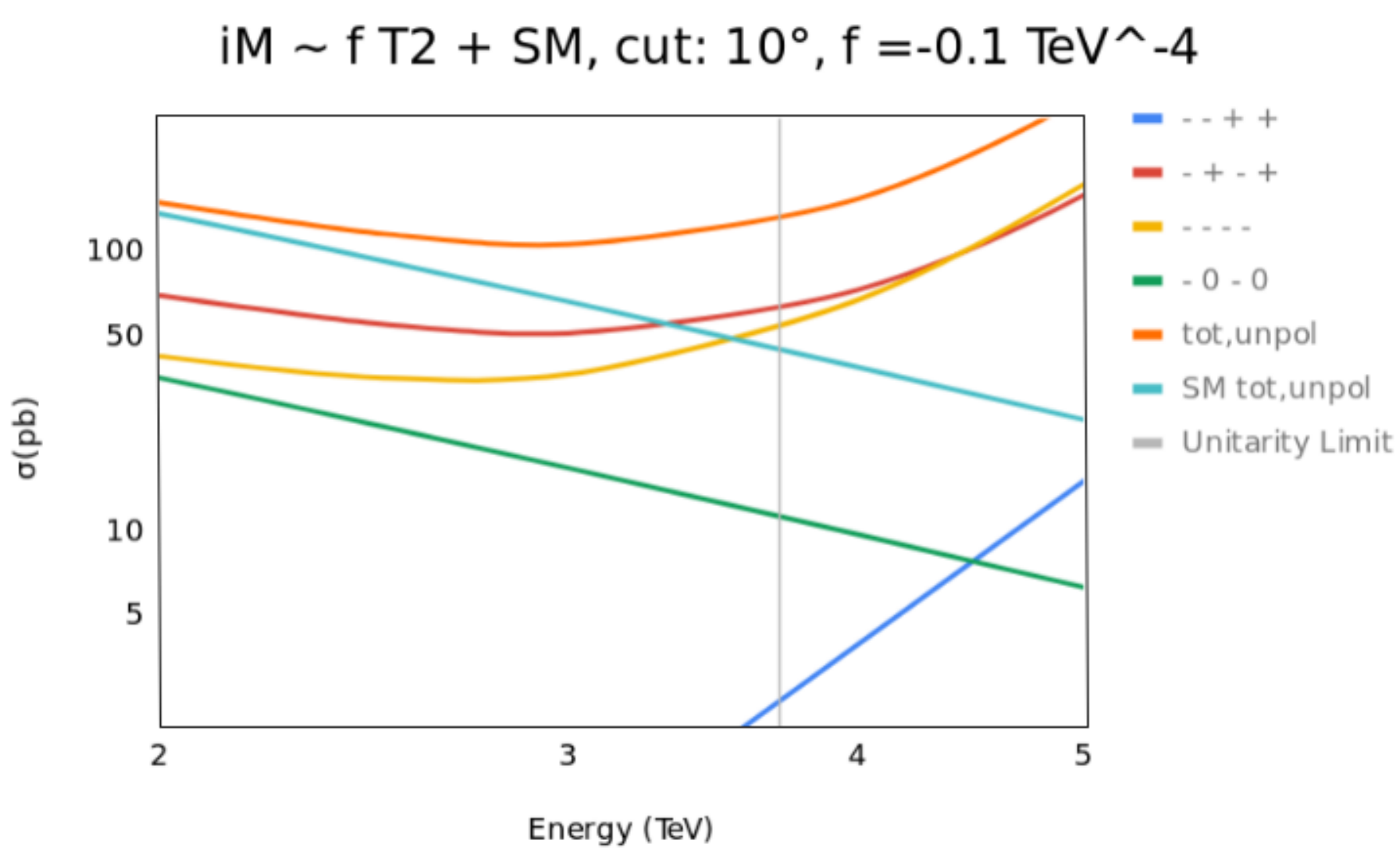} 
& \includegraphics[width=0.5\linewidth]{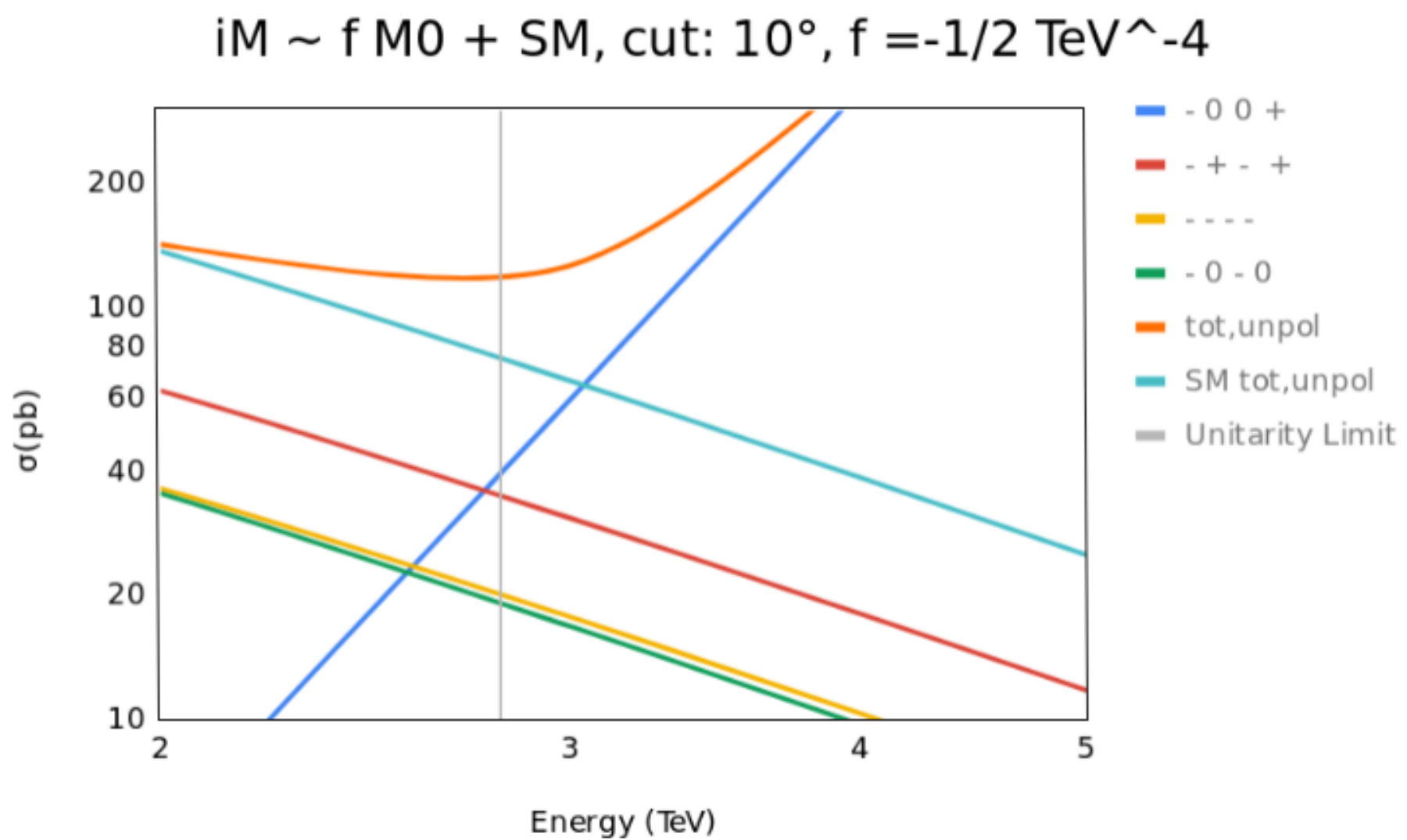} \\
\includegraphics[width=0.5\linewidth]{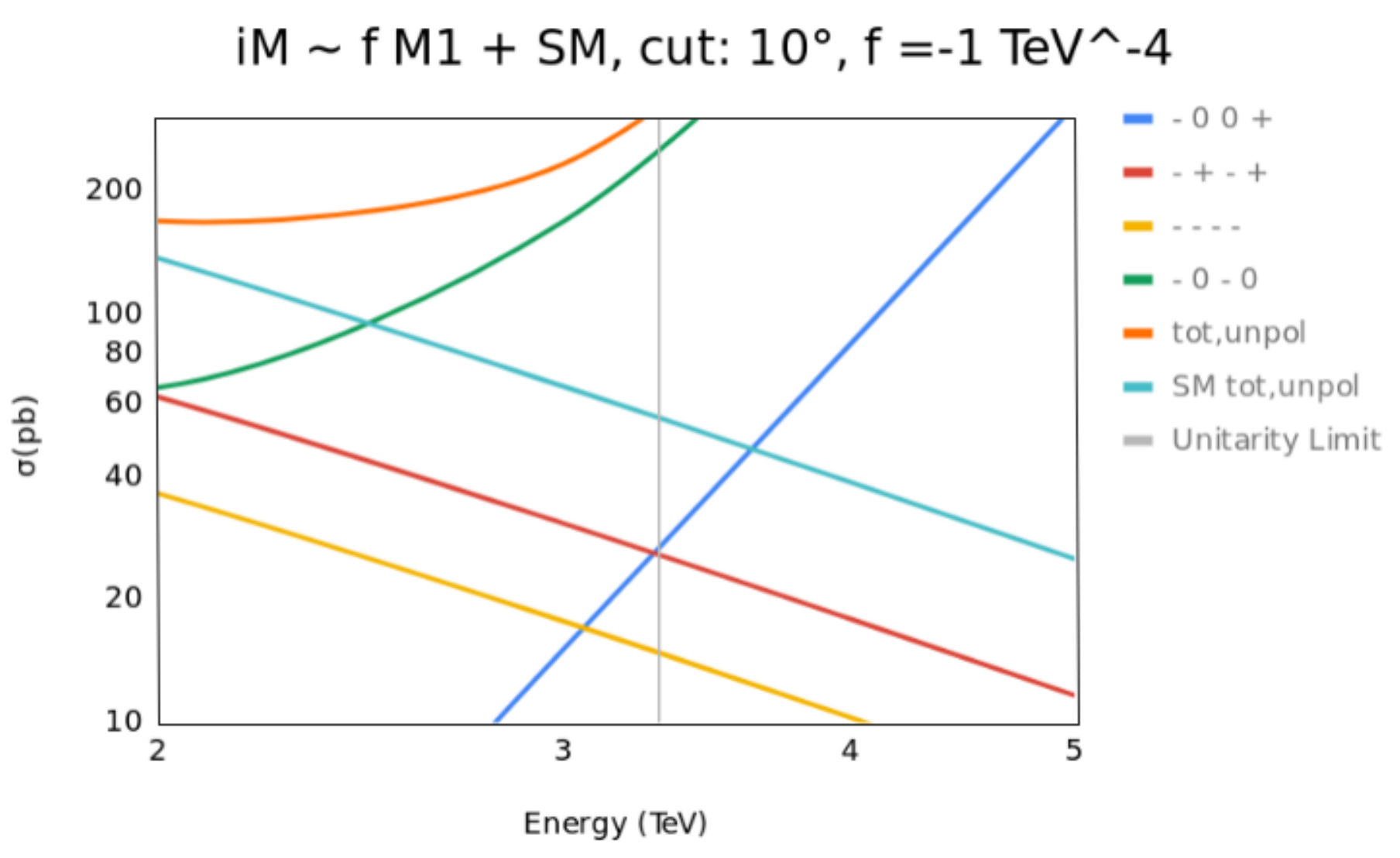} 
& \includegraphics[width=0.5\linewidth]{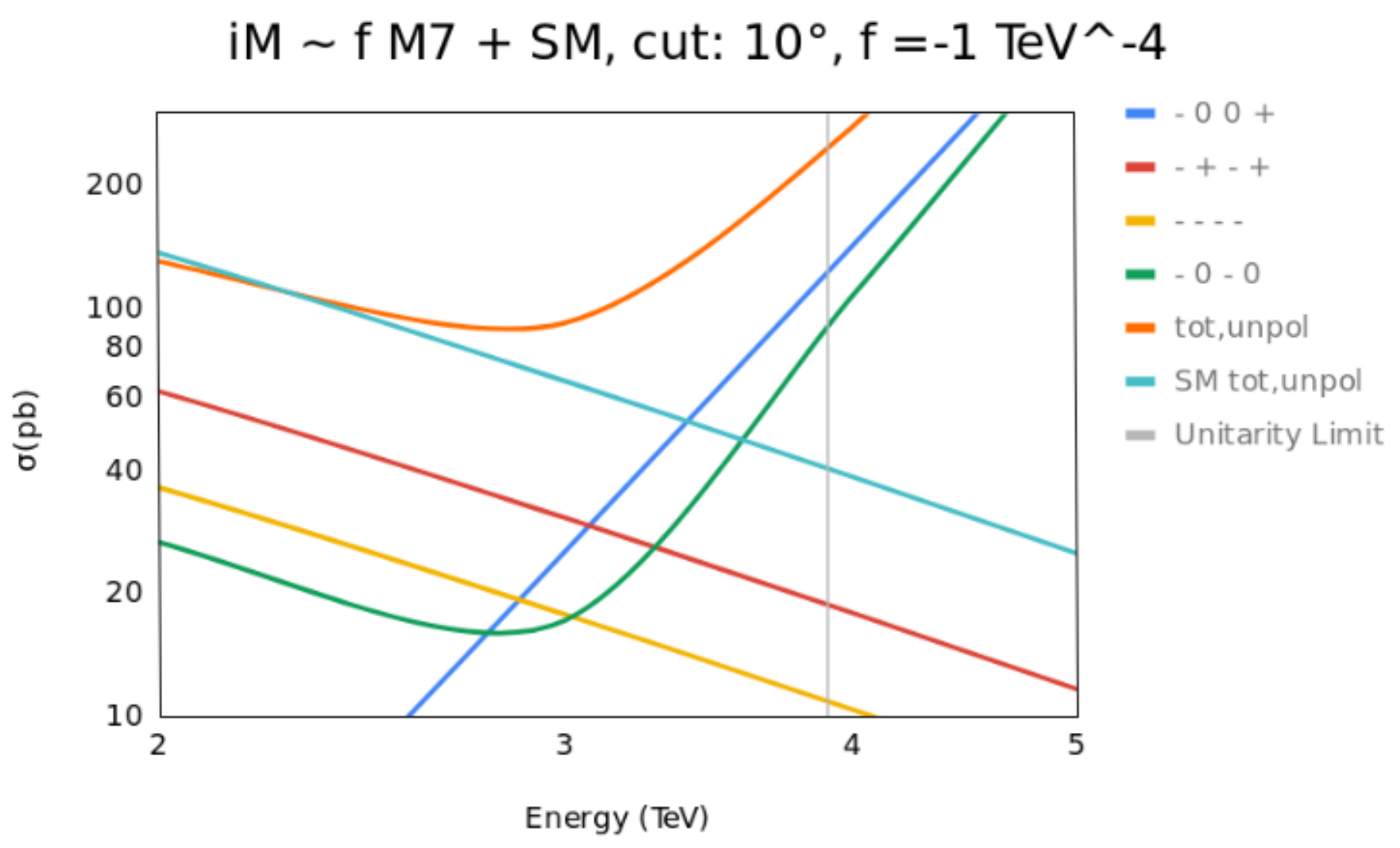} 
  \end{tabular}
\caption{Contributions of the polarized cross sections (multiplicity taken into account) as functions of the center-of-mass collision energy
($E_{CM} \equiv \sqrt{s}$, in TeV) for chosen values of $f_i<0$. The remaining (not shown) polarized cross sections are negligibly small. In each plot shown is in addition the total cross section of a EFT ''model'' and the total cross section in the SM.}
\label{fig:polsNeg}
\end{figure}

In the presence of $n=8$ operators some of the saturating helicities grow with $s$, maximally as $s^2$. 
The corresponding case for each ``EFT model'' studied is shown in figs.~\ref{fig:polsPos} and~\ref{fig:polsNeg} for $f_i>0$ and $f_i<0$, respectively.
In fact, for each ``EFT model'' there is at least one polarization configuration with the asymptotic $s^2$ energy dependence providing  dominant contribution to the unpolarized cross section at $M_{WW}=M^U$. In particular, in the case of ``EFT models'' with scalar operators ($S$)  only  the amplitude with all $W$ bosons polarized longitudinally grows as $s^2$. In the case of transverse  operators ($T$)   some  amplitudes with all $W$ polarized transversally grow as $s^2$, while  for the case of mixed operators ($M$) it happens for amplitudes with two longitudinal and  two  transverse polarizations. It  follows from $D_\mu \Phi$ and $W_{\mu\nu}$ building blocks of BSM operators  which project mostly on the longitudinal and transverse modes, respectively. It is interesting to notice, however,  that for different $S,\, T$  and $M$  distinct  polarization configurations of the outgoing $W$'s  dominate the total cross section at large $M_{WW}$. Measurement of final state $W$ polarizations would give an insight to the dynamics of their interactions.

\begin{figure}[h] 
  \begin{tabular}{cc}
      \includegraphics[width=0.5\linewidth]{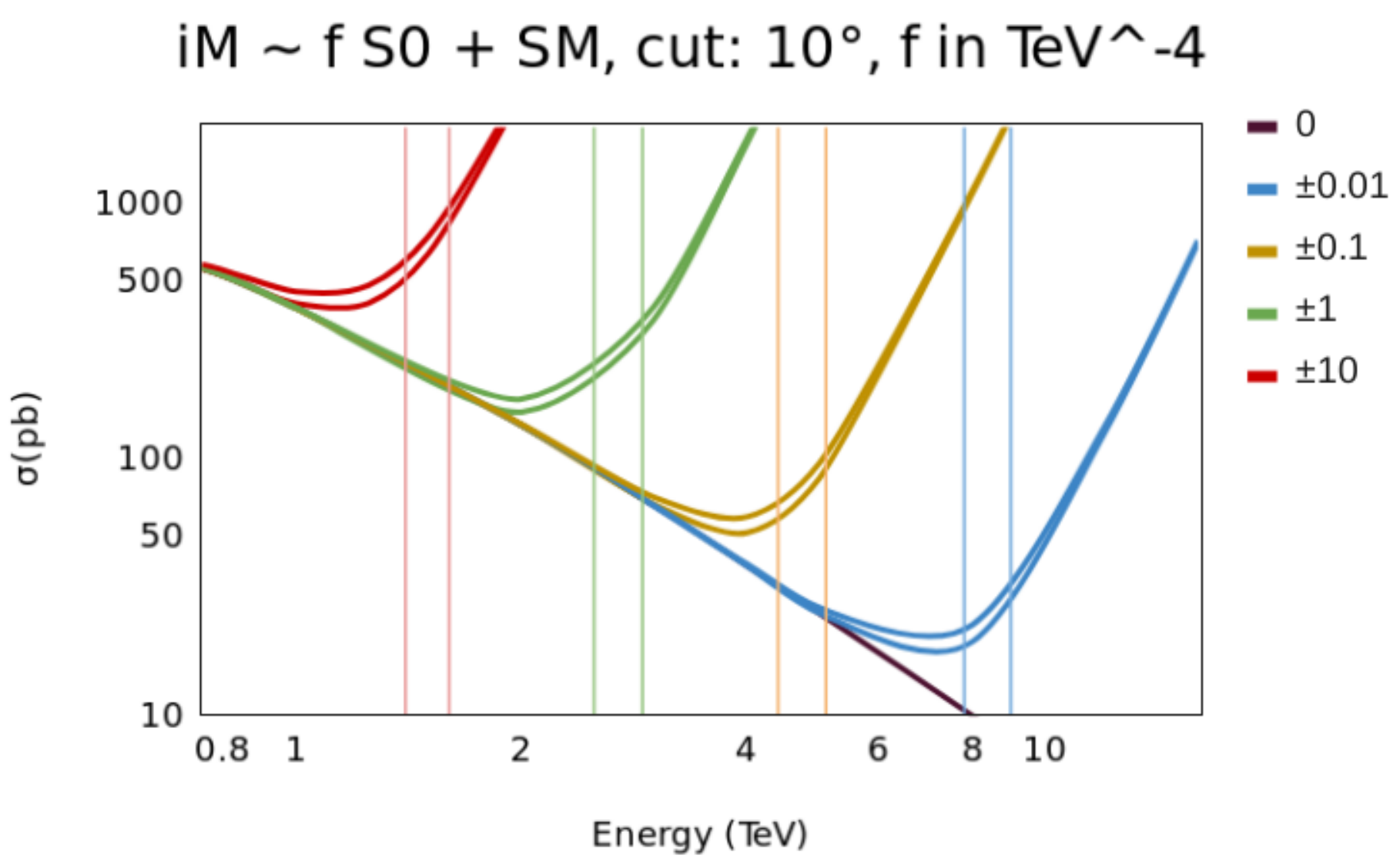} 
& \includegraphics[width=0.5\linewidth]{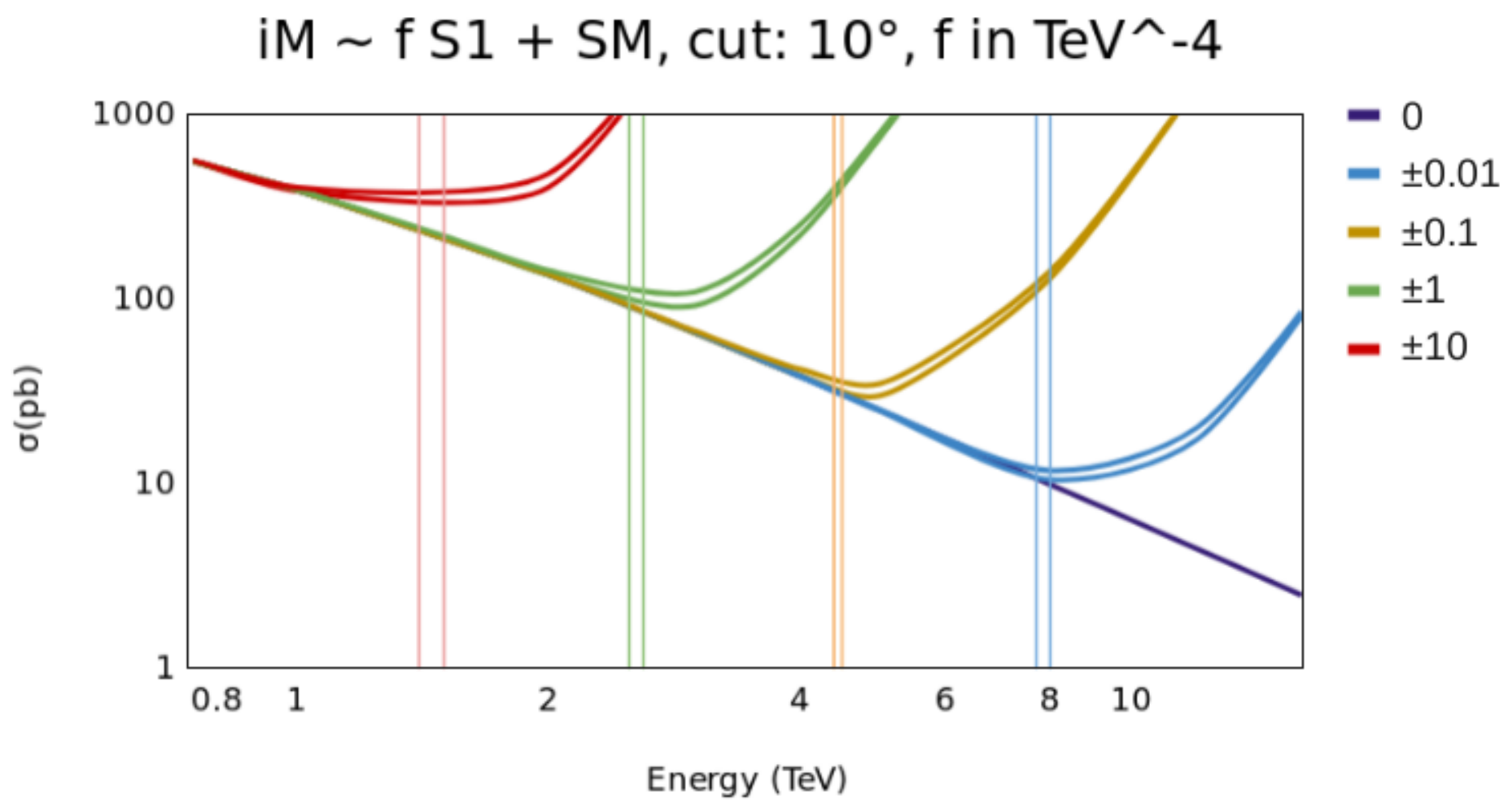} \\
\includegraphics[width=0.5\linewidth]{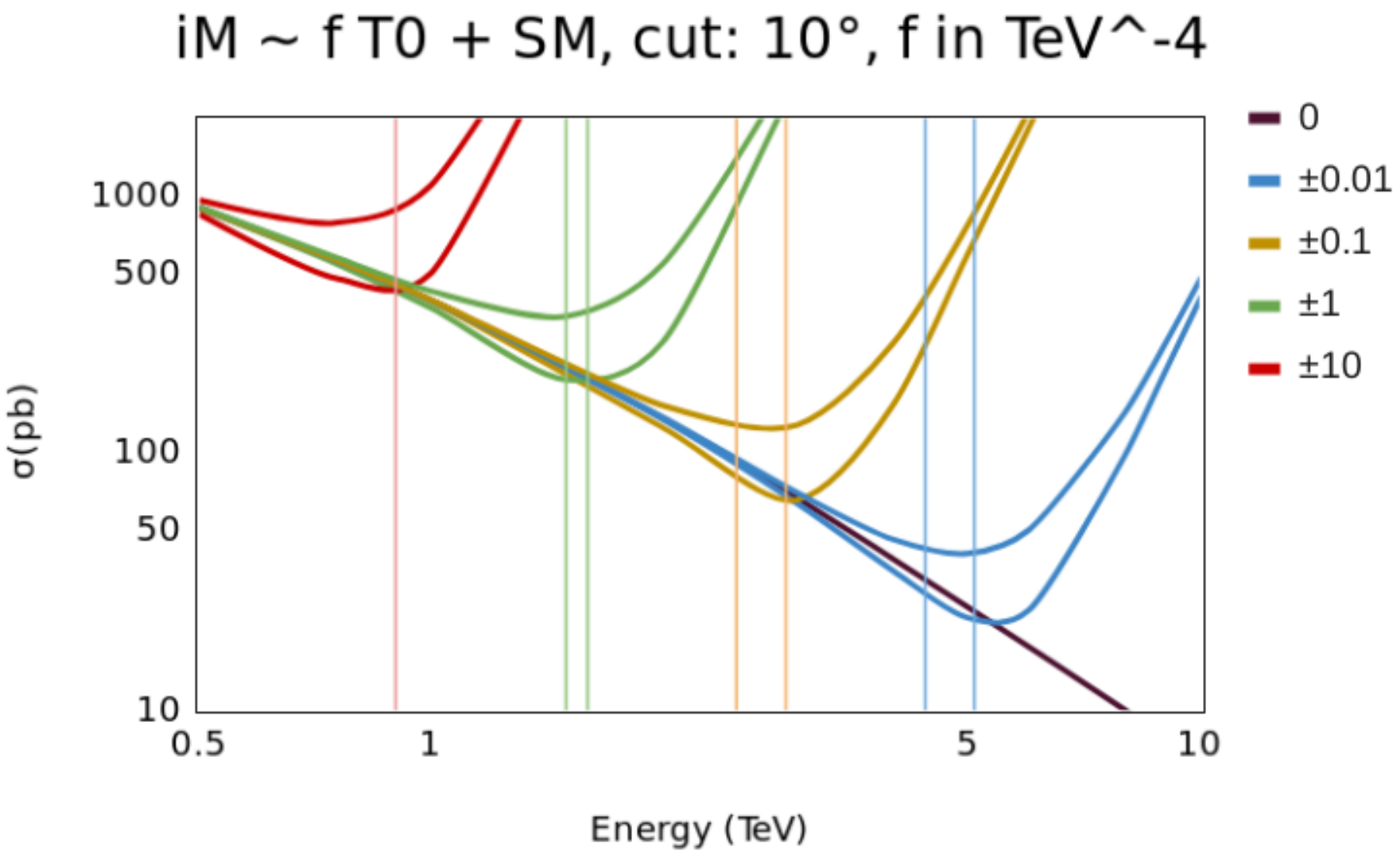} 
& \includegraphics[width=0.5\linewidth]{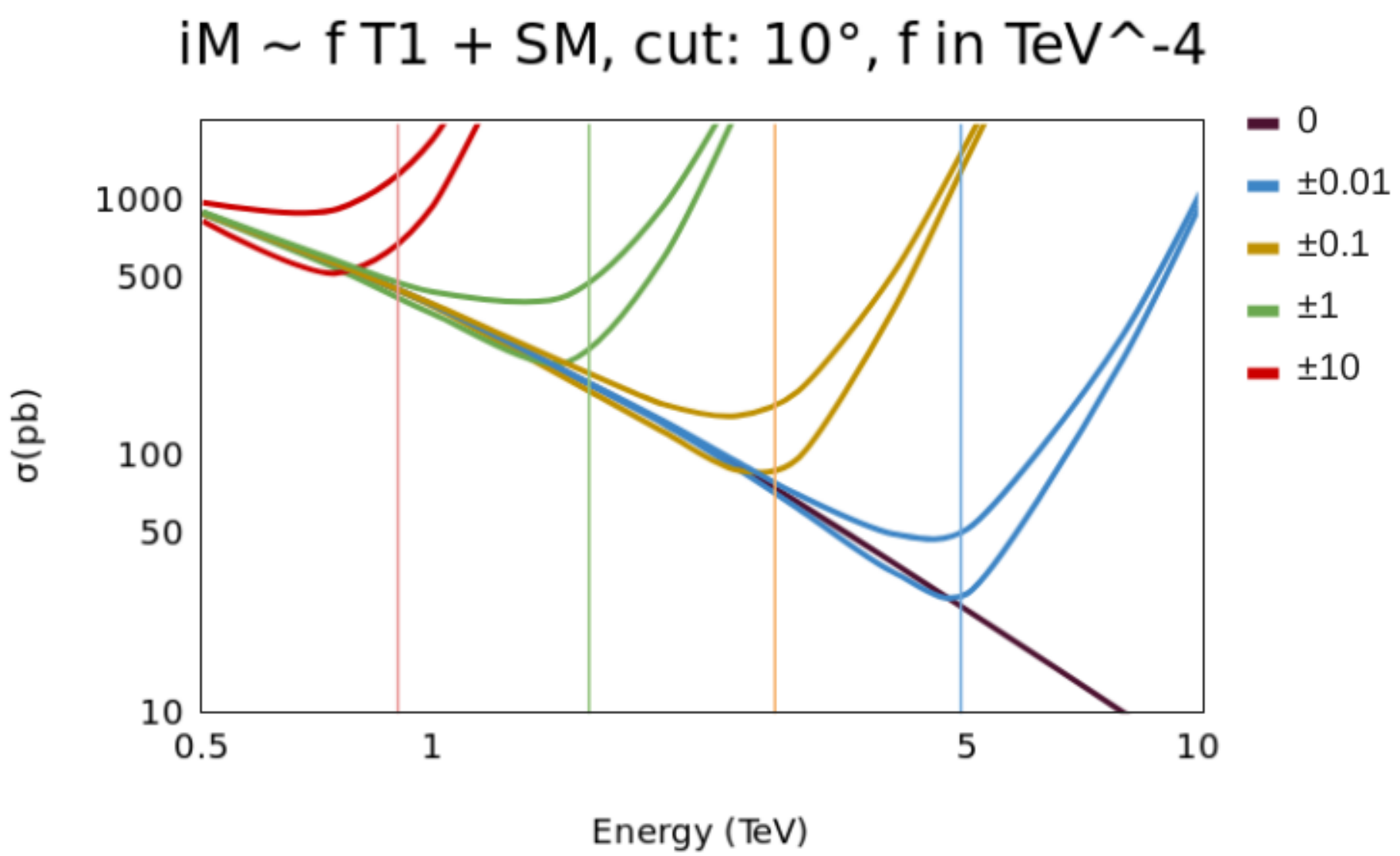} \\
\includegraphics[width=0.5\linewidth]{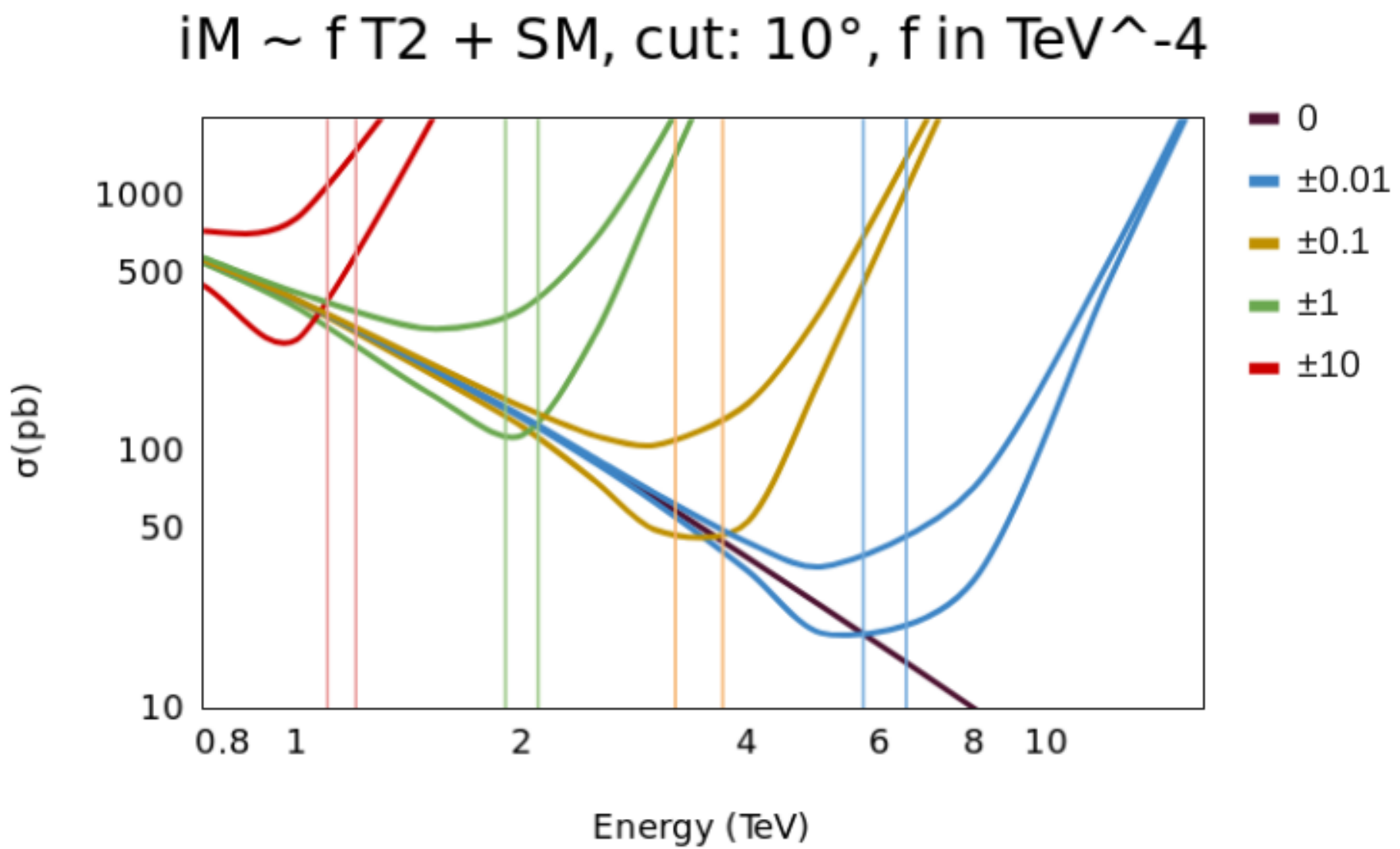} 
& \includegraphics[width=0.5\linewidth]{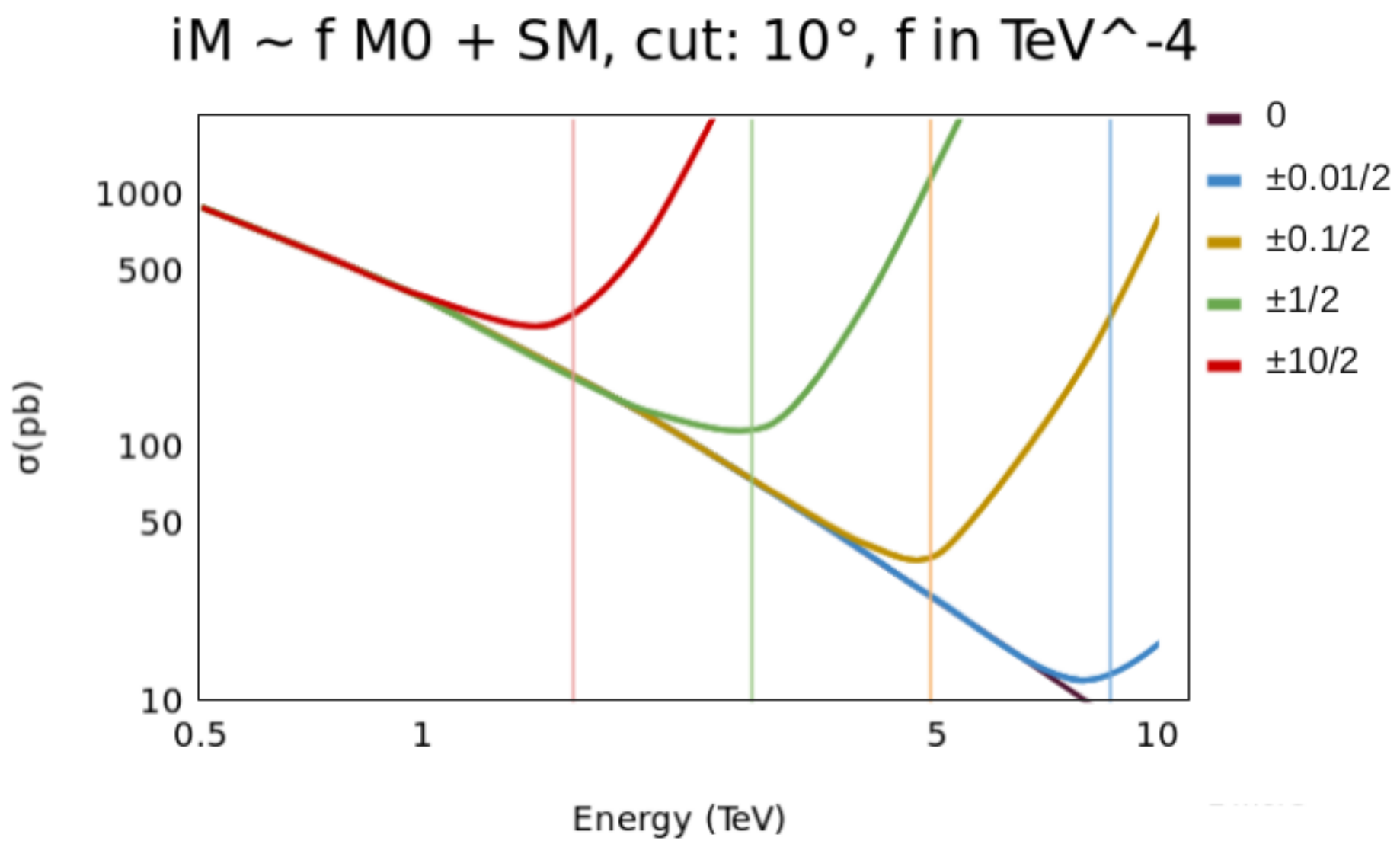} \\
\includegraphics[width=0.5\linewidth]{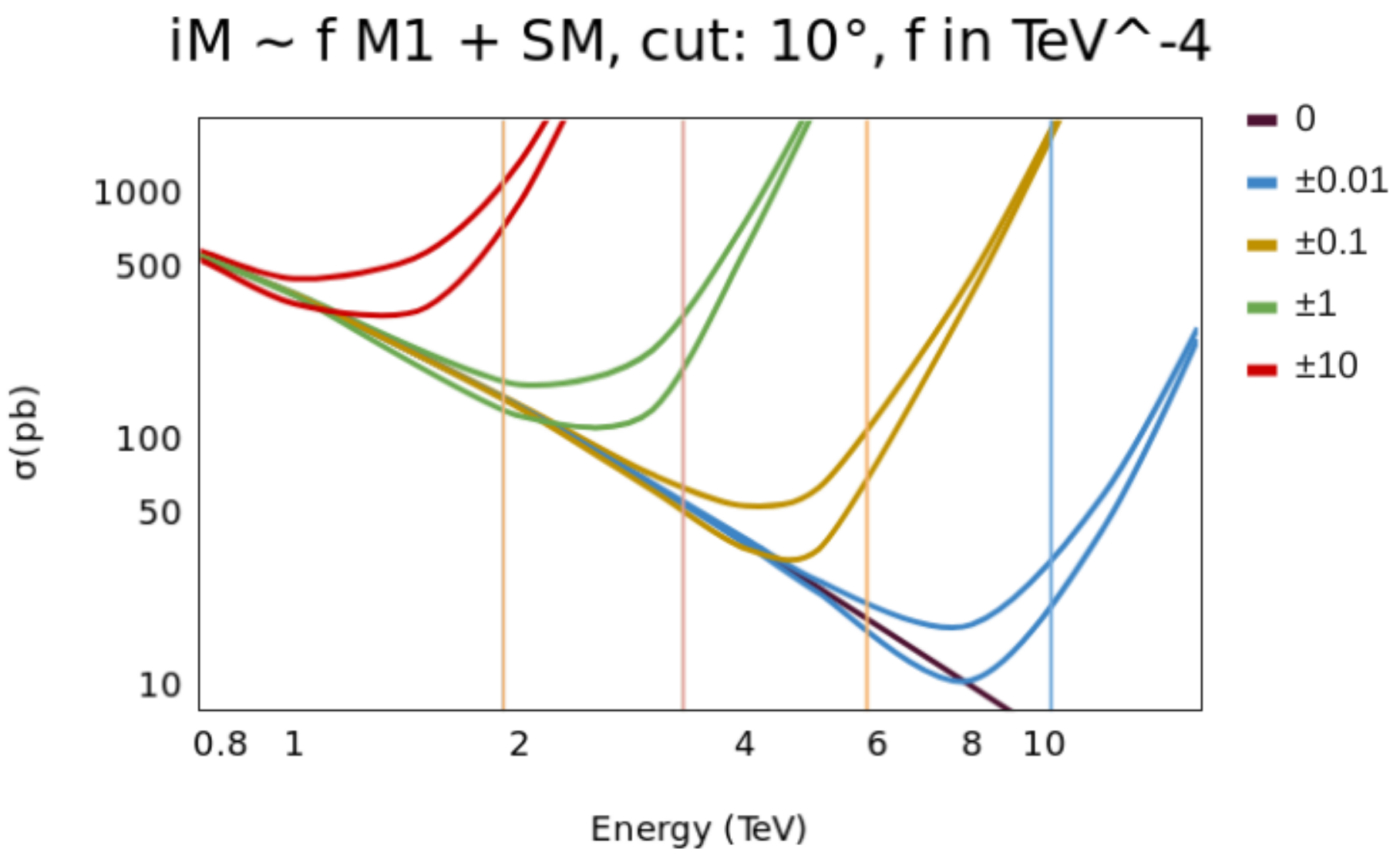} 
& \includegraphics[width=0.5\linewidth]{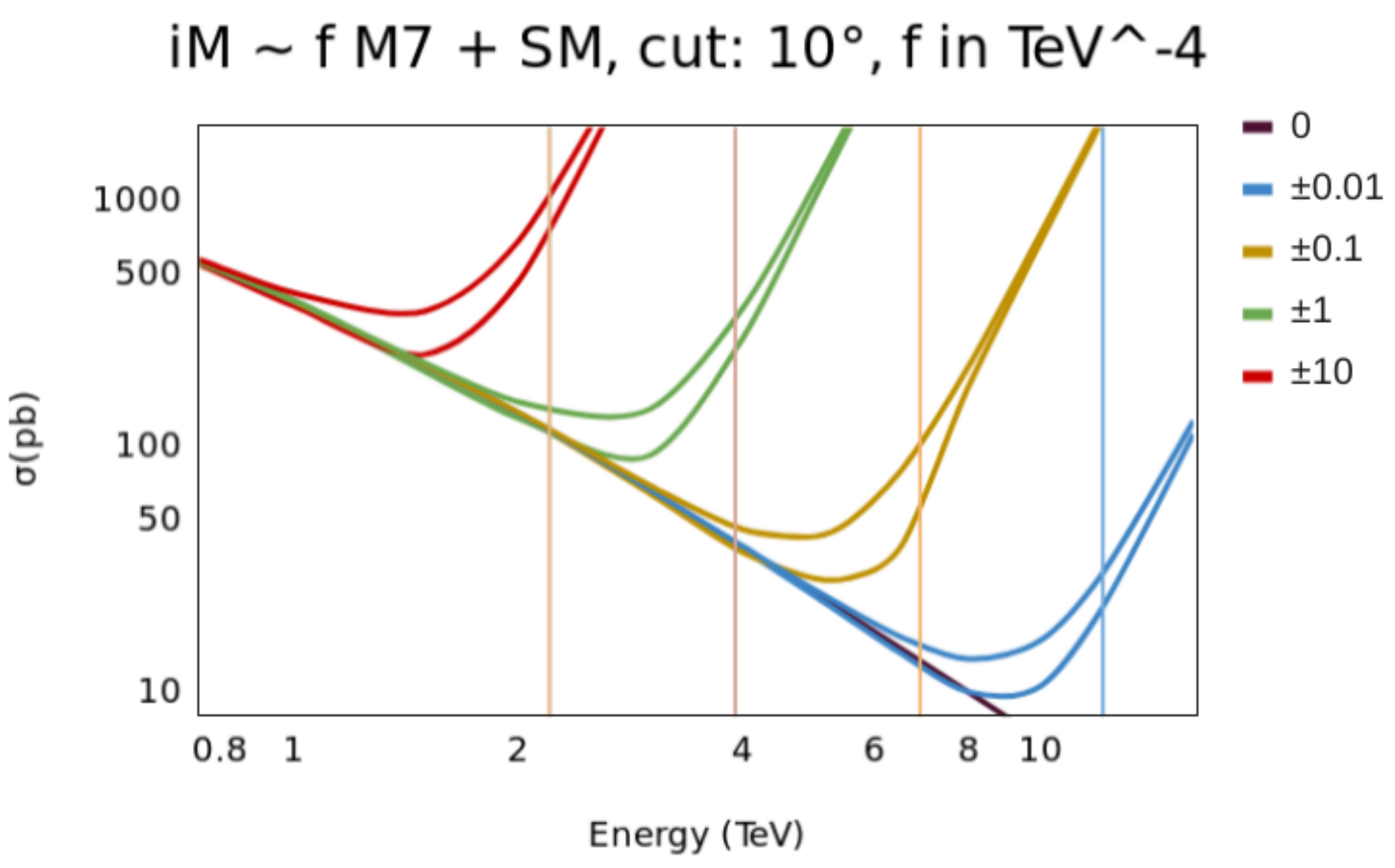} 
  \end{tabular}
\caption{Energy dependence of the total unpolarized $W^+W^+$ cross sections ($E_{CM} \equiv \sqrt{s}$, in TeV) for a chosen set of $f_i$ values.  Vertical lines denote the unitarity bound $\sqrt{s^U}$ (color correspondence).  There is no color distinction between the signs: except for $M1$ and $M7$, upper cross section curves correspond to $f<0$; in $S0,\, S1\,  (T0,\,T2)$ stronger unitarity limits correspond to $f<0$ ($f>0$).}
\label{fig:unpols}
\end{figure}

Since helicity is  an observable for the on-shell $WW$ scattering reaction, different helicity configurations do not interfere among themselves. The total unpolarized elastic on-shell $WW$ cross sections as a function of the center of mass $WW$ energy and its dependence on the $f_i$ sign is shown in Fig.~\ref{fig:unpols}. 
The sign dependence of the total unpolarized cross section, most visible for $T0,\, T2$ and also present for $T1,\, M1,\, M7,\, S0,\, S1$, is due to 
the interference terms in eq.~\eqref{eq:pk3}. More precisely, the dependence on the sign of the $f_i$ is determined by the magnitude of 
SM-BSM terms relative to the BSM$^2$ ones in the region $E\lesssim\Lambda\leq M^U$. While there are always BSM$^2$ 
terms that asymptotically behave as $s^4/\Lambda^8$, the earlier growth as 
$\sim s^2/\Lambda^4$ of the interference terms is not necessarily visible in each of the ``EFT models''.  If the helicity configurations for which the amplitude depends on energy as $s^2$ are not among the saturating helicities of the SM, extra suppression factor(s) of $v/\Lambda<<1$ with respect to the opposite case, will be present in the SM-BSM terms. The latter means suppressed sign dependence of the unpolarized cross section, i.e. suppressed interference. It  can be inferred from the  polarization decomposition plots in figs.~\ref{fig:polsPos},~\ref{fig:polsNeg} that it is the case for the $M0$ operator, and indeed  in Fig.~\ref{fig:unpols} it is seen that the interference effect is practically invisible for this operator.

Although for the off-shell bosons the helicities are not observable and amplitudes with different helicity configurations may interfere,  their interference will be  dumped by different structures of fermionic currents to which they are coupled. 

\subsection{The unitarity bound}
\label{app:unitarity}
The dominating polarization configurations in the total unpolarized cross section can be read from figs.~\ref{fig:polsPos},~\ref{fig:polsNeg}.  However, the helicity combination that determines the $M^U$, i.e. that yields strongest unitarity bound,   is not necessarily  among them. The reason is as follows. The partial wave expansion of  helicity amplitude starts with $J_{\mathrm{min}}=\mathrm{max\{|\lambda_{1}-\lambda_{2}|,|\lambda_{3}-\lambda_{4}|\}}$, where $\lambda_{1,2}$ and  $\lambda_{3,4}$ correspond  initial and final $W$ polarizations,  and it is the $J=J_{\mathrm{min}}$ partial wave that yields the strongest unitarity limit. It has been checked that  for the same-sign $WW$ helicity amplitude that depends on energy as $s^2$ for the case of $M$ operators $J_{\mathrm{min}}=1$, while for  the $S$ and $T$ operators $J_{\mathrm{min}}=0$.  It would imply then that the unitarity limit for the $M$ operators would be weaker than for $S$ and $T$, especially if only the $J=0$ partial waves were considered. 
However, the same operators affect both the same-sign and opposite-sign $WW$ scattering processes, and both processes should be 
considered for the  determination of  the unitarity bounds. In the case of the latter reaction the number of independent helicity 
configurations is 17 resulting from the fact that { one can use all three $\mathcal{C}$, $\mathcal{P}$ and $\mathcal{T}$ symmetries; however one }cannot use the Bose symmetry. In this case for each  ``EFT model'', including the $M$ ones,  there exists a helicity configuration 
that depends on energy as $s^2$ and has $J_{\mathrm{min}}=0$. As a result, for the $M$-type ``EFT model''  the  unitarity limit is  
considerably stronger as compared to the limit derived  from same-sign $WW$ partial wave expansion. This should be kept in mind  
in particular when using a VBFNLO calculator to determine the unitarity bounds that both same- and opposite-sign $WW$ scattering 
processes  are looked at.  
The helicity combination yielding the strongest helicity partial wave unitarity limits for each operator are summed up in Table~\ref{tab:pk}.

\begin{table}
\vspace{8mm}
\begin{center}
\begin{tabular}{|c|c|c|c|c|c|c|c|c|}
\hline
 $i=$ & $S0$ & $S1$ & $T0$ & $T1$ & $T2$ & $M0$ & $M1$ & $M7$ \\ \hline
$f_i>0$ & $0000$ & $0000$ & $----$ and $--++$ & $--++$ & $----$ & $--00$ & $--00$ & $--00$ \\
\hline
$f_i<0$ & $0000$ & $0000$ & $--++$ & $--++$ & $--++$ & $--00$ & $--00$ & $--00$ \\ \hline
\end{tabular}
\end{center}
\vspace{3mm}
\caption{The helicity combinations yielding the strongest helicity partial waves unitarity limits for each operator in case of each sign of $f_i$. It is always a J=0 partial wave that yields the strongest  unitarity limits}
\label{tab:pk}
\end{table}
\section{Extracting the BSM coupling from the discovery regions found}
\label{subsec:matching}
%%%%%%%%%%%%%%%%%%%%%%%%%%%%%%%%%%%%%%%%%%%%%%%%%%%%%%%%%%%%%%%%%%
Assuming that  the departure form the SM predictions is indeed observed at the HL/HE-LHC we turn to the question what can be said about the couplings of higher dimension operators that defined 
the ``EFT model''.  While the probed $\Lambda$ scale can be read off directly from figs.~\ref{fig:compPos},~\ref{fig:compNeg}, the matching between fundamental parameters $C_i$ of a deeper BSM physics and the Wilson coefficients $f_i$ of the low energy approximation is needed to extract the information about couplings.

Let us start with operators that contain the stress tensor $W_{\mu\nu}$. The $W$ bosons, being fundamental SU(2) gauge bosons, would couple  to the to-be-integrated-out BSM states via gauge coupling $g$. Therefore from the corresponding $f_i$ one can factor out $g^2$ for each $W_{\mu\nu}$.  The  Naive Dimensional Analysis (NDA)~\cite{NDApaper} suggests then the following matching  

\begin{equation}
\begin{array}{ll}
	\mathcal{L}\supset & f_{i}\mathcal{O}_{i}\equiv c_{i}\cdot 2\frac{g^2}{\Lambda^4}\mathcal{O}_{i}, \qquad i=M0,M1\\ \mbox{}\\
	&f_{i}\mathcal{O}_{i}\equiv c_{i}\cdot 2^2\frac{g^4}{16\pi^2\Lambda^4}\mathcal{O}_{i}, \qquad i=T0,T1,T2\\ \mbox{}\\
		&f_{i}\mathcal{O}_{i}\equiv c_{i}\cdot 2^2\frac{g^2}{\Lambda^4}\mathcal{O}_{i}, \qquad i=M6,M7
\end{array}
\label{eq:NDA}
\end{equation}
The factor  2   follows from the relation 
$\mathrm{Tr}\hat{W}_{\alpha\beta}\hat{W}_{\mu\nu}=\frac{1}{2}W^i_{\alpha\beta}W^i_{\mu\nu}$  since in NDA the stress tensor  $W^i_{\mu\nu}$ is used for counting purposes rather than the matrix form $\hat{W}_{\mu\nu}$.  Extra factor 2  in case of $M6,\, M7$ operators is due to differences in the SU(2) structure, as can be seen from the relation $\mathcal{O}_{M7}=\frac{1}{2}\mathcal{O}_{M1} + \ldots,$ while $\frac{1}{16\pi^2}$ in front of $T$ operators is a single loop suppression factor suggested by the 4-th power of the electroweak coupling factored out.

The remaining dimensionless factors $c_i$  could be combinations of 
\begin{equation}
\frac{g_\ast}{4\pi}, \;\; \frac{y_\ast}{4\pi}, \; \;\frac{\lambda_\ast}{16\pi^2},
\label{eq:ci}
\end{equation}
where $g_\ast$, $y_\ast$ and $\lambda_\ast$ are some gauge, Yukawa and scalar couplings  of the BSM sector of full theory, respectively. In the weakly-coupled theory one would expect that $c_i$ are naturally of order 1. 

Using~\eqref{eq:NDA} it is straightforward to find the range of $c_i$ corresponding to the discovery regions shown in Fig.~\ref{fig:compPos},~\ref{fig:compNeg}. 
The numerical values, presented in Table~\ref{tab:cminCmax} and~\ref{tab:cminCmax14} for 27 and 14 TeV case, respectively, are found to be roughly consistent. 
\begin{table}[ht]
\vspace{8mm}
\begin{center}
\begin{tabular}{|c|c|c|c|c|c|c|}
\hline
  $f_i>0$ & $T0$ & $T1$ & $T2$ & $M0$ & $M1$ & $M7$ \\ \hline
$c_{min}$--$c_{max}$ & 130.--770.  & 120.--1300. & 670.--2200. & 23.--32. & 45.--133. & 33.--140. \\ \hline
%$g_\ast^{min}$--$g_\ast^{max}$  &--& --& --& --& --& --\\ \hline%2.2--3.4 & 2.2--3.9 & 3.3--4.5 & 6.3--6.8 & 7.4--9.7 & 6.9--9.8 \\ \hline
\end{tabular}
%& g*: 8.4--10. c_i: 0.2--0.46 dla f_i=s0>0 & 27 
\begin{tabular}{|c|c|c|c|c|c|c|c|}
\hline
  $f_i<0$ & $T0$ & $T1$ & $T2$ & $M1$ & $M7$ \\ \hline
$c_{min}$--$c_{max}$  & 110.--1500.  & 140.--2600. & 410.--4500.  & 48.--130. & 54.--270. \\ \hline
%$g_\ast^{min}$--$g_\ast^{max}$  & -- & -- & -- & -- & -- \\ \hline%& 2.2--4.7 & 2.9--5.3  & 7.6--9.7 & 6.5--9.7 \\ \hline
\end{tabular}
\end{center}
\vspace{3mm}
\caption{For each ``EFT model'' characterized by a choice of a single $n=8$ operator shown are the ranges of the overall coefficients $c_i$ in eq.~\ref{eq:NDA} that correspond to the discovery regions found in the 27 TeV study.}
\label{tab:cminCmax}
\end{table}
\begin{table}[ht]
\vspace{8mm}
\begin{center}
\begin{tabular}{|c|c|c|c|c|c|c|}
\hline
  $f_i>0$ & $T0$ & $T1$ & $T2$ & $M0$ & $M1$ & $M7$ \\ \hline
$c_{min}$--$c_{max}$  & 137.--790.  & 76.--1300. & 280.--2200. & 23.-33. & 38.-140. & 24.-130. \\ \hline%& 137.--790.  & 76.--1300. & 280.--2200. & 23.-33. & 38.-140. & 24.-130. \\ \hline
%$g_\ast^{min}$--$g_\ast^{max}$  & -- & -- & -- & -- & -- & -- 
%& 2.4--3.5 & 1.9--3.9 & 2.7--4.5 & 6.2--6.9 & 7.1-9.9 & 6.4--9.8 \\ \hline
\end{tabular}
%& c_i=s0 0.21--0.44; g* 8.5--10.
\begin{tabular}{|c|c|c|c|c|c|c|c|}
\hline
  $f_i<0$ & $T0$ & $T1$ & $T2$ & $M0$ & $M1$ & $M7$\\ \hline
$c_{min}$--$c_{max}$  & 510.--1400.  & 170.--1200. & 700.--4100. & 23.-33. & 45.-140. & 24.-140. \\ \hline
%$g_\ast^{min}$--$g_\ast^{max}$  & -- & -- & -- & -- & -- & -- 
%& 2.6--4.0 & 2.4--3.9 & 3.3--5.2 & 6.2--6.9 & 7.4-9.9 & 6.3--9.8 \\ \hline
\end{tabular}
%& c_i 0.061--0.25; g* 6.2--8.9
\end{center}
\vspace{3mm}
\caption{See Table~\ref{tab:cminCmax} for description; 14 TeV case.}
\label{tab:cminCmax14} 	
\end{table}
However, instead of being of order 1 they are much larger. It suggests that in case of linearly realized spontaneous breaking of $SU(2)\times U(1)$ symmetry our method of probing BSM physics  is sensitive only to strong dynamics. Alternatively one could consider   relaxed assumptions concerning the symmetry breaking mechanism{ For the analysis on ``EFT triangles" in the so-called Higgs Effective Field Theory~\cite{Alonso:2012px}, see~\cite{wwpaperHEFT}.}

We turn now to the discussion of the $S0$ and $S1$ operators. We assume these are generated at loop level in the BSM. Otherwise they would come associated with $n=6$ operators, which are neglected in our analysis. Then, the NDA suggests 
\begin{equation}
	\mathcal{L}\supset f_{i}\mathcal{O}_{i}\equiv c_i\cdot\frac{g_\ast^4}{16\pi^2\Lambda^4}\mathcal{O}_{i},\qquad i=S0,S1 \\ \mbox{}
\label{eq:NDA2}
\end{equation}
Again, $c_i$ are some combinations of BSM couplings and naturally expected to be of order 1. 
If we set $c_i=1$ in eq.~\eqref{eq:NDA2}, then for $f_{S0}>0$ we find that that $g_\ast\in(8.5;10.)$ and $g_\ast\in(8.4;10.)$ in the 14 and 27 TeV case respectively. For $f_{S0}<0$  we find $g_\ast\in(6.2;8.9)$ and $g_\ast\in(7.3;8.8)$ for the 14 and 27 TeV case,  respectively. The coupling is large, but  interestingly it satisfies $g_\ast<4\pi$.
For $f_{S1}$ the discovery regions are empty for both 14 and 27 TeV cases  for both signs. \\

\clearpage


\begin{thebibliography}{nn}

\bibitem{wwpaper}
  J.~Kalinowski, P.~Koz\'ow, S.~Pokorski, J.~Rosiek, M.~Szleper, S.~Tkaczyk, 
  %``Vices and virtues of Higgs effective field theories at large energy,''
	Eur.\ Phys.\ J.\ C\ (2018)\ 78:\ 403,
  doi: 10.1140/epjc/s10052-018-5885-y
  [arXiv:1802.02366 [hep-ph]].

\bibitem{Alboteanu:2008my}
  A.~Alboteanu, W.~Kilian and J.~Reuter,
  %``Resonances and Unitarity in Weak Boson Scattering at the LHC,''
  JHEP {\bf 0811}, 010 (2008)
  doi:10.1088/1126-6708/2008/11/010
  [arXiv:0806.4145 [hep-ph]].
  %%CITATION = doi:10.1088/1126-6708/2008/11/010;%%
 
%\cite{Kilian:2014zja}
 \bibitem{Kilian:2014zja}
  W.~Kilian, T.~Ohl, J.~Reuter and M.~Sekulla,
  %``High-Energy Vector Boson Scattering after the Higgs Discovery,''
  Phys.\ Rev.\ D {\bf 91}, 096007 (2015)
  doi:10.1103/PhysRevD.91.096007
  [arXiv:1408.6207 [hep-ph]].
  %%CITATION = doi:10.1103/PhysRevD.91.096007;%%

%\cite{Eboli:2006wa}
\bibitem{Eboli:2006wa} 
  O.~J.~P.~Eboli, M.~C.~Gonzalez-Garcia and J.~K.~Mizukoshi,
  %``p p ---> j j e+- mu+- nu nu and j j e+- mu-+ nu nu at O( alpha(em)**6) and O(alpha(em)**4 alpha(s)**2) for the study of the quartic electroweak gauge boson vertex at CERN LHC,''
  Phys.\ Rev.\ D {\bf 74}, 073005 (2006)
  doi:10.1103/PhysRevD.74.073005
  [hep-ph/0606118].
  %%CITATION = doi:10.1103/PhysRevD.74.073005;%%
  %132 citations counted in INSPIRE as of 24 Sep 2019
	
 
 

\bibitem{Degrande:2013rea}
  C.~Degrande {\it et al.},
  %``Monte Carlo tools for studies of non-standard electroweak gauge boson interactions in multi-boson processes: A Snowmass White Paper,''
  [arXiv:1309.7890 [hep-ph]].
  %%CITATION = ARXIV:1309.7890;%%
  %28 citations counted in INSPIRE as of 21 Oct 2017 
	
		
\bibitem{MadGraph}
  J.~Alwall {\it et al.},
  % ``The automated computation of tree-level and next-to-leading order differential cross sections, and their matching to parton shower simulations''
  JHEP {\bf 07} (2014) 079,
  doi:10.1007/JHEP07(2014)079
  [arXiv:1405.0301 [hep-ph]].
	
\bibitem{Pythia6}
  T.~Sjostrand, S.~Mrenna, P.~Skands,
  % PYTHIA 6.4 Physics and Manual
  JHEP {\bf 0605} (2016) 026,
  doi:10.1088/1126-6708/2006/05/026
  [arXiv:hep-ph/0603175].
	
\bibitem{VBFNLO}
  K.~Arnold {\it et al.},
  % ``VBFNLO: A parton level Monte Carlo for processes with electroweak bosons''
  Comput.\ Phys.\ Commun.\ {\bf 180} (2009) 1661,
  doi:10.1016/j.cpc.2009.03.006
  [arXiv:0811.4559 [hep-ph]];
  J.~Baglio {\it et al.}
  % ``VBFNLO: A parton level Monte Carlo for processes with electroweak bosons -- Manual for Version 2.7.0''
  arXiv:1107.4038 [hep-ph],
  % ``Release Note - VBFNLO 2.7.0''
  arXiv:1404.3940 [hep-ph].

\bibitem{Sirunyan:2017ret} 
  A.~M.~Sirunyan {\it et al.} [CMS Collaboration],
  %``Observation of electroweak production of same-sign W boson pairs in the two jet and two same-sign lepton final state in proton-proton collisions at $\sqrt{s} = $ 13 TeV,''
  Phys.\ Rev.\ Lett.\  {\bf 120}, no. 8, 081801 (2018)
  doi:10.1103/PhysRevLett.120.081801
  [arXiv:1709.05822 [hep-ex]].
  %%CITATION = doi:10.1103/PhysRevLett.120.081801;%%
  %77 citations counted in INSPIRE as of 12 Dec 2019
	
\bibitem{Aaboud:2019nmv} 
  M.~Aaboud {\it et al.} [ATLAS Collaboration],
  %``Observation of electroweak production of a same-sign $W$ boson pair in association with two jets in $pp$ collisions at $\sqrt{s}=13$ TeV with the ATLAS detector,''
  Phys.\ Rev.\ Lett.\  {\bf 123}, no. 16, 161801 (2019)
  doi:10.1103/PhysRevLett.123.161801
  [arXiv:1906.03203 [hep-ex]].
  %%CITATION = doi:10.1103/PhysRevLett.123.161801;%%
  %13 citations counted in INSPIRE as of 12 Dec 2019
	


\bibitem{Snowmass2013}
  C.~Degrande, J.~L.~Holzbauer, S.-C.~Hu, A.~V.~Kotwal, S.~Li, M.~Marx, O.~Mattelaer, J.~Metcalfe, M.-A.~Pleier, C.~Pollard, M.~Rominsky, D.~Wackeroth,
  %``Studies of Vector Boson Scattering and Triboson Production with DELPHES Parametrized FastSimulation for Snowmass 2013''
  arXiv:1309.7452 [physics.comp-ph].



%\cite{Biedermann:2017bss}
\bibitem{Biedermann:2017bss} 
  B.~Biedermann, A.~Denner and M.~Pellen,
  %``Complete NLO corrections to W$^{+}$W$^{+}$ scattering and its irreducible background at the LHC,''
  JHEP {\bf 1710}, 124 (2017)
  doi:10.1007/JHEP10(2017)124
  [arXiv:1708.00268 [hep-ph]].
  %%CITATION = doi:10.1007/JHEP10(2017)124;%%
  %30 citations counted in INSPIRE as of 26 Sep 2019

\bibitem{Azzi:2019yne} 
  A. Denner and M. Pellen, Section 4.2.2. in P.~Azzi {\it et al.} [HL-LHC Collaboration and HE-LHC Working Group],
  %``Standard Model Physics at the HL-LHC and HE-LHC,''
  arXiv:1902.04070 [hep-ph].
  %%CITATION = ARXIV:1902.04070;%%
  %44 citations counted in INSPIRE as of 17 Nov 2019


%\cite{Zhang:2018shp}
\bibitem{Zhang:2018shp} 
Q.~Bi, C.~Zhang and S.~Y.~Zhou,
  %``Positivity constraints on aQGC: carving out the physical parameter space,''
  JHEP {\bf 1906}, 137 (2019)
  doi:10.1007/JHEP06(2019)137
  [arXiv:1902.08977 [hep-ph]].
  %%CITATION = doi:10.1007/JHEP06(2019)137;%%
  %2 citations counted in INSPIRE as of 03 Oct 2019



\bibitem{Maina:1710}
  A.~Ballestrero, E.~Maina and G.~Pelliccioli,
  %``W boson polarization in vector boson scattering at the LHC''
	J.\ High\ Energ.\ Phys.\ (2018)\ 2018:\ 170,
	doi: 10.1007/JHEP03(2018)170
  [arXiv:1710.09339 [hep-ph]].
  %%CITATION = %%
  %

  %%CITATION = doi:10.1103/PhysRevD.91.055029;%%
  %73 citations counted in INSPIRE as of 05 Dec 2017


\bibitem{NDApaper}
  B.~M.~Gavela, E.~E.~Jenkins, A.~V.~Manohar and L.~Merlo,
  %``Analysis of General Power Counting Rules in Effective Field Theory,''
  Eur.\ Phys.\ J.\ C {\bf 76} (2016) no.9,  485
  doi:10.1140/epjc/s10052-016-4332-1
  [arXiv:1601.07551 [hep-ph]].
  %%CITATION = doi:10.1140/epjc/s10052-016-4332-1;%%
  %51 citations counted in INSPIRE as of 16 Jan 2019

\bibitem{Alonso:2012px} 
  R.~Alonso, M.~B.~Gavela, L.~Merlo, S.~Rigolin and J.~Yepes,
  %``The Effective Chiral Lagrangian for a Light Dynamical "Higgs Particle",''
  Phys.\ Lett.\ B {\bf 722}, 330 (2013)
  Erratum: [Phys.\ Lett.\ B {\bf 726}, 926 (2013)]
  doi:10.1016/j.physletb.2013.04.037, 10.1016/j.physletb.2013.09.028
  [arXiv:1212.3305 [hep-ph]].
  %%CITATION = doi:10.1016/j.physletb.2013.04.037, 10.1016/j.physletb.2013.09.028;%%
  %150 citations counted in INSPIRE as of 16 Jan 2019

%\cite{Kozow:2019txg}
\bibitem{wwpaperHEFT} 
  P.~Koz\'ow, L.~Merlo, S.~Pokorski and M.~Szleper,
  %``Same-sign WW Scattering in the HEFT: Discoverability vs. EFT Validity,''
  JHEP {\bf 1907}, 021 (2019)
  doi:10.1007/JHEP07(2019)021
  [arXiv:1905.03354 [hep-ph]].
  %%CITATION = doi:10.1007/JHEP07(2019)021;%%
  %2 citations counted in INSPIRE as of 26 Sep 2019

%
%%\cite{Azatov:2016sqh}
%\bibitem{Azatov:2016sqh} 
  %A.~Azatov, R.~Contino, C.~S.~Machado and F.~Riva,
  %%``Helicity selection rules and noninterference for BSM amplitudes,''
  %Phys.\ Rev.\ D {\bf 95}, no. 6, 065014 (2017)
  %doi:10.1103/PhysRevD.95.065014
  %[arXiv:1607.05236 [hep-ph]].
  %%%CITATION = doi:10.1103/PhysRevD.95.065014;%%
  %%38 citations counted in INSPIRE as of 16 Dec 2018

%\cite{Gavela:2016bzc}

	
	%\cite{Alonso:2012px}

	

  %``Analysis of General Power Counting Rules in Effective Field Theory,''
	%\cite{Alboteanu:2008my}
	
 
 


\end{thebibliography}
\end{document}